\documentclass[article]{jss}\usepackage[]{graphicx}\usepackage[]{xcolor}
\makeatletter
\def\maxwidth{ %
  \ifdim\Gin@nat@width>\linewidth
    \linewidth
  \else
    \Gin@nat@width
  \fi
}
\makeatother

\definecolor{fgcolor}{rgb}{0.345, 0.345, 0.345}
\newcommand{\hlnum}[1]{\textcolor[rgb]{0.686,0.059,0.569}{#1}}%
\newcommand{\hlsng}[1]{\textcolor[rgb]{0.192,0.494,0.8}{#1}}%
\newcommand{\hlcom}[1]{\textcolor[rgb]{0.678,0.584,0.686}{\textit{#1}}}%
\newcommand{\hlopt}[1]{\textcolor[rgb]{0,0,0}{#1}}%
\newcommand{\hldef}[1]{\textcolor[rgb]{0.345,0.345,0.345}{#1}}%
\newcommand{\hlkwa}[1]{\textcolor[rgb]{0.161,0.373,0.58}{\textbf{#1}}}%
\newcommand{\hlkwb}[1]{\textcolor[rgb]{0.69,0.353,0.396}{#1}}%
\newcommand{\hlkwc}[1]{\textcolor[rgb]{0.333,0.667,0.333}{#1}}%
\newcommand{\hlkwd}[1]{\textcolor[rgb]{0.737,0.353,0.396}{\textbf{#1}}}%

\usepackage{framed}
\makeatletter
\newenvironment{kframe}{%
 \def\at@end@of@kframe{}%
 \ifinner\ifhmode%
  \def\at@end@of@kframe{\end{minipage}}%
  \begin{minipage}{\columnwidth}%
 \fi\fi%
 \def\FrameCommand##1{\hskip\@totalleftmargin \hskip-\fboxsep
 \colorbox{shadecolor}{##1}\hskip-\fboxsep
     \hskip-\linewidth \hskip-\@totalleftmargin \hskip\columnwidth}%
 \MakeFramed {\advance\hsize-\width
   \@totalleftmargin\z@ \linewidth\hsize
   \@setminipage}}%
 {\par\unskip\endMakeFramed%
 \at@end@of@kframe}
\makeatother

\definecolor{shadecolor}{rgb}{.97, .97, .97}
\definecolor{messagecolor}{rgb}{0, 0, 0}
\definecolor{warningcolor}{rgb}{1, 0, 1}
\definecolor{errorcolor}{rgb}{1, 0, 0}
\newenvironment{knitrout}{}{} 

\usepackage{alltt}

\usepackage{framed}
\usepackage{alltt}
\IfFileExists{upquote.sty}{\usepackage{upquote}}{}
\makeatletter
\@ifundefined{knitrout}{\newenvironment{knitrout}{}{} }{}
\@ifundefined{kframe}{%
\newenvironment{kframe}{%
\def\at@end@of@kframe{}%
\ifinner\ifhmode%
\def\at@end@of@kframe{\end{minipage}}%
\begin{minipage}{\columnwidth}%
\fi\fi%
\def\FrameCommand##1{\hskip\@totalleftmargin \hskip-\fboxsep
\colorbox{shadecolor}{##1}\hskip-\fboxsep
    \hskip-\linewidth \hskip-\@totalleftmargin \hskip\columnwidth}%
\MakeFramed{\advance\hsize-\width
\@totalleftmargin\z@ \linewidth\hsize
\@setminipage}}%
{\par\unskip\endMakeFramed%
\at@end@of@kframe}}{}
\makeatother

\usepackage{booktabs}
\usepackage{bm}
\usepackage{paralist}
\usepackage{lmodern}
\usepackage{rotating}

\newcommand{\tn}[1]{\textnormal{#1}}

\newcommand{\bfx}{\mathbf{x}}

\newcommand{\bfz}{\mathbf{z}}

\newcommand{\mcZ}{\mathcal{Z}}

\newcommand{\drm}{\mathrm{d}}

\newcommand{\tz}{\ensuremath{t_z}}
\newcommand{\tlag}{t_{\text{lag}}}
\newcommand{\tlead}{t_{\text{lead}}}
\newcommand{\tw}[1]{\ensuremath{\mathcal{T}_{#1}(t)}}

\newcommand{\Rlang}{\textbf{\textsf{R}}}

\usepackage{array}
\newcolumntype{L}[1]{>{\raggedright\let\newline\\\arraybackslash\hspace{0pt}}m{#1}}
\newcolumntype{C}[1]{>{\centering\let\newline\\\arraybackslash\hspace{0pt}}m{#1}}
\newcolumntype{R}[1]{>{\raggedleft\let\newline\\\arraybackslash\hspace{0pt}}m{#1}}

\usepackage[markup=underlined]{changes}

\usepackage{amsmath,amssymb}
\usepackage{tikz}
\usepackage{array}
\usepackage{geometry}
\usepackage{longtable}
\usepackage{caption}
\usepackage{prodint}
\usepackage{todonotes}
\newcommand{\yes}{$\boldsymbol{+}$}
\newcommand{\no}{$\boldsymbol{-}$}
\newcommand{\lmark}{$\circ$}
\newcommand{\CIFk}{\operatorname{CIF}_k}

\newtheorem{rexample}{\Rlang-chunk}

\author{Johannes Piller\\Munich Center for\\Machine Learning,\\LMU Munich 
  \And Fabian Scheipl\\Munich Center for\\Machine Learning,\\LMU Munich 
  \And Philipp Kopper\\Munich Center for\\Machine Learning,\\LMU Munich
  \And Andreas Bender\\Munich Center for\\Machine Learning,\\LMU Munich}
\Plainauthor{Johannes Piller, Fabian Scheipl, Philipp Kopper, Andreas Bender}

\title{\pkg{pammtools}: Piecewise Exponential Additive Mixed Modeling Tools}
\Plaintitle{pammtools: Piecewise Exponential Additive Mixed Modeling Tools}

\Abstract{ 
\noindent Piecewise exponential additive mixed models (PAMMs) provide a flexible framework for analyzing censored and truncated time-to-event data, bridging classical hazard-based modeling with modern regression techniques. 
They enable the estimation of complex covariate effects, including non-linear and time-varying (cumulative) effects, and naturally incorporate time-varying covariates. 
Moreover, PAMMs are applicable across a wide range of survival settings, including non-proportional hazards, recurrent events, competing risks, and multi-state analyses. 
This article introduces the \pkg{pammtools} package, which facilitates data transformation, estimation, and interpretation for PAMMs within a unified workflow. 
The package provides a comprehensive, user-friendly, and extensible interface covering the full modeling pipeline, from data transformation to estimation and visualization.
In addition, simulation-based inference allows the calculation of confidence intervals for arbitrary quantities of interest such as covariate dependent hazards, survival probabilities, restricted mean survival times and transition probabilities.
} 
\Keywords{survival analysis, competing risks, event-history analysis, time-varying effects, time-varying covariates, cumulative effects, distributed lags, exposure-lag-response associations} 
\Plainkeywords{survival analysis, competing risks, event-history analysis, time-varying effects, time-varying covariates, cumulative effects, exposure-lag-response associations} 

\Address{
  Johannes Piller\\
  Department of Statistics\\
  LMU Munich\\
  Ludwigstrasse 33\\
  80539 Munich, Germany\\
  E-mail: \email{johannes.piller@lmu.de}
}

\IfFileExists{upquote.sty}{\usepackage{upquote}}{}
\begin{document}


\newpage
\section{Introduction}

\begin{figure}[!ht]
    \centering
\includegraphics[width=1\linewidth]{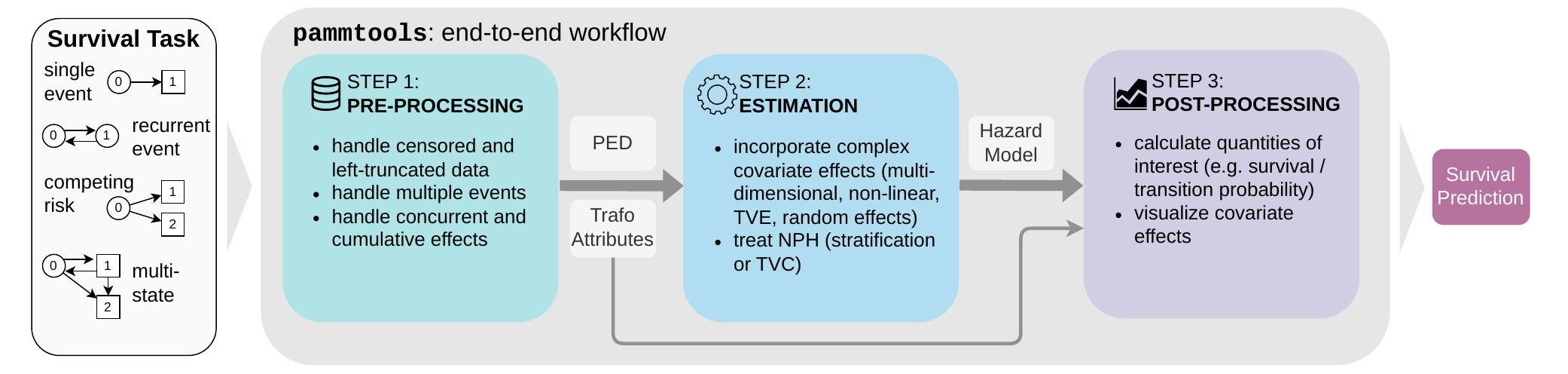}
    \caption{Schematic representation of the data analysis workflow with \pkg{pammtools}. \pkg{pammtools} is able to process diverse types of survival data as input, transforms the data into piecewise-exponential data (PED), then estimates complex covariate-conditional hazards, and finally calculates quantities of interest and offers convenience functions to visualize previous results.
    \label{fig:pammtools}}
\end{figure}

This article introduces the \pkg{pammtools} package (\href{https://adibender.github.io/pammtools/}{https://adibender.github.io/pammtools/}).
It provides functions to facilitate the estimation and interpretation of a flexible model class for time-to-event and event history analysis (cf. Figure~\ref{fig:pammtools}), which we refer to as \emph{P}iecewise exponential \emph{A}dditive \emph{M}ixed \emph{M}odel \citep[PAMM; ][]{Bender2018b}.

PAMMs provide a flexible and unified framework for survival analysis,
\begin{compactitem}
  \item facilitating flexible modeling of complex covariate effects, including non-linear, multi-dimensional, random, and spatial effects;
  \item accommodating departures from proportional hazards through stratification and time-varying effects (including non-linear specifications)
  \item supporting time-varying covariates with cumulative (distributed lag) and/or non-linear effects;
  \item handling right-censored and left-truncated data, as well as -- via multiple imputation -- interval-censored single-event and competing risks data;
  \item enabling the estimation of (transition) hazards and any quantities derived from them across a wide range of survival settings, including single-event, recurrent event, competing risks, and multi-state analyses.
\end{compactitem}

From a survival analysis point of view, PAMMs are multiplicative hazards models like the Cox model \citep{cox1972regressionmodels}, however, the baseline hazard is estimated explicitly, without making (strong) parametric assumptions about the distribution of event times.
From a statistical modeling point of view, PAMMs can be represented as generalized additive mixed models (GAMM), enabling the use of established tools for estimation and inference that are highly performant and robust and continuously being developed further.
This combination of flexibility and computational tractability makes PAMMs particularly attractive for modern applications involving complex time-dependent structures and large-scale data.

Several \proglang{R} packages address related survival analysis tasks: \pkg{flexsurv} \citep{Jackson2016flexsurv} offers parametric and spline-based models; \pkg{mstate} \citep{dewreede2011mstate} focuses on multi-state data; \pkg{rstpm2} \citep{Liu2018rstpm2} supports penalized spline hazards; \pkg{casebase} \citep{Bhatnagar2022casebase} provides a case-base sampling approach with GAM-based smoothing; and \pkg{dlnm} \citep{Gasparrini2011} specializes in distributed lag non-linear models.
\pkg{pammtools} complements these by offering a unified framework that integrates all three stages of the modeling pipeline across a broad range of survival settings, while uniquely combining support for cumulative (weighted exposure) effects, a fully \pkg{tidyverse}-compliant UI, and simulation-based confidence intervals for arbitrary functionals of the hazard (cf.\ Figure~\ref{fig:pammtools}):
\begin{compactenum}
\item \textbf{Data pre-processing}: This can be more or less involved, depending on the events to be modeled (single-event, competing risks, multi-state, etc.; cf.\ Section~\ref{sec:practical_workflow}) and on the type of effects to be estimated (e.g.,\ time-varying covariates and cumulative effects; cf.\ Section~\ref{sec:extended_topics}).
\item \textbf{Estimation}: \pkg{pammtools} is optimized for use with \pkg{mgcv} \citep{Wood2011}, leveraging its established infrastructure for smooth effect estimation, regularization, and scalable computation. However, any method that can optimize a Poisson likelihood with offsets can be used, including e.g., model-based boosting via \pkg{mboost} \citep{Hothorn2006, Hofner2012} or tree-based boosting via \pkg{xgboost} \citep{xgboost}
\item \textbf{Model post-processing}: This includes visualization of covariate effects, calculation of estimated hazard rates or cumulative hazards as well as survival and transition probabilities, which all need to take into account the specific data structure of PAMMs. Currently, most post-processing and visualization functions in \pkg{pammtools} are customized to work with \code{mgcv::gam} objects. However, due to its modularity, the package can easily be extended as shown in recent package extensions supporting shape-contrained additive models via \pkg{scam} \citep{pya2015scam} or the package vignettes on \href{https://adibender.github.io/pammtools/articles/bayesian.html}{Bayesian PAMM inference} and \href{https://adibender.github.io/pammtools/articles/xgboost-backend.html}{PAMMs using gradient-boosted trees.}
\end{compactenum}

The remainder of this article is organized as follows: Section~\ref{sec:pamm} briefly defines PAMMs and introduces the notation for single-event  (Section~\ref{sub:singleevent}), competing risk (Section~\ref{sub:cr}), and event history analyses (Section~\ref{sub:eha}) used throughout this article.
Section~\ref{sec:practical_workflow} illustrates the general workflow of fitting PAMMs with \pkg{pammtools}, focusing on practical applications and comparison to estimates obtained from other established models.
Section~\ref{sec:extended_topics} demonstrates the data transformations necessary to fit PAMMs in different scenarios, i.e., for data with time-constant covariates (TCCs, Section~\ref{sub:tcc}), with time-varying covariates (TVCs, Section~\ref{sub:tvc}) and cumulative effects of TVCs (Section~\ref{sub:elra}), by showcasing examples on real and simulated data to illustrate the estimation, visualization and interpretation of the different effect types, facilitated by convenience functions provided in \pkg{pammtools}; in addition, Section~\ref{sub:ic} demonstrates the estimation of PAMMs for interval-censored event times via multiple imputation. Section~\ref{sec:comparison} provides a detailed feature comparison of packages, which partially overlap in scope. Finally, Section~\ref{sec:pammtools:discussion} summarizes \pkg{pammtools}'s functionality, discusses limitations, and gives a brief outlook on potential future extensions.

For the code examples, the following packages will be used:

\begin{knitrout}\small
\definecolor{shadecolor}{rgb}{0.961, 0.961, 0.961}\color{fgcolor}\begin{kframe}
\begin{alltt}
\hlkwd{library}\hldef{(}\hlsng{"dplyr"}\hldef{);} \hlkwd{library}\hldef{(}\hlsng{"etm"}\hldef{);} \hlkwd{library}\hldef{(}\hlsng{"ggplot2"}\hldef{);} \hlkwd{library}\hldef{(}\hlsng{"mgcv"}\hldef{);}
\hlkwd{library}\hldef{(}\hlsng{"mstate"}\hldef{);} \hlkwd{library}\hldef{(}\hlsng{"pammtools"}\hldef{);} \hlkwd{library}\hldef{(}\hlsng{"purrr"}\hldef{);} \hlkwd{library}\hldef{(}\hlsng{"survival"}\hldef{);}
\hlkwd{library}\hldef{(}\hlsng{"tidyr"}\hldef{)}
\end{alltt}
\end{kframe}
\end{knitrout}

\section{Piecewise exponential additive mixed models}
\label{sec:pamm}
\begin{figure}[!ht]
      \centering
  \includegraphics[width=.6\linewidth]{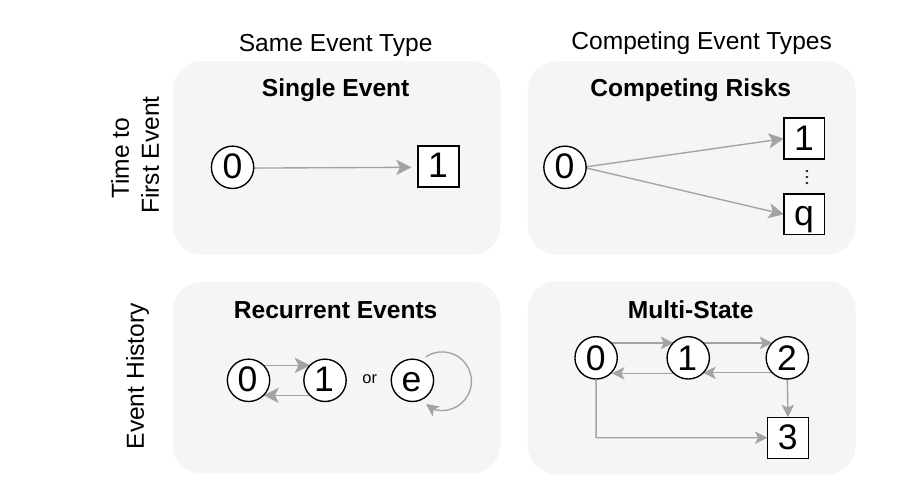}
      \caption{Schematic representation of different model types. Circles indicate transient states (i.e., states that can be exited), while squares represent absorbing states (from a data collection or modeling perspective).
  }
      \label{fig:msm-state-diagram}
  \end{figure}

In this article, we consider models for survival analysis and event-history analysis as depicted schematically in Figure~\ref{fig:msm-state-diagram}, where different models are distinguished by the type of transitions allowed between different transient (circles) and absorbing (squares) states. 
While all models can be considered special cases of the multi-state process, the reduction to special cases usually leads to simpler model representation, estimation and interpretation. Details of individual models will be discussed in Sections~\ref{sub:singleevent}, \ref{sub:cr} and~\ref{sub:eha}. Recurrent events analysis in the PAMM framework is treated in depth by \cite{ramjith2024recurrent} and is omitted here; it is naturally handled as a multi-state process, particularly when terminal competing events are present.

Across all of these settings, estimation rests on the same approximation argument: any hazard function can be approximated arbitrarily well by a piecewise constant function given sufficient data \citep{Whitehead1980, Holford1980}; these piecewise constant rates can then be estimated by fitting a Poisson model to suitably transformed data; and while such estimates can be sensitive to the number and placement of interval cut-points, penalized spline estimation stabilizes the estimates and improves computational efficiency \citep{Bender2018b}. The following sections develop this argument for each model class in turn.

\subsection{Single event analysis}
\label{sub:singleevent}

Let $(t_i, \delta_i)$ be the tuple of the observed time and status indicator for subject $i=1,\ldots,n$.
In this section, we consider models with hazard rates given by \eqref{eq:hazard} and in the log-linear form by \eqref{eq:loghazard}

\begin{align}
h_i(t; \bfx_i, \mcZ_i, \ell_i)
  & = h_0(t)\exp\left[\sum_{p=1}^P f_p(x_{i,p}, t) +
    \sum_{m=1}^M g(\bfz_{i,m}, t) + b_{\ell_i}\right]\label{eq:hazard}&\ \\
\log(h_i(t; \bfx_i, \mcZ_i, \ell_i))
  & = \beta_0 + f_0(t) + \sum_{p=1}^P f_p(x_{i,p}, t) +
    \sum_{m=1}^M g(\bfz_{i,m}, t) + b_{\ell_i},\label{eq:loghazard}
\end{align}

where $\log(h_0(t)) = \beta_0 + f_0(t)$ is the (log-)baseline hazard, $\bfx_i$ is the vector of time-constant covariates $x_{i,p},p=1,\ldots,P$ and $\mcZ_i = \{\bfz_{i,m}:m=1\ldots,M\}$ is the set of $M$ time-varying covariate vectors (exposure histories) $\bfz_{i,m}=\{z_{i,m}(t_{z_m,q_m}):q_{m}=1,\ldots, Q_{m}\}$, where $t_{z_m, q_m}$ denotes the (exposure) time points at which covariate $z_m$ was observed and $Q_m$ is the number of observed time-points for the $m$-th time-varying covariate.
It is important to stress the difference between $t$, which denotes the time scale on which the event times are observed and $t_z$, which denotes the time scale on which time-varying covariate $z$ is observed.
The time scales of $t$ and $t_z$ do not need to be identical nor even overlap, nor do they have to be measured in the same units (see Section~\ref{sub:elra} for examples).
\\

In the following paragraphs, we briefly describe the individual components in \eqref{eq:loghazard}.
A tutorial-style exposition of the model without the $g(\bfz,t)$ terms is given in \cite{Bender2018b} and a general framework for models with cumulative effects $g(\bfz,t)$ is described \cite{Bender2018a}.
\\

The terms $f_p(x_{i,p}, t)$ denote (potentially time-varying) effects (TVEs) of time-constant covariates $\bfx_{.,p}$, and our notation subsumes the entire range of effects of this kind, i.e., from time-constant linear effects all the way to non-linearly time-varying non-linear effect surfaces and everything in between.
A selection of possible TVEs along with their specification for estimation with \code{mgcv::gam} are summarized in Table~\ref{tab:tvEffectsPAM}.
Note that models with multiple time-varying effects may need to impose additional identifiability constraints \citep[Ch. 5.6.3]{Wood2017a}, see \code{?mgcv::ti} and the examples in the \pkg{pammtools} \href{https://adibender.github.io/pammtools/articles/tveffects.html}{vignette on time-varying effects}.
Also note that (non-linear, non-linearly time-varying) interaction effects of multiple covariates can be specified and estimated in the same way in this framework.\\

\begin{table}[!htbp]
\caption[TVE specification]{Selection of possible $f(x_{i,p},t)$ effect
specifications in PAMMs including \Rlang{} code for \code{mgcv::gam} fits.
We write \code{x} for any covariate of interest in the data set and \code{t} for
time in each interval. The latter can be 
interval end-points $t_j := \kappa_{j}$ or mid-points
$t_j := \kappa_{j-1} + \tfrac{1}{2}(\kappa_{j}-\kappa_{j-1}) = \tfrac{1}{2}(\kappa_{j-1} + \kappa_{j})$ (e.g.).}
\label{tab:tvEffectsPAM}
\begin{center}
  \resizebox{\textwidth}{!}{%
    \begin{tabular}{lll}
    \toprule
      $f(x_{i,p}, t) =$ & Effect Type & \code{mgcv::gam} Call\\
    \midrule
      $\beta_p x_{i,p}$ & linear and time-constant  & \code{... + x + ...} \\
      $f_p(x_{i,p})$     & smooth nonlinear and time-constant & \code{... + s(x) + ...}\\
      $\beta_p x_{i,p} + \beta_{p:t}\cdot x_{i,p} \cdot t$ & linear and linearly
        time-varying  & \code{... + x + x:t ... }\\
      $f_p(x_{i,p})\cdot t$ & smooth and linearly time-varying & \code{... + s(x, by=t) + ...}\\
      $x_{i,p}\cdot f_p(t)$ & linear and smoothly time-varying &\code{... + s(t,by=x) + ...}\\
      $f_p(x_{i,p}, t)$     & smooth and smoothly time-varying &\code{... + te(x,t) + ...}\\
    \bottomrule
    \end{tabular}%
  }
\end{center}
\end{table}

The terms $g(\bfz_{i,m},t)$ are potentially (non-linearly) time-varying, potentially cumulative effects of time-varying covariates $\bfz_{.,m}$.
Such terms are discussed in more detail in Sections \ref{subsub:tvc:prep} (pre-processing) and \ref{subsub:tvc:mod} (modeling) as well as the \href{https://adibender.github.io/pammtools/articles/tdcovar.html#analysis-of-the-pbc-data}{package vignette on time-varying covariates}.
\\

The term $b_{\ell_i}$ denotes random effects (a log-normal frailty) associated with group $\ell=1,\ldots,L$ to which subject $i$ belongs.
Extensions to more complex random effect structures for nested or crossed groups or spatial effects are possible within the presented framework as well \citep[e.g.,][]{Wood2017c,pedersenHierarchicalGeneralizedAdditive2019}.
For examples of random effect models estimated via PAMMs see the \href{https://adibender.github.io/pammtools/articles/frailty.html}{package vignette on frailty effects}.
In the following, we omit the random effect terms to focus on time-varying and cumulative effects rather than hierarchical models.

To estimate model \eqref{eq:hazard} using PAMMs, the time under risk is divided into $J$ intervals with interval cut points $\kappa_0 < \ldots < \kappa_J$ that define intervals $(\kappa_{j-1}, \kappa_{j}], j=1\ldots, J$.
The hazard rate $h(t)$ is approximated by piecewise constant hazards $h(t) \equiv h(t_j) \,\forall\, t\in (\kappa_{j-1}, \kappa_{j}]$ where $t_j \in (\kappa_{j-1}, \kappa_j]$ denotes any fixed time point in the $j$-th interval, (typically $t_j:=\kappa_j$), such that for $i = 1,\ldots,n$
\begin{align}
\log(h_i(t; \bfx_i, \mcZ_i)) & \approx
  \log(h_{ij}):=\log(h_i(t_j;\bfx_i, \mcZ_i)) \label{eq:lambda_ij}\\
& = \beta_0 + f_0(t_j) + \sum_{p=1}^P f_p(x_{i,p}, t_j) +
    \sum_{m=1}^M g(\bfz_{i,m}, t_j). \nonumber
\end{align}
Piecewise constant hazard rates imply a piecewise exponential distribution of event times (thus: PEM and PAMM), but note that any shape of the conditional hazard rate can be approximated arbitrarily well by a suitably defined step function.

\begin{figure}[!h]
\begin{knitrout}\small
\definecolor{shadecolor}{rgb}{0.961, 0.961, 0.961}\color{fgcolor}

{\centering \includegraphics[width=\maxwidth]{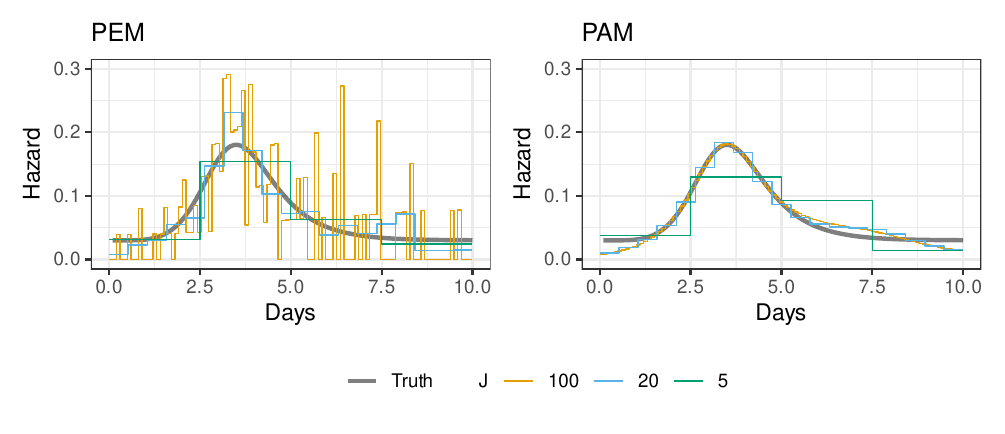} 

}

\end{knitrout}
\caption{\label{fig:pemvspam-haz-illustration} PEM baseline hazard estimates (left panel) vs. PAM baseline hazard estimates (right panel) for a different number of cut-points, i.e., 5 (green), 20 (blue), and 100 (red), compared with the true hazard (black).}
\end{figure}

In the classical PEM \citep{Whitehead1980}, the number of intervals $J$ as well as the positioning of cut points $\kappa_j$ are important parameters that affect the quality of the approximation \citep{Demarqui2008}.
This is less important for PAMMs as long as $J$ is not too small and $\kappa_j$ are sufficiently dense in areas where $h(t; \bfx, \mcZ)$ varies more quickly. To illustrate this behavior qualitatively,  Figure~\ref{fig:pemvspam-haz-illustration} compares a PEM and PAM on a simple synthetic dataset with $n=250$ event times for example. A light-weight simulation study, which investigates this behavior for multiple repetitions can be found on the \href{https://adibender.github.io/simpamm/pem_vs_pam.html}{\pkg{simpamm} package homepage}.

For $J = 5$, both PEM and PAM baseline hazards lack flexibility.
For PEMs (left panel), increasing $J$ leads to increasingly unstable and implausible baseline hazard estimates.
For PAMs (right panel), this pattern cannot be observed.
In this illustrative example, the estimates obtained using 20 and 100 cut-points are closely aligned and provide similar approximations of the true baseline hazard.

In agreement with \cite{Whitehead1980}, for small to medium-sized data sets we recommend to use the unique observed event times as cut-points, which automatically leads to improved approximation with increasing $n$ and high $\kappa_j$ density in the relevant parts of the follow-up.
This is the default in \pkg{pammtools}.
For large data sets a random sub sample of the unique observed event times can be used to reduce the data size (while maintaining high density of cut points in areas with many events), typically with little loss in estimation quality; the machine-learning benchmark study of \citet{benderGeneralMachineLearning2021a} uses such subsampling at scale, though it does not quantify the approximation error induced by cut-point subsampling for the estimands considered here, so the appropriate subsample size should be checked for the application at hand.
GAMMs for big data (cf. \cite{Wood2015} and \code{?mgcv::bam}) can also be very useful in this context in order to reduce both memory load and computation time.

In addition to $J$, the flexibility of the spline depends on the number of basis functions $M$, which controls the maximum degrees of freedom of the spline. Normally, $M\ll J$, but should be large enough to capture the complexity of the hazard rate. When in doubt, see \code{?mgcv::gam.check} for diagnostics regarding the choice of $M$.

Regardless of the interval splitting scheme, once the split points $\kappa_j$ are chosen, the data has to be expanded to what we call the piecewise exponential data (PED) format (cf. \cite{Bender2018b} and the \href{https://adibender.github.io/pammtools/articles/data-transformation.html}{data-transformation vignette}), exemplary showcased in Table~\ref{tab:data-partitioning} with

\begin{compactitem}
  \item\label{it:detal_ij} interval specific event indicators $\delta_{ij} =
  \begin{cases}
    1, &\text{ if } t_i \in (\kappa_{j-1}, \kappa_j] \wedge \delta_i = 1\\
    0, &\text{ else }
  \end{cases}$, and
  \item\label{it:o_ij} offsets $o_{ij} = \log(t_{ij})$,
  where $t_{ij} = \min(\kappa_j-\kappa_{j-1}, t_i-\kappa_{j-1}).$
\end{compactitem}

\renewcommand{\arraystretch}{1.5}
\begin{table}[!ht]
\caption{Illustration of survival data transformation to PED format. A standard survival dataset with a single row per subject $i$ and columns observed time, status indicator, and features (here: baseline age) is expanded into long format by splitting follow-up at $\kappa_0=0$, $\kappa_1=1$, $\kappa_2=1.5$, and $\kappa_3=3$.}
\label{tab:data-partitioning}
\centering
\begin{tabular}{
  @{}p{0.25\textwidth}@{}
  @{}p{0.15\textwidth}@{}
  @{}p{0.60\textwidth}@{}
}
\begin{minipage}{\linewidth}
\centering
\begin{tabular}{|c|c|c|c|}
\hline
\textbf{i} & \textbf{t}$_i$ & $\boldsymbol{\delta}_{i}$ & \textbf{age}$_i$ \\ \hline
1 & 1.3 & 0 & 31 \\ \hline
2 & 0.5 & 0 & 67 \\ \hline
3 & 2.6 & 1 & 42 \\ \hline
\end{tabular}
\end{minipage}
&
\begin{minipage}{\linewidth}
\centering
\scriptsize
PED\\
Transformation\\[1.2em]
\hspace{0pt}\centering
\scalebox{4}{$\rightarrow$} \\
$\kappa = (0, 1, 1.5, 3)$
\end{minipage}
&
\begin{minipage}{\linewidth}
\centering
\begin{tabular}{|c|c|p{1.2cm}|c|c|c|c|c|}
\hline
\textbf{i} & \textbf{j} & \centering \textbf{I}$_j$ & $\boldsymbol{\delta}_{ij}$ & \textbf{t}$_{ij}$ & \textbf{o}$_{ij}$ & $\boldsymbol{\kappa}_j$ & \textbf{age}$_i$ \\ \hline
1 & 1 & (0, 1]   & 0 &   1 &  0     & 1     & 31 \\ \hline
1 & 2 & (1, 1.5]   & 0 & 0.3 & -1.2   & 1.5   & 31 \\ \hline
2 & 1 & (0, 1]   & 0 & 0.5 & -0.7   & 1     & 67 \\ \hline
3 & 1 & (0, 1]   & 0 &   1 &  0     & 1     & 42 \\ \hline
3 & 2 & (1, 1.5]   & 0 & 0.5 & -0.7   & 1.5   & 42 \\ \hline
3 & 3 & (1.5, 3]   & 1 & 1.1 &  0.1   & 3     & 42 \\ \hline
\end{tabular}
\end{minipage}
\end{tabular}
\end{table}
\renewcommand{\arraystretch}{1}

After this data transformation, the model can be estimated using Poisson regression with offsets $o_{ij}$ under the working assumption $\delta_{ij}\stackrel{i.i.d.}{\sim}Po(\mu_{ij})$ with $\mu_{ij} = h_{ij}t_{ij}$ and $h_{ij}$ as defined in \eqref{eq:lambda_ij}.
See \citet{Holford1980,Whitehead1980,Laird1981,Friedman1982} for the original justification of the PEM and \citet{Cai2002,Kauermann2005a,Argyropoulos2015,Bender2018a} for penalized and mixed model based approaches.

\subsection{Competing risk analysis}
\label{sub:cr}

The basic assumption for standard models for right-censored data is that the censoring distribution and the event time distribution are independent.
When competing risks are involved, naively treating the events from competing causes as censored observations leads to biased estimators \citep[c.f.][]{coemans2022biascenscompeting, gooley1999kmbiascompeting}.
Cause-specific hazard models, which are implemented in \pkg{pammtools} and feature transition intensities for all absorbing states among individuals under risk, provide a principled way to avoid treating competing events as ordinary independent censoring \citep{putter2007competing}.
Which quantity is appropriate is, however, estimand-dependent rather than universally resolved: cause-specific hazards target the instantaneous rate of each event type among those still at risk and are natural for studying covariate effects on the event-specific hazard, whereas absolute-risk questions are answered by the cumulative incidence function (CIF), which \pkg{pammtools} derives from the estimated cause-specific hazards (Section~\ref{sub:cr-workflow}). The subdistribution-hazard (Fine--Gray) formulation \citep{finegray1999} instead parametrizes the CIF directly; \pkg{pammtools} follows the cause-specific route (see Section~\ref{sub:cr-workflow} and the \href{https://adibender.github.io/pammtools/articles/competing-risks.html}{vignette on competing risks} for examples).

\renewcommand{\arraystretch}{1.5}
\begin{table}[!ht]
\caption{Illustration of competing risk survival data transformation to PED format. A standard competing risk survival dataset with a single row per subject $i$ and columns observed time $t$, event indicator $\delta$ with event types $\{1, 2\}$, and features (here: baseline age) is expanded into PED format by splitting follow-up at $\kappa_0=0$, $\kappa_1=1$, $\kappa_2=1.5$, and $\kappa_3=3$ and separating the event indicator into cause $k \in \{1, 2\}$ and status $\delta \in \{0, 1\}$.}
\label{tab:data-partitioning-cr}
\centering
\begin{tabular}{
  @{}p{0.25\textwidth}@{}
  @{}p{0.15\textwidth}@{}
  @{}p{0.60\textwidth}@{}
}
\begin{minipage}{\linewidth}
\centering
\begin{tabular}{|c|c|c|c|}
\hline
\textbf{i} & \textbf{t}$_i$ & $\boldsymbol{\delta}_{i}$ & \textbf{age}$_i$ \\ \hline
1 & 1.3 & 0 & 31 \\ \hline
2 & 0.5 & 1 & 67 \\ \hline
3 & 2.6 & 2 & 42 \\ \hline
\end{tabular}
\end{minipage}
&
\begin{minipage}{\linewidth}
\centering
\scriptsize
PED\\
Transformation\\[1.2em]
\scalebox{4}{$\rightarrow$}\\
$\kappa = (0, 1, 1.5, 3)$
\end{minipage}
&
\begin{minipage}{\linewidth}
\centering
\begin{tabular}{|c|c|c|c|c|c|c|c|c|}
\hline
\textbf{i} & \textbf{j} & \textbf{k} & \centering \textbf{t}$_j$ & $\boldsymbol{\delta}_{ij}$ & $\mathbf{t}_{ij}$ & $\mathbf{o}_{ij}$ & $\boldsymbol{\kappa}_j$ & \textbf{age}$_i$ \\ \hline
1 & 1 & 1 & (0,  1]   & 0  &   1 &  0     & 1     & 31 \\ \hline
1 & 2 & 1 & (1,1.5]   & 0  & 0.3 & -1.2   & 1.5   & 31 \\ \hline
2 & 1 & 1 & (0,  1]   & 1  & 0.5 & -0.7   & 1     & 67 \\ \hline
3 & 1 & 1 & (0,  1]   & 0  &   1 &  0     & 1     & 42 \\ \hline
3 & 2 & 1 & (1,1.5]   & 0  & 0.5 & -0.7   & 1.5   & 42 \\ \hline
3 & 3 & 1 & (1.5,3]   & 0  & 1.1 &  0.1   & 3     & 42 \\ \hline
1 & 1 & 2 & (0,  1]   & 0  &   1 &  0     & 1     & 31 \\ \hline
1 & 2 & 2 & (1,1.5]   & 0  & 0.3 & -1.2   & 1.5   & 31 \\ \hline
2 & 1 & 2 & (0,  1]   & 0  & 0.5 & -0.7   & 1     & 67 \\ \hline
3 & 1 & 2 & (0,  1]   & 0  &   1 &  0     & 1     & 42 \\ \hline
3 & 2 & 2 & (1,1.5]   & 0  & 0.5 & -0.7   & 1.5   & 42 \\ \hline
3 & 3 & 2 & (1.5,3]   & 1  & 1.1 &  0.1   & 3     & 42 \\ \hline
\end{tabular}
\end{minipage}
\end{tabular}
\end{table}
\renewcommand{\arraystretch}{1}

To estimate cause-specific hazards with PAMMs, the PED transformation from Section~\ref{sub:singleevent} is extended to handle competing events.
Such data either include separate columns for each cause or encode event types in a single indicator $k \in \{0, 1, \ldots, K\}$.
The transformation is illustrated in Table~\ref{tab:data-partitioning-cr}.
As in the single-event case, the follow-up is divided into $J$ intervals with cut points $\kappa_0 < \ldots < \kappa_J$, and the data are expanded accordingly.
In a nutshell, the same procedure as in the single-event case (cf. Table~\ref{tab:data-partitioning}) is applied for each event type $k \in \{1, \ldots, K\}$, considering only the event of type $k$ as event and everything else as censored (as is usual in cause-specific hazard models).
Subsequently, the individual tables are stacked and the indicator $k$ is added as a column to indicate event types.
Thus, the cause-$k$ event indicator in interval $j$ for subject $i$ is given by

\begin{equation}
\delta_{ijk} = \begin{cases}
1, &\text{ if } t_i \in (\kappa_{j-1},\kappa_j] \wedge \delta_i = k,\\
0, &\text{ otherwise.}
\end{cases}
\end{equation}

With the adapted PED transformation for competing risks, the transition-specific hazard can be approximated by piecewise-constant hazards in each interval, and reads $\forall\, t\in (\kappa_{j-1}, \kappa_j], i = 1,\ldots,n$
\begin{align}
\log(h_{ik}(t; \bfx_{ik}, \mcZ_{ik}, \ell_i)) & \approx
  \log(h_{ijk}):= \log(h_{ik}(t_j;\bfx_{ik}, \mcZ_{ik}, \ell_i))\nonumber\\ 
& = \beta_{0k} + f_{0k}(t_j) + \sum_{p=1}^P f_{pk}(x_{ikp}, t_j) +
    \sum_{m=1}^M g_{mk}(\bfz_{ikm}, t_j) + b_{\ell_ik} &\ \label{eq:cr-logpamm2}
\end{align}
where the baseline and (potentially non-linear) covariate effects are learned from the data and can be either assumed to be identical for different transitions $k$ (i.e., $f_{pk} \equiv f_p\;\forall k$ etc) or to be transition-specific. The latter is achieved by simply introducing interaction terms between the (categorical) cause variable $k$ and other variables in the model, such that main terms represent shared effects and interaction terms represent cause-specific deviations from shared effects. Note that the full interaction model could also be estimated by fitting separate models to the data sets for each transition, but for post-processing it is more convenient to fit one combined model. 

In addition to the hazard, the cumulative incidence function (CIF), the absolute (rather than relative) risk of experiencing an event of type $k$ up to time $t$, is an important quantity in competing risks analysis.
The CIF is defined as
\begin{equation}
\CIFk(t) = \int_0^t h_k(s)\, S(s)\, ds.
\end{equation}
Consider the contribution of interval $j$ to this integral, i.e.
\begin{equation}
\int_{\kappa_{j-1}}^t h_k(s)\, S(s)\, ds \quad \text{for } t \in (\kappa_{j-1}, \kappa_j].
\end{equation}
Within interval $j$, the all-cause hazard is constant at $h_j$, so survival decays from $S(\kappa_{j-1})$ as
\begin{equation}
S(s) = S(\kappa_{j-1}) \cdot \exp\left(-\int_{\kappa_{j-1}}^s h_j\, du\right)
= S(\kappa_{j-1})\, e^{-h_j (s - \kappa_{j-1})}.
\end{equation}

Substituting this expression for $S(s)$, and noting $h_k(s) = h_{kj}$ is also constant within interval $j$ gives the recursive formula

\begin{equation}
\CIFk(t) = \CIFk(\kappa_{j-1}) + \frac{h_{kj}}{h_j}\, S(\kappa_{j-1}) \cdot \left(1 - e^{-h_j (t - \kappa_{j-1})}\right).
\label{eq:cr-cif}
\end{equation}

\subsection{Event-history analysis}
\label{sub:eha}

Following \cite{reulenBoostingMultistateModels2015,sennhenn-reulenStructuredFusionLasso2016,Piller2024flexible}, the PAMM framework naturally extends to general event-history analysis, including multi-state data and its special cases such as recurrent events and competing risks (see Figure~\ref{fig:msm-state-diagram}).
Building on the competing-risk PED transformation described in Section~\ref{sub:cr}, the multi-state survival problem can likewise be reformulated as a Poisson regression task, as illustrated in Table~\ref{tab:data-partitioning-msm}.

Consider the general multi-state setting with observed data
\begin{align*}
(t^{\text{entry}}_{ike}, t^{\text{exit}}_{ike}, \delta_{ike}, \mathbf{x}_{ike}).
\end{align*}

where $t^{\text{entry}}_{ike}$ is the entry time of subject $i$ into the risk set for transition $k=1,\ldots,K$ in episode $e=1,\ldots,E$ and $t^{\text{exit}}_{ike}$ the respective exit time, either due to the transition occurring or censoring.
Here, $\delta_{ike}$ is the corresponding status indicator and $\mathbf{x}_{ike}^\top = (x_{ike, 1} \cdots x_{ike, p})$ is the covariate row-vector, which can contain information about the subject's past (e.g., number and timing of previous transitions and episodes) along with the usual subject-specific data.

\renewcommand{\arraystretch}{1.5}
\begin{table}[!ht]
\caption{Illustration of multi-state survival data (according to Figure~\ref{fig:msm-state-diagram} transformation to PED format. A survival multi-state dataset with a single row per transition $i$ and columns ID, observed time, event indicator, and transition is expanded into long format by splitting the follow-up at the event-times $t_0=0$, $t_1=0.5$, $t_2=1.3$, and $t_3=2.6$ and adding censoring due to competing events.}
\label{tab:data-partitioning-msm}
\centering
\begin{tabular}{
  @{}p{0.26\textwidth}@{}
  @{}p{0.14\textwidth}@{}
  @{}p{0.6\textwidth}@{}
}
\begin{minipage}{\linewidth}
\centering
\begin{tabular}{|c|c|c|c|c|}
\hline
\textbf{i} & \textbf{t}$^{\text{entry}}_{i}$ & \textbf{t}$^{\text{exit}}_{i}$ & \textbf{k} \\ \hline
1 & 0 & 1.3 & 0 \\ \hline
2 & 0 & 0.5 & 1 \\ \hline
2 & 0.5 & 2.6 & 2 \\ \hline
\end{tabular}
\end{minipage}
&
\begin{minipage}{\linewidth}
\centering
\scriptsize
PED\\
Transformation\\[1.2em]
\scalebox{4}{$\rightarrow$}\\
$\kappa = (0, 0.5, 1.3, 2.6)$
\end{minipage}
&
\begin{minipage}{\linewidth}
\centering
\begin{tabular}{|c|c|c|c|c|c|c|c|}
\hline
\textbf{i} & \textbf{j} & \textbf{k} & \centering \textbf{I}$_j$ & $\boldsymbol{\delta}_{ij}$ & \textbf{t}$_{ij}$ & \textbf{o}$_{ij}$ & $\boldsymbol{\kappa}_j$ \\ \hline
1 & 1 & $0 \rightarrow 1$ & (0, 0.5]     & 0 & 0.5 & -0.7   & 0.5  \\ \hline
1 & 1 & $0 \rightarrow 3$ & (0, 0.5]     & 0 & 0.5 & -0.7   & 0.5  \\ \hline
1 & 2 & $0 \rightarrow 1$ & (0.5, 1.3]   & 0 & 0.8 & -0.2   & 1.3  \\ \hline
1 & 2 & $0 \rightarrow 3$ & (0.5, 1.3]   & 0 & 0.8 & -0.2   & 1.3  \\ \hline
2 & 1 & $0 \rightarrow 1$ & (0, 0.5]     & 1 & 0.5 & -0.7   & 0.5  \\ \hline
2 & 1 & $0 \rightarrow 3$ & (0, 0.5]     & 0 & 0.5 & -0.7   & 0.5  \\ \hline
2 & 2 & $1 \rightarrow 2$ & (0.5, 1.3]   & 0 & 0.8 & -0.2   & 1.3  \\ \hline
2 & 2 & $1 \rightarrow 0$ & (0.5, 1.3]   & 0 & 0.8 & -0.2   & 1.3  \\ \hline
2 & 3 & $1 \rightarrow 2$ & (1.3, 2.6]   & 1 & 1.3 &  0.3   & 2.6  \\ \hline
2 & 3 & $1 \rightarrow 0$ & (1.3, 2.6]   & 0 & 1.3 &  0.3   & 2.6  \\ \hline
\end{tabular}
\end{minipage}
\end{tabular}

\end{table}
\renewcommand{\arraystretch}{1}

To appropriately account for a delayed entry into the risk set, cut-points $\kappa_j$ are defined at the transition times.
Let $\mathcal{T}$ denote the set of all transition times, given by the union of exit times $t^{\text{exit}}_{ike}, \forall k, e$ over all subjects $i$, that is $\mathcal{T} = \bigcup_{i} \bigcup_{k,e} \{t^{\text{exit}}_{ike}\}$.
Using these transition times as cut-points, i.e., $\{\kappa_0, ..., \kappa_J\} = \text{sort}(\mathcal{T})$, with $J = \lvert \mathcal{T} \rvert - 1$, we refer to $\kappa_{j-1}$ as the state entry and $\kappa_j$ as the state exit time of interval $j$.
Delayed entry arises naturally in multi-state processes and is commonly referred to as \emph{process-induced} or \emph{internal} left-truncation \citep{Andersen1993_ModelSpecification}.
In other words, subjects enter states at different times and they are not at risk for transitions out of a state before they enter into it, which induces left-truncated event times.
In the PED format, left truncation is handled by simply omitting the intervals before a subject becomes at risk, both for truncation induced by the process itself as here or other causes of delayed entry: a subject contributes rows only for intervals after its (state) entry time, and the exposure offset of the interval in which it enters is computed from the actual entry time rather than the interval boundary. This is specified through the counting-process response \code{Surv(entry, exit, status)} and reproduces the partial-likelihood estimates of the corresponding Cox model; see Appendix~\ref{app:as-ped-dispatch} for details.

To account for censoring due to competing events, adding rows for all transition times and transitions, the status indicator is adapted as
\begin{equation}
    \delta_{ijke} = I(\delta_{ike} = 1 \wedge t_{ike} \in (\kappa_{j-1},\kappa_{j}]) \in \{0,1\} \sim Po(\mu_{ijke})
\end{equation}
and again used as outcome in a Poisson model.
With this transformation, the transition-specific hazard can be approximated by piecewise-constant hazards in each interval, and reads
\begin{align}
\log(h_{ike}(t; \bfx_{ike}, \mcZ_{ike}, \ell_i)) & \approx
  \log(h_{ijke}):=\log(h_{ike}(t_j;\bfx_{ike}, \mcZ_{ike}, \ell_i))\ \label{eq:msm-logpamm1}\\
& = \beta_{0ke} + f_{0ke}(t_j) + \sum_{p=1}^P f_{pke}(x_{ikep}, t_j) +
    \sum_{m=1}^M g_{mke}(\bfz_{ikem}, t_j) + b_{\ell_ike} &\ \label{eq:msm-logpamm2}
\end{align}
where the baseline and (potentially non-linear) covariate effects are learned from the data, and can be different for different transitions and/or episodes (once again using interaction terms between the categorical transition and/or episode number variable and the respective covariate). 
Shared frailty effects ($b_{\ell_iek} \equiv b_{\ell_i}$, e.g.) are usually the most parsimonious choice and tend to capture the bulk of unobserved subject-level heterogeneity; transition-specific frailties are mostly useful when transitions are believed to involve distinct latent mechanisms (e.g.,\ progression vs.\ recovery) and the data are informative enough to estimate them separately.

Naturally, multi-state models can be specified on two different time scales for the transition hazards: the calendar-time scale (\emph{clock-forward}), where time $t$ represents time since study entry, and the gap-time scale (\emph{clock-reset}), where time $t$ is reset to zero upon each transition \citep{andersenMSM2002}.  The clock-reset approach is appropriate if transition hazards depend on the time spent in the current state, e.g., recovery time after surgery or time since last relapse, whereas the clock-forward approach is more natural when transition hazards depend on overall disease duration, age, or other cumulative processes \citep{hougaard1999timescales}. 

A distinction between two classes of multi-state models is useful to choose the appropriate time scale and to simplify eq. \eqref{eq:msm-logpamm2}:

First, for competing risks models in which only one transition occurs for each subject as well as for multi-state models in which back-transitions are not allowed and for memoryless multi-state models, the episode index $e$ can be omitted entirely. 
Such temporal-order-dependent processes are fully characterized by the current time, determining which states were visited and in what order. This suggests the use of calendar-time as the default time scale in these so-called \emph{progressive} models. 

Second, in Renewal Processes \citep{AalenRenewal1991}, special cases of multi-state 
models in which a single type of event recurs repeatedly for the same subject, e.g., recurrence of staphylococcus aureus and malaria in children \citep{ramjith2024recurrent}. By construction of these models, only the episode but not the transition index exists. Consequently, the hazard of the next event depends on the time elapsed since the last event rather than on the total time since study entry, which is why it is useful to switch from calendar-time to gap-time.
However, when the primary interest lies in summarizing treatment effects over the full course of the process, the calendar-time scale is generally preferred, as the gap-time approach can yield misleading results in the presence of subject-level heterogeneity \citep{hougaard2022recurrent}. 
An example analysis using a gap-time timescale is available in the \pkg{pammtools} \href{https://adibender.github.io/pammtools/articles/recurrent-events.html}{vignette on recurrent events}. Readers interested in the analysis of recurrent events with \pkg{pammtools} are referred to \cite{ramjith2024recurrent} for a more detailed introduction and discussion of this topic.

Transition probabilities at time $t$ are quantities which can be directly derived from hazards, specifically from the increments of the multi-state model's cumulative hazards, $dH_{ke}(t) = H_{ke}(t) - H_{ke}(t- \Delta t)$, with $H_{ke}(t) = \int^t_0 h_{ke}(s) ds$ \citep{beyersmann2012}.
In case of piecewise constant functions and $\mathcal{T}$ being sufficiently large, for a timepoint $t = \kappa_j \in \mathcal{T}$ one can simplify the increment as $dH_{ke}(t) = H_{ke}(\kappa_j) - H_{ke}(\kappa_{j-1})$.
Let $v, w, v \neq w$ be the entry and exit states of transition $k$, then the matrix element $\left( \textbf{dH}(t)\right)_{v,w} = dH_{ke}(t)$, and define the diagonal with $\left( \textbf{dH}(t)\right)_{v,v} = - \sum_{w} \left(\textbf{dH}(t)\right)_{v,w}$.
Transition probabilities can then be obtained by successive products of past increments \citep{aalenjohansen1978, andersenMSM2002} given as
\begin{align}
    \mathbf{P(s, t)} = \prod_{u \in [s, t) \cap \mathcal{T}} \left( \textbf{I} + \textbf{dH}(u)\right)
    \label{eq:msm-transprob}
\end{align}

In complex multi-state settings, particularly when back-transitions are possible, transition probabilities can be difficult to interpret.
This arises because each entry $(v, w)$ of the transition probability matrix $\mathbf{P(s, t)}$ represents the probability of being in state $w$ at time $t$, given that the process was in state $v$ at time $s$, regardless of the intermediate states visited between $s$ and $t$. As a result, transition probabilities summarize net movement between states over an interval but do not distinguish between different underlying paths.
For example, in a simple illness–death model (cf. Example in Section~\ref{sub:eha-workflow}, Figure~\ref{fig:prothr-state-diagram}), the estimated transition probabilities from the ``healthy'' state to death include not only individuals who die while they are ``healthy'', but also all patients with more complex trajectories of infection followed by subsequent recoveries and relapses who die eventually.

State occupation probabilities serve as an alternative or complementary statistic for transition probabilities in multi-state analyses.
To obtain state occupation probabilities, multiply the transition probabilities with the initial state distribution vector $\boldsymbol\pi(s)$ at time $s$ of the multi-state process with the transition probability matrix $\mathbf{P(s, t)}$ \citep{andersenMSM2002}, i.e.,
\begin{align*}
    \boldsymbol\pi(t) = \boldsymbol\pi(s) \mathbf{P(s, t)}.
\end{align*}

State occupation probabilities directly quantify the probability of being in a particular state at a given time (see Section~\ref{sub:eha-workflow} or \href{https://adibender.github.io/pammtools/articles/multi-state.html}{vignette on multi-state models} for examples).

Note that Equation \eqref{eq:msm-logpamm1} together with either \eqref{eq:cr-cif} or \eqref{eq:msm-transprob} yield cumulative incidence functions or transition probabilities computed as functionals of the estimated hazards which depend on (potentially non-linear) covariate effects learned from the data.
Consequently, the flexibility of the hazard specification achieved through generalized additive (mixed) models propagates to the post-processed quantities of interest, allowing for complex and data-driven effect structures in the resulting estimates. 

In \pkg{pammtools}, for instance, non-Markovian history dependence can be accommodated by including (time-varying, transition-specific, or history-dependent) covariates in the hazard specification.

\subsection{Inference}
\label{sub:inference}

Using PAMMs to model the (log-)hazard, the log-hazard specifications in \eqref{eq:loghazard}, \eqref{eq:cr-logpamm2}, and \eqref{eq:msm-logpamm2} can be written in unified form as $\boldsymbol{\eta} = \mathbf{X}\boldsymbol{\beta}$,
where $\mathbf{X}$ is the full design matrix and $\boldsymbol{\beta}$ the corresponding coefficient vector, comprising the mean baseline hazard $\beta_{0ke}$, (spline basis function) coefficients for $f_{0ke}, f_{1ke}, \ldots, f_{Pke}$ and $g_{1ke}, \ldots, g_{Mke}$ as well as all random effects $b_{\ell ke}$.
Since $\boldsymbol{\hat{\beta}} \stackrel{as}{\sim} \mathcal{N}(\boldsymbol{\beta}, \mathbf{V}_{\beta})$ \citep{Wood2006, Wood2017a}, linear transformations of $\boldsymbol{\hat{\beta}}$ are also asymptotically Gaussian, yielding the (approximate) distribution of the estimated log-hazard \begin{equation*}
  \boldsymbol{\hat\eta} \stackrel{as}{\sim} \mathcal{N}(\mathbf{X} \boldsymbol{\beta}, \mathbf{X} \mathbf{V}_{\beta} \mathbf{X}^\top)
\end{equation*}

In event history modeling, the focus is usually on non-linear transformations of $\eta$ such as hazard rates, cumulative hazards, CIFs, or transition probabilities. To compute confidence intervals for such functionals $g\left(\hat\eta\right)$, \pkg{pammtools} implements three different methods:

First, for a monotone transformation $g(\cdot)$ of a \emph{single} linear-predictor entry $\hat\eta_j$ we can simply apply $g$ to the lower and upper confidence bounds of $\hat\eta_j$ at each time interval $j = 1, \dots, J$; for the hazard $h_j = \exp(\hat\eta_j)$ this yields a valid pointwise $(1-\alpha)$ interval. \pkg{pammtools} extends this transform to the cumulative hazard and survival probability by summing the per-interval hazard bounds:
\begin{align*}
    h_j &\in \left[\exp\left(\hat{\eta}_j \mp z_{1-\alpha/2}\,\widehat{\mathrm{se}}(\hat{\eta}_j)\right)\right], \\[4pt]
    \Lambda(t) &\in \left[\sum_{j=1}^{J} \Delta_j \cdot h_j^{\pm}\right], \\[4pt]
    S(t)       &\in \left[\exp\!\left(-\sum_{j=1}^{J} \Delta_j \cdot h_j^{\pm}\right)\right],
\end{align*}
where $z_{1-\alpha/2}$ is the $(1-\alpha /2)$ quantile of the standard normal distribution, $\hat{\eta}_j = \mathbf{X}_j \boldsymbol{\hat\beta}$ is the estimated log-hazard in interval $j$, $\widehat{\mathrm{se}}(\hat{\eta}_j)$ its pointwise standard error, $h_j^{\pm}$ denotes the lower and upper hazard bounds respectively and $\Delta_j = \kappa_j - \kappa_{j-1}$ is the length of the $j$-th interval. Since $S(t)$ is decreasing in the cumulative hazard, the bounds pair up crosswise: the lower survival bound results from the upper hazard bounds $h_j^{+}$ and vice versa.
Unlike the interval-wise hazard bound, however, $\Lambda(t)$ and $S(t)$ are functions of the whole vector $(\hat\eta_1,\dots,\hat\eta_J)$, and summing the per-interval upper (lower) hazard bounds places \emph{every} interval's hazard simultaneously at its pointwise upper (lower) limit, ignoring the cross-interval covariances $\mathrm{Cov}(\hat\eta_i,\hat\eta_j)$. The resulting cumulative-hazard and survival bands are therefore a deliberately \emph{conservative envelope} whose pointwise coverage typically exceeds the nominal $1-\alpha$, rather than an exact pointwise interval. This transform is nonetheless the default for \code{add\_hazard} and \code{add\_surv\_prob} because it is fast, respects monotonicity and non-negativity by construction, and errs towards wider intervals; the delta method or posterior simulation below should be preferred when calibrated pointwise (or simultaneous) intervals for $\Lambda(t)$ or $S(t)$ are required.

Second, the Delta method explicitly constructs the Jacobian matrix of each functional $g(\cdot)$ and propagates uncertainty via $\mathbf{V}(g(\boldsymbol{\hat\beta})) \approx \nabla_\beta g(\boldsymbol{\hat\beta})\, \mathbf{V}_{\beta}\, \nabla_\beta g(\boldsymbol{\hat\beta})^\top$ \citep{Wood2017a}, then forms symmetric confidence intervals on the response scale $g(\boldsymbol{\hat{\beta}}) \pm z_{1-\alpha/2} \cdot \widehat{\mathrm{se}}$, where $\widehat{\mathrm{se}}$ are the respective square-root diagonal entries of $\mathbf{V}(g(\boldsymbol{\hat\beta}))$.\\
To make this explicit for hazards, cumulative hazards and survival probabilities, denote the $J \times J$ lower-triangular matrix of interval lengths as
\begin{equation*}
    \boldsymbol{\Delta} = 
    \begin{pmatrix}
        \Delta_1 & \cdots & 0 \\
        \vdots   & \ddots & \vdots \\
        \Delta_1 & \cdots & \Delta_J
    \end{pmatrix},
\end{equation*}
then the Jacobians for the three quantities are \citep{Carstensen2005}:
\begin{align*}
    \nabla_\beta h(\boldsymbol{\beta})        &= \mathrm{diag}(\hat{\mathbf{h}})\, \mathbf{X}, \\[4pt]
    \nabla_\beta \Lambda(t; \boldsymbol{\beta}) &= \boldsymbol{\Delta}\, \mathrm{diag}(\hat{\mathbf{h}})\, \mathbf{X}, \\[4pt]
    \nabla_\beta
    S(t; \boldsymbol{\beta})      &= -\mathrm{diag}(\hat{S}(t))\, \boldsymbol{\Delta}\, \mathrm{diag}(\hat{\mathbf{h}})\, \mathbf{X},
\end{align*}
where $\hat{\mathbf{h}} = \exp\left(\boldsymbol{\hat\eta}\right)$ is the vector of estimated hazard rates in each time interval.

Third, for arbitrary non-linear functionals, in particular CIFs and transition probabilities, confidence intervals are obtained by drawing $N$ coefficient vectors from their empirical Bayesian posterior,
\begin{equation}
    \boldsymbol{\beta}^{(n)} \sim 
    \mathcal{N}\!\left(\boldsymbol{\hat{\beta}},\; \mathbf{V}_{\boldsymbol{\beta}}\right), 
    \quad n = 1, \ldots, N,
\end{equation}
propagating each draw through the functional of interest $g\left(\boldsymbol{\beta}^{(n)}\right)$, and taking the $(1-\alpha)$ confidence interval as the empirical $\alpha/2$ and $(1-\alpha/2)$ quantiles of the resulting distribution \citep{Argyropoulos2015}.
For the \emph{smooth components} of the GAM, credible intervals derived from this posterior have been shown to attain close-to-nominal frequentist coverage \citep{Wahba1983, Nychka1988, Marra2012}. That guarantee concerns the estimated smooth terms under the conditions of those references; it does not automatically carry over to the arbitrary non-linear survival functionals to which the same draws are propagated -- CIFs, transition probabilities, RMST differences, or the MI-pooled interval-censored estimands of Section~\ref{sub:ic} -- whose calibration should be assessed empirically for the application at hand.

In practice, the choice between these three approaches involves a trade-off between computational cost, generality, and interval shape. The pointwise transform is the fastest and simplest and produces asymmetric intervals by construction; it is exact only for the interval-wise hazard and, as discussed above, conservative for the cumulative hazard and survival probability, which is why it is used as the (safe) default in \code{add\_hazard} and \code{add\_surv\_prob} (cf. Section~\ref{sub:singleevent}). The delta method yields symmetric intervals on the response scale and applies to any differentiable functional, but it is a first-order approximation (exact only for linear functionals of $\boldsymbol{\hat\beta}$), requires an explicit Jacobian, and can be slow when $J$ and the number of coefficients are both large. Simulation from the posterior is the most general approach and handles arbitrary non-linear functionals such as CIFs and transition probabilities without requiring analytical derivatives. The main cost is computational, scaling with the number of draws. In practice, $200$ to $500$ draws usually keep the Monte-Carlo error of the resulting quantile-based bounds small relative to their width, but not negligible: for approximately Gaussian functionals, the simulation error of the empirical $2.5\%/97.5\%$ quantiles at $N = 200$ ($N = 500$) is still on the order of $10\%$ ($6\%$) of the interval half-width, so increasing $N$ and checking that reported bounds are stable is cheap and advisable. Note that this concerns only the simulation error of the bounds for a given posterior and is a separate question from their frequentist calibration, which is inherited from the posterior itself (see above).

All three approaches assume that event times are observed exactly or right-censored, as the equivalence of the PEM likelihood to a Poisson likelihood breaks down under interval censoring.
For interval-censored data, \pkg{pammtools} instead provides a multiple imputation workflow (Section~\ref{sub:ic}) that re-fits the model on multiply completed data sets and pools coefficient estimates \citep[via Rubin's rules;][]{rubin1987} as well as the per-fit posterior draws described above, such that the resulting intervals also reflect the imputation uncertainty.

\section{Practical workflow}
\label{sec:practical_workflow}

In the following sections, we illustrate the practical use of PAMMs with \pkg{pammtools} focusing on a diverse set of applications -  Section~\ref{sub:singleevent-workflow} for estimation in single event scenarios, Section~\ref{sub:cr-workflow} for the estimation procedure for competing risks data including the built-in calculation of cumulative incidence functions, and Section~\ref{sub:eha-workflow} for multi-state data workflows including the built-in calculation of transition probabilities and state occupation probabilities. 
In addition, vignettes \href{https://adibender.github.io/pammtools/articles/competing-risks.html}{on competing risks} and \href{https://adibender.github.io/pammtools/articles/multi-state.html}{multi-state models} are available for additional practical advice and code examples.

This section presents details on how to transform the data for the respective use case (single-event, competing risks, multi-state), fit models under different assumptions about the data generating process, and visualize and interpret the results.
Where applicable, we also include alternative modeling approaches to demonstrate that PAMMs not only match established models but provide a much more flexible framework that unifies many different applications. 

We emphasise that these comparisons illustrate the qualitative behaviour on one realization; they are not a substitute for a proper simulation study of bias, coverage, and efficiency across replicates, misspecified hazard shapes, and inspection densities, which would be needed to establish these properties in general and which we regard as outside the scope of this software article \citep[cf.][on such evaluations]{morris2019simulation}.

The general workflow using \code{pammtools} consists of the following steps:

\begin{itemize}
  \item transform the data to a format suitable for the intended model with \code{as\_ped}
  \item estimate the model using \code{pamm} (and interpret the model summary)
  \item create a new data set for which predictions should be generated with \code{make\_newdata}
  \item append predictions (and CIs) for quantities of interest to this new data set using utility functions like \code{add\_hazard}, \code{add_surv_prob}, \code{add\_trans\_prob}, ...
\end{itemize}

Further steps usually include visualization of the estimated hazards, hazard ratios, survival probabilities (or differences between survival probabilities), and so on.
By separating these steps in our implementation, we provide users with maximal transparency, control and the flexibility to adapt each individual step to their specific use cases and needs.

\subsection{Single-event analysis}
\label{sub:singleevent-workflow}

\subsubsection{Baseline hazard and marginal survival probability}

For the single-event setting, consider the \code{tumor} data set \citep{schiergensThirtydayMortalityLeads2015a} included in the \pkg{pammtools} package (\Rlang-chunk \ref{rchunk:tumor-data}).
The data contains time-to-death information (in days) of 776 patients after tumor removal.
Additional confounders include a comorbidity score (\code{charlson\_score}), \code{age}, \code{sex}, and indicators for whether \code{complications} occurred, whether a blood \code{transfusion} was necessary and whether the tumor developed \code{metastases}, see Table~\ref{tab:tumor-example}.

\begin{table}

\caption{\label{tab:tumor-example}Excerpt of the \texttt{tumor} dataset. Each row contains one subject with information on event time (days), event status (status, 0 in case of censoring), and covariates age and sex.}
\centering
\begin{tabular}[t]{rrrl}
\toprule
days & status & age & sex\\
\midrule
569 & 1 & 69 & male\\
201 & 1 & 62 & female\\
5 & 1 & 78 & male\\
854 & 0 & 65 & male\\
165 & 1 & 41 & male\\
\bottomrule
\end{tabular}
\end{table}

Follow-up continued after release from hospital, thus censoring times can be assumed to be  independent of the event times.

\begin{knitrout}\small
\definecolor{shadecolor}{rgb}{0.961, 0.961, 0.961}\color{fgcolor}\begin{kframe}
\begin{rexample}\label{rchunk:tumor-data}\hfill{}\begin{alltt}
\hlkwd{data}\hldef{(}\hlsng{"tumor"}\hldef{,} \hlkwc{package} \hldef{=} \hlsng{"pammtools"}\hldef{)}
\hldef{ped_tumor} \hlkwb{=} \hldef{tumor |>} \hlkwd{as_ped}\hldef{(}\hlkwd{Surv}\hldef{(days, status)} \hlopt{~} \hldef{.)}
\end{alltt}
\end{rexample}\end{kframe}
\end{knitrout}

The \code{as\_ped} call in \Rlang-chunk \ref{rchunk:tumor-data} transforms the survival data (cf. Table~\ref{tab:tumor-example}) into PED (cf. Table~\ref{tab:tumor-ped-example})
\begin{compactitem}
  \item the left hand side (LHS) of the \code{formula} specifies the event time (\code{time\_var} attribute)
  and status information (\code{status\_var} attribute).
  \item the right hand side (RHS) of the \code{formula} specifies covariates
  that should be kept after data transformation. This can be useful when the data
  contains many variables but only a few will be used to estimate the hazard.
  As usual, dot notation (\code{~ . }) can be used to include all variables.
  \item the follow up is partitioned at the split points $\kappa_j$
  provided through the \code{cut} argument. If the \code{cut} argument is not specified, the function uses all observed event times as interval cut points (cf. \ref{sub:singleevent-workflow}, 326 in this case - see \code{breaks} attribute). The start (\code{tstart}) and stop
  (\code{tend}) times are created as well as an \code{interval} column.
  \item the $\delta_{ij}$, which will serve as the outcome of the Poisson
  regression, are stored in the column \code{ped\_status} and are 1 only in the
  interval in which the subject experienced an event (if uncensored), which is
  also the final interval for that subject.
  \item the offset variable is calculated, e.g., subject \code{id = 132}
    was transitioned at 5 days, therefore
    $o_{i=3,j=3} = \log(5-3) \approx 0.69$.
\end{compactitem}

We start with the estimation of the baseline hazard and resulting marginal survival probabilities.
\Rlang-chunk \ref{rchunk:pam-tumor} estimates the baseline hazard $h(t) = \exp(\beta_0 + f_0(t))$ from the PED using \code{pamm}, a thin wrapper around \code{mgcv::gam} that automatically sets \code{family = poisson()} and \code{offset = offset}, ensuring the model is correctly specified without requiring the user to do so manually.
Since the returned model object is a standard \code{mgcv::gam} object, the full \code{mgcv} ecosystem is immediately available, e.g., all functions for plotting and prediction, and the output can be interpreted accordingly.
For instance, the summary output gives the value of $\beta_0$ (\code{(Intercept)}) and indicates that the estimated smooth function $f_0(t)$ is non-linear (\code{edf = 3.653}).

\begin{table}

\caption{\label{tab:tumor-ped-example}PED of ID 132 derived from the tumor data with covariates age and sex.}
\centering
\begin{tabular}[t]{rrrlrrrl}
\toprule
id & tstart & tend & interval & offset & ped\_status & age & sex\\
\midrule
132 & 0 & 1 & (0,1] & 0.00 & 0 & 78 & male\\
132 & 1 & 2 & (1,2] & 0.00 & 0 & 78 & male\\
132 & 2 & 3 & (2,3] & 0.00 & 0 & 78 & male\\
132 & 3 & 5 & (3,5] & 0.69 & 1 & 78 & male\\
\bottomrule
\end{tabular}
\end{table}

\begin{knitrout}\small
\definecolor{shadecolor}{rgb}{0.961, 0.961, 0.961}\color{fgcolor}\begin{kframe}
\begin{rexample}\label{rchunk:pam-tumor}\hfill{}\begin{alltt}
\hldef{pam_tumor} \hlkwb{=} \hlkwd{pamm}\hldef{(ped_status} \hlopt{~} \hlkwd{s}\hldef{(tend),} \hlkwc{data} \hldef{= ped_tumor)}
\hlkwd{summary}\hldef{(pam_tumor)}
\end{alltt}
\begin{verbatim}

Family: poisson 
Link function: log 

Formula:
ped_status ~ s(tend)

Parametric coefficients:
            Estimate Std. Error z value Pr(>|z|)    
(Intercept) -7.54985    0.05534  -136.4   <2e-16 ***
---
Signif. codes:  0 '***' 0.001 '**' 0.01 '*' 0.05 '.' 0.1 ' ' 1

Approximate significance of smooth terms:
          edf Ref.df Chi.sq  p-value    
s(tend) 3.874   4.81  24.69 0.000167 ***
---
Signif. codes:  0 '***' 0.001 '**' 0.01 '*' 0.05 '.' 0.1 ' ' 1

R-sq.(adj) =  -0.00157   Deviance explained = 0.578%
UBRE = -0.96843  Scale est. = 1         n = 147896
\end{verbatim}
\end{rexample}\end{kframe}
\end{knitrout}

The most programmatic and flexible way to generate predictions and visualization using \pkg{pammtools} is to create a new data with the desired (covariate) specifications using \code{make\_newdata} and to add the quantities of interest via one of the \code{add\_*} functions. Alternative convenience functions are discussed in Section~\ref{sub:postprocessing}.
An example is given in \Rlang-chunk \ref{rchunk:pam-tumor-ndf}: First, a data set is created containing each unique value of \code{tend} (the only covariate in the model) and then the predicted hazard, cumulative hazard, and survival probabilities are computed (including approximate confidence intervals).

\begin{knitrout}\small
\definecolor{shadecolor}{rgb}{0.961, 0.961, 0.961}\color{fgcolor}\begin{kframe}
\begin{rexample}\label{rchunk:pam-tumor-ndf}\hfill{}\begin{alltt}
\hldef{ndf_tumor} \hlkwb{=} \hldef{ped_tumor |>}
  \hlkwd{make_newdata}\hldef{(}\hlkwc{tend} \hldef{=} \hlkwd{unique}\hldef{(tend)) |>}
  \hlkwd{add_hazard}\hldef{(pam_tumor) |>}
  \hlkwd{add_cumu_hazard}\hldef{(pam_tumor)|>}
  \hlkwd{add_surv_prob}\hldef{(pam_tumor)}
\end{alltt}
\end{rexample}\end{kframe}
\end{knitrout}

\begin{table}
\centering
\caption{\label{tab:tumor-ndf-example}New created data set containing the desired (covariate) specifications and added quantities of interest based on the model pam\_tumor for arbitrarily chosen time points.}
\centering
\resizebox{\ifdim\width>\linewidth\linewidth\else\width\fi}{!}{
\begin{tabular}[t]{rrrrrrr}
\toprule
tend & cumu\_lower & cumu\_hazard & cumu\_upper & surv\_lower & surv\_prob & surv\_upper\\
\midrule
0 & 0.00 & 0.00 & 0.00 & 1.00 & 1.00 & 1.00\\
5 & 0.00 & 0.00 & 0.01 & 0.99 & 1.00 & 1.00\\
43 & 0.03 & 0.03 & 0.04 & 0.96 & 0.97 & 0.97\\
139 & 0.08 & 0.10 & 0.12 & 0.88 & 0.90 & 0.92\\
818 & 0.35 & 0.41 & 0.50 & 0.61 & 0.66 & 0.71\\
\bottomrule
\end{tabular}}
\end{table}

\begin{figure}[!h]
\begin{knitrout}\small
\definecolor{shadecolor}{rgb}{0.961, 0.961, 0.961}\color{fgcolor}

{\centering \includegraphics[width=\maxwidth]{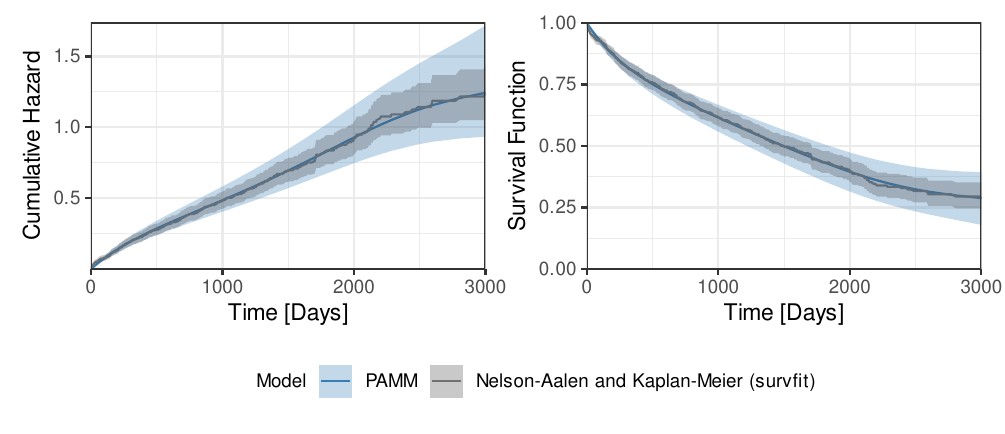} 

}

\end{knitrout}
\caption{\label{fig:tumor-comb-viz} Panels show the estimated cumulative hazards (left), and survival probabilities (right) and compare the \pkg{pammtools} estimates and point-wise CIs (blue) to the non-parametric \pkg{survival} estimates (grey).}
\end{figure}

For visualizing the cumulative hazard, and the survival probability in Figure~\ref{fig:tumor-comb-viz} we use \pkg{ggplot2}, but the advantage of this approach is that any package and software can be used once the data is generated.

\subsubsection{Stratified hazards}
So far, we have ignored covariates.
In the \code{tumor} dataset, the variable \code{complications} is particularly relevant: patients experiencing complications during surgery are expected to have higher hazards immediately post-operation and, consequently, lower survival probabilities.
This effect is likely to diminish over time, i.e., it is time-varying, which violates the PH assumption of a standard Cox model.

One common solution is to use stratified models (e.g., \citet[Ch. 9.3]{Klein1997}), which allow different baseline hazards for each level of a categorical covariate.
In \Rlang-chunk \ref{rchunk:pam-tumor-complications}, we estimate such a model again using the \code{pamm} function.

Note that \code{complications} appears twice in the model: once in the smooth term via the \code{by} argument and once as a main effect to adjust the intercept.
The latter is required whenever the \code{by} variable is a factor, because smooth terms in \code{mgcv} are centered around zero by default.

\begin{knitrout}\small
\definecolor{shadecolor}{rgb}{0.961, 0.961, 0.961}\color{fgcolor}\begin{kframe}
\begin{rexample}\label{rchunk:pam-tumor-complications}\hfill{}\begin{alltt}
\hldef{pam_tumor_comp} \hlkwb{=} \hlkwd{pamm}\hldef{(ped_status} \hlopt{~} \hlkwd{s}\hldef{(tend,} \hlkwc{by} \hldef{= complications)} \hlopt{+} \hldef{complications,}
                      \hlkwc{data} \hldef{= ped_tumor)}
\hlkwd{summary}\hldef{(pam_tumor_comp)}
\end{alltt}
\begin{verbatim}

Family: poisson 
Link function: log 

Formula:
ped_status ~ s(tend, by = complications) + complications

Parametric coefficients:
                 Estimate Std. Error z value Pr(>|z|)    
(Intercept)      -8.05832    0.09845 -81.856  < 2e-16 ***
complicationsyes  0.95596    0.13504   7.079 1.45e-12 ***
---
Signif. codes:  0 '***' 0.001 '**' 0.01 '*' 0.05 '.' 0.1 ' ' 1

Approximate significance of smooth terms:
                           edf Ref.df Chi.sq p-value    
s(tend):complicationsno  8.197  8.826  20.11  0.0174 *  
s(tend):complicationsyes 5.730  6.872 101.33  <2e-16 ***
---
Signif. codes:  0 '***' 0.001 '**' 0.01 '*' 0.05 '.' 0.1 ' ' 1

R-sq.(adj) =  -0.00118   Deviance explained = 3.56%
UBRE = -0.96923  Scale est. = 1         n = 147896
\end{verbatim}
\end{rexample}\end{kframe}
\end{knitrout}

Example \Rlang-chunk \ref{rchunk:pam-tumor-ndf-strata} extends the example given in \Rlang-chunk \ref{rchunk:pam-tumor-ndf} for this stratified model.
Again, this data set builds the basis for visualizations using \pkg{ggplot2}, e.g.,
Figure~\ref{fig:tumor-strata-viz}, where we added an additional dimension for the covariate \code{complications}. As the cumulative hazards are summed up over all covariate combinations of the new data set, we need to group the data by \code{complications}.

Note that this grouping is required whenever \code{add\_cumu\_hazard()} (or any other cumulative \code{add\_*} function) is applied across groups: the input data must be grouped by the covariates prior to the call, since the cumulative quantity would otherwise be accumulated across distinct covariate profiles. As of \pkg{pammtools} 0.8.1, the cumulative \code{add\_*} functions abort with an informative error in this case instead of returning silently wrong results.

\begin{knitrout}\small
\definecolor{shadecolor}{rgb}{0.961, 0.961, 0.961}\color{fgcolor}\begin{kframe}
\begin{rexample}\label{rchunk:pam-tumor-ndf-strata}\hfill{}\begin{alltt}
\hldef{ndf_tumor_strata} \hlkwb{=} \hldef{ped_tumor |>}
  \hlkwd{make_newdata}\hldef{(}
    \hlkwc{tend} \hldef{=} \hlkwd{unique}\hldef{(tend),}
    \hlkwc{complications}\hldef{=}\hlkwd{unique}\hldef{(complications)}
    \hldef{) |>}
  \hlkwd{group_by}\hldef{(complications) |>}
  \hlkwd{add_hazard}\hldef{(pam_tumor_comp) |>}
  \hlkwd{add_cumu_hazard}\hldef{(pam_tumor_comp)|>}
  \hlkwd{add_surv_prob}\hldef{(pam_tumor_comp)}
\end{alltt}
\end{rexample}\end{kframe}
\end{knitrout}

\begin{figure}[!h]
\begin{knitrout}\small
\definecolor{shadecolor}{rgb}{0.961, 0.961, 0.961}\color{fgcolor}

{\centering \includegraphics[width=\maxwidth]{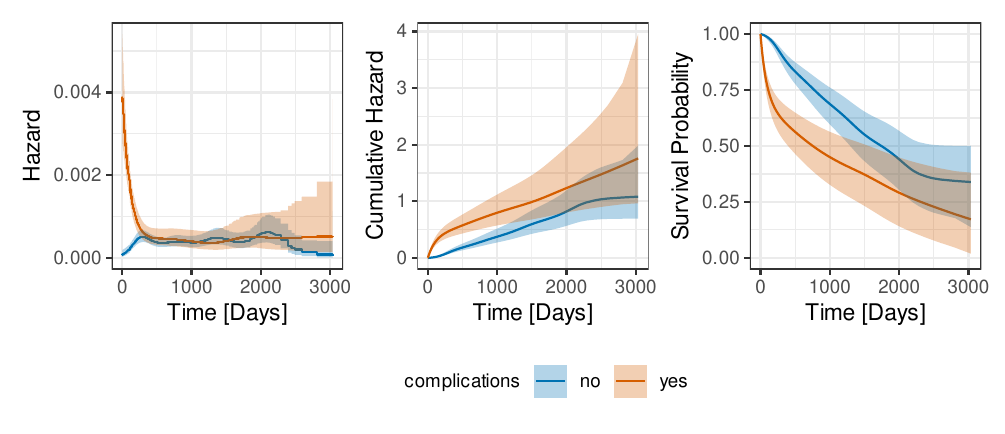} 

}

\end{knitrout}
\caption{\label{fig:tumor-strata-viz} Panels show estimated hazard rates (left), cumulative hazards (middle), and survival probabilities (right) stratified by \code{complications}, computed using \pkg{pammtools}'s \code{add\_*}-functions on a suitable \code{make\_newdata()}-output.}
\end{figure}

Figure~\ref{fig:tumor-strata-viz} shows that, immediately after a major surgery, people with complications have a much higher hazard than people without complications.
With increasing time, this hazard difference shrinks and becomes much more uncertain.
Accordingly, the survival probability drops substantially faster and further for patients with complications than for those without.
The extended \pkg{pammtools} functionality, presented in Section~\ref{sub:rmst}, demonstrates how differences in survival can be assessed using restricted mean survival time (RMST) calculations.

Note that standard survival-specific residual diagnostics — such as martingale, Cox-Snell, or Schoenfeld residuals \citep{therneau2000residuals} — are defined at the subject level using original event times and censoring indicators, and therefore do not transfer naturally to the interval-row structure of the PED format.
Model fit can be assessed by comparing PAMM-based hazard and survival estimates to their non-parametric counterparts (Nelson-Aalen, Kaplan-Meier), by inspecting time-varying covariate effects as implicit tests of the proportional hazards assumption, or via proper scoring rules such as the (integrated) Brier score and time-dependent C-index \citep{graf1999assessment, antolini2005timedep, sonabend2024properness}.

\subsection{Competing risk analysis}
\label{sub:cr-workflow}

We use the \code{mgus2} data set from the \pkg{survival} package \citep{kyle2002mgus} for the competing risk setting.
The data includes 1341 patients with monoclonal gammopathy of undetermined significance (MGUS), see Table~\ref{tab:mgus2-example}.
The primary endpoint is progression to multiple myeloma or other plasma-cell cancer (PCM) with death by any cause as the competing risk.

\begin{table}[!h]
\centering
\caption{\label{tab:mgus2-example}Data slice of the \texttt{mgus2} data set, containing progression to PCM time \texttt{ptime} and death time \texttt{futime} as well as respective event indicators \texttt{pstat} and \texttt{death}. Column \texttt{mspike} contains level of M-Spike proteins.}
\centering
\begin{tabular}[t]{rrlrrrrrrrr}
\toprule
id & age & sex & dxyr & hgb & creat & mspike & ptime & pstat & futime & death\\
\midrule
1 & 88 & F & 1981 & 13.1 & 1.3 & 0.5 & 30 & 0 & 30 & 1\\
2 & 78 & F & 1968 & 11.5 & 1.2 & 2.0 & 25 & 0 & 25 & 1\\
3 & 94 & M & 1980 & 10.5 & 1.5 & 2.6 & 46 & 0 & 46 & 1\\
4 & 68 & M & 1977 & 15.2 & 1.2 & 1.2 & 92 & 0 & 92 & 1\\
5 & 90 & F & 1973 & 10.7 & 0.8 & 1.0 & 8 & 0 & 8 & 1\\
\bottomrule
\end{tabular}
\end{table}

\cite{kylemgusbounds2010} and \cite{mgusnbackground2011} point out that high levels of M-Spike proteins (\code{mspike}, e.g., $> 1.5 g/dL$) or low levels of Hemoglobin (\code{hgb}, e.g., $< 10g/dL$) could have an effect on progression to PCM or Death.
For this reason, \Rlang-Chunk \ref{rchunk:mgus2-cr-preprocessing} builds the biomarker of high or low \code{mspike} named \code{bm_mspike}. Here, for the sake of simplicity of the illustration, we also excluded subjects with missing \code{mspike} values. The competing risks status variable \code{status} is recoded such that a value of 1 indicates progression to PCM and a value of 2 indicates death, based on the original \code{pstat} and \code{death} indicators; the event time \code{time} is then taken as the earlier of the two respective observation times, \code{ptime} and \code{futime}.

\begin{knitrout}\small
\definecolor{shadecolor}{rgb}{0.961, 0.961, 0.961}\color{fgcolor}\begin{kframe}
\begin{rexample}\label{rchunk:mgus2-cr-preprocessing}\hfill{}\begin{alltt}
\hlkwd{data}\hldef{(}\hlsng{"cancer"}\hldef{,} \hlkwc{package} \hldef{=} \hlsng{"survival"}\hldef{)}
\hldef{mgus_cr} \hlkwb{=} \hldef{mgus2 |>}
  \hlkwd{mutate}\hldef{(}
    \hlkwc{status} \hldef{=} \hlkwd{ifelse}\hldef{(pstat} \hlopt{==} \hlnum{1}\hldef{,} \hlnum{1}\hldef{,} \hlnum{2} \hlopt{*} \hldef{death) |>} \hlkwd{factor}\hldef{(),}
    \hlkwc{time} \hldef{=} \hlkwd{pmin}\hldef{(ptime, futime),}
    \hlkwc{bm_mspike} \hldef{=} \hlkwd{ifelse}\hldef{(mspike} \hlopt{>} \hlnum{1.5}\hldef{,} \hlsng{"Hi"}\hldef{,} \hlsng{"Lo"}\hldef{) |>} \hlkwd{factor}\hldef{() |>}
      \hlkwd{relevel}\hldef{(}\hlkwc{ref} \hldef{=} \hlsng{"Lo"}\hldef{),}
  \hldef{) |>}
  \hlkwd{filter}\hldef{(}\hlopt{!}\hlkwd{is.na}\hldef{(bm_mspike))}
\end{alltt}
\end{rexample}\end{kframe}
\end{knitrout}

To fit PAMMs for competing risks, the required data transformation closely mirrors that used for single-event analyses.
The \code{as\_ped}-function facilitates this by recognizing competing risk structures when the \code{status} variable includes at least two distinct event categories in addition to the censoring category.
To indicate the specific type of event each subject experiences, \code{as\_ped} adds an additional column named \code{cause}.

\begin{knitrout}\small
\definecolor{shadecolor}{rgb}{0.961, 0.961, 0.961}\color{fgcolor}\begin{kframe}
\begin{rexample}\label{rchunk:mgus-cr-example_ped}\hfill{}\begin{alltt}
\hldef{mgus_cr_ped} \hlkwb{=} \hldef{mgus_cr |>} \hlkwd{as_ped}\hldef{(}\hlkwd{Surv}\hldef{(time, status)}\hlopt{~}\hldef{.)}
\end{alltt}
\end{rexample}\end{kframe}
\end{knitrout}

We can use the \pkg{pammtools} convenience function \code{pamm} for this competing risk setting similarly to the single-event examples in \Rlang-chunks \ref{rchunk:pam-tumor}, \ref{rchunk:pam-tumor-complications}.
To estimate different baseline hazards for different endpoints, we need to stratify inference by \code{cause}.
To speed up inference for this large PED, we use \code{mgcv::gam} with its big data extension \code{bam} \cite{Wood2015}.
As before, \code{pamm} automatically sets \code{family = poisson()} and \code{offset = offset} for the model to be estimated correctly.

\begin{knitrout}\small
\definecolor{shadecolor}{rgb}{0.961, 0.961, 0.961}\color{fgcolor}\begin{kframe}
\begin{rexample}\label{rchunk:mgus-cr-pamm-output}\hfill{}\begin{alltt}
\hldef{pam_mgus_cr} \hlkwb{=} \hlkwd{pamm}\hldef{(}
  \hldef{ped_status} \hlopt{~} \hlkwd{s}\hldef{(tend,} \hlkwc{by} \hldef{= cause)} \hlopt{+} \hldef{cause} \hlopt{*} \hldef{bm_mspike,}
  \hlkwc{data} \hldef{= mgus_cr_ped,} \hlkwc{engine}\hldef{=}\hlsng{"bam"}\hldef{)}
\hlkwd{summary}\hldef{(pam_mgus_cr)}
\end{alltt}
\begin{verbatim}

Family: poisson 
Link function: log 

Formula:
ped_status ~ s(tend, by = cause) + cause * bm_mspike

Parametric coefficients:
                   Estimate Std. Error z value Pr(>|z|)    
(Intercept)         -7.3831     0.1281 -57.626  < 2e-16 ***
cause2               2.3708     0.1341  17.678  < 2e-16 ***
bm_mspikeHi          0.9336     0.1879   4.969 6.72e-07 ***
cause2:bm_mspikeHi  -1.0127     0.2050  -4.939 7.86e-07 ***
---
Signif. codes:  0 '***' 0.001 '**' 0.01 '*' 0.05 '.' 0.1 ' ' 1

Approximate significance of smooth terms:
                 edf Ref.df Chi.sq p-value    
s(tend):cause1 1.003  1.006  4.826  0.0283 *  
s(tend):cause2 7.116  8.136 54.792  <2e-16 ***
---
Signif. codes:  0 '***' 0.001 '**' 0.01 '*' 0.05 '.' 0.1 ' ' 1

R-sq.(adj) =  0.00274   Deviance explained = 6.64%
fREML = 5989.6  Scale est. = 1         n = 242902
\end{verbatim}
\end{rexample}\end{kframe}
\end{knitrout}

Based on the results in \Rlang-chunk \ref{rchunk:mgus-cr-pamm-output}, we see that patients with high M-spike levels showed a significantly increased risk of progression to PCM compared with those with low M-spike levels (reference group).
In contrast, this strong effect of M-spike cannot be observed for the progression to death.
Further, note that the baseline hazard for death displayed a highly nonlinear time course (\code{edf} $>> 1$), suggesting that the hazard for death changes more dynamically over follow-up than the linearly increasing hazard for progression.

\begin{knitrout}\small
\definecolor{shadecolor}{rgb}{0.961, 0.961, 0.961}\color{fgcolor}\begin{kframe}
\begin{rexample}\label{rchunk:mgus-cr-visualization-short}\hfill{}\begin{alltt}
\hldef{pam_cr_long} \hlkwb{=} \hldef{mgus_cr_ped |>}
  \hlkwd{make_newdata}\hldef{(}
    \hlkwc{tend} \hldef{=} \hlkwd{unique}\hldef{(tend),}
    \hlkwc{cause} \hldef{=} \hlkwd{unique}\hldef{(cause),}
    \hlkwc{bm_mspike} \hldef{=} \hlkwd{unique}\hldef{(bm_mspike)}
  \hldef{) |>}
  \hlkwd{group_by}\hldef{(bm_mspike, cause) |>}
  \hlkwd{add_cif}\hldef{(pam_mgus_cr)}
\end{alltt}
\end{rexample}\end{kframe}
\end{knitrout}

To present the findings in a more interpretable way, we again create a new 
dataset with the desired covariate combinations using \code{make\_newdata}.
We then add the quantities of interest via one of the \code{add\_*} functions.
As shown in \Rlang-chunk \ref{rchunk:mgus-cr-visualization-short}, the 
constructed dataset includes each unique combination of \code{tend}, 
\code{cause}, and \code{bm\_mspike} (the model's covariates, restricted to the 
observed follow-up range to avoid extrapolation) and groups by these variables (as noted previously, when computing cumulative quantities across groups the input data must be grouped beforehand).
For these combinations, we compute the corresponding CIFs along with 
approximate point-wise confidence intervals.

\begin{figure}[!hbtp]
\begin{knitrout}\small
\definecolor{shadecolor}{rgb}{0.961, 0.961, 0.961}\color{fgcolor}

{\centering \includegraphics[width=\maxwidth]{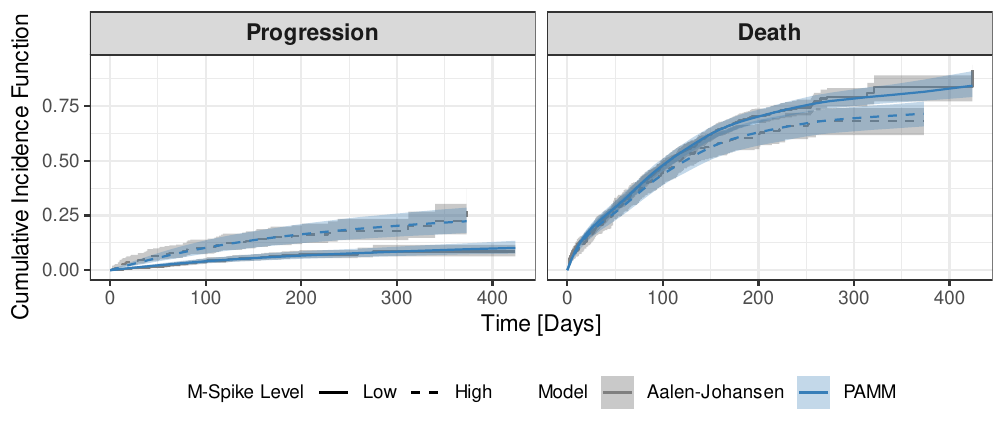} 

}

\end{knitrout}
\caption{\label{fig:mgus-cr-viz} Panels show estimated cumulative incidence functions for each transition (facet), model/estimator (color) for high and low levels of Mspike (linetype).}
\end{figure}

Figure~\ref{fig:mgus-cr-viz} compares the \pkg{pammtools} results with the corresponding non-parametric Aalen-Johansen estimates using the \pkg{etm} package \citep{etm2011}.\footnote{Confidence intervals for the AJ-estimated transition probabilities are computed with the unexported helper \code{etm:::ci\_transform}, reproduced verbatim in the \code{.Rnw} preamble of this article so that the comparison code can be executed without relying on \pkg{etm} internals. It is used only for the reference estimator and is not part of \pkg{pammtools}.}
Both methods are in close agreement. The estimated probability of transitioning to PCM is higher for elevated \code{mspike} levels; estimated differences in the probability of death by \code{mspike} level are small in this model but ultimately result in lower probability of death for higher \code{mspike} levels.

\subsection{Event-history analysis}
\label{sub:eha-workflow}

We use the long-format data set \code{prothr} of 488 patients with liver cirrhosis from a randomized trial comparing prednisone and placebo, available in the \pkg{mstate} package \citep{dewreede2011mstate} and originally provided by \cite{Andersen1993_ModelSpecification} for the multi-state setting.

\begin{figure}[!ht]
\begin{knitrout}\small
\definecolor{shadecolor}{rgb}{0.961, 0.961, 0.961}\color{fgcolor}

{\centering \includegraphics[width=\maxwidth]{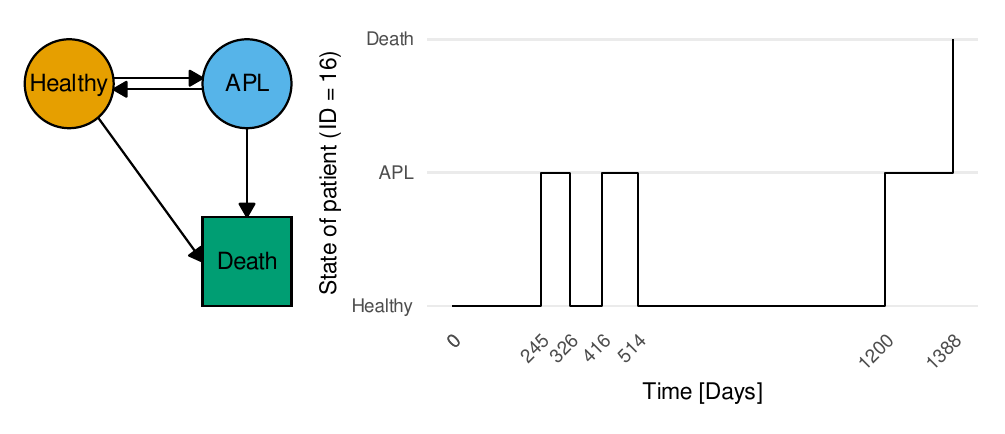} 

}

\end{knitrout}
\caption{Left column: State-diagram of illness-death model in the \code{prothr} data example. Transient states (healthy, abnormal prothrombin (APL)) are marked by circles, and the absorbing state (death) by a square. Arrows show possible transitions, including recovery from abnormal prothrombin levels, i.e., transition from abnormal prothrombin levels to healthy. Right column: Transition plot for ID 16. Patient starts with \texttt{APL}, transitions between \texttt{Healthy} and \texttt{APL} multiple times, and finally dies after 1388 hours in the study.
}
\label{fig:prothr-state-diagram}
\end{figure}

This data is a classical example for an illness-death multi-state process, because patients can repeatedly transition between the transient states healthy and ill (here: abnormal prothrombin level, abbr.
APL) or transition into the absorbing state death, cf. Figure~\ref{fig:prothr-state-diagram}.
Besides the state transition information \code{from}, \code{to}, and \code{status}, the data contain a treatment variable \code{treat} indicating whether the patient received the placebo or medication, see Table~\ref{tab:prothr-example}.

\begin{table}

\caption{\label{tab:prothr-example}Data excerpt of the \texttt{prothr} data set, containing progression to healthy (\texttt{to = 1}), APL (\texttt{to = 2}) and death (\texttt{to = 3}) as well as the respective event indicator \texttt{status}. Column \texttt{treat} contains treatment group \texttt{placebo} or \texttt{prednisone}. Here, transitions that did not occur due to another event are censored with \texttt{status = 0}}
\centering
\begin{tabular}[t]{rrrrrrrl}
\toprule
id & from & to & trans & Tstart & Tstop & status & treat\\
\midrule
16 & 1 & 2 & 1 & 0 & 245 & 1 & Placebo\\
16 & 1 & 3 & 2 & 0 & 245 & 0 & Placebo\\
16 & 2 & 1 & 3 & 245 & 326 & 1 & Placebo\\
16 & 2 & 3 & 4 & 245 & 326 & 0 & Placebo\\
16 & 1 & 2 & 1 & 326 & 416 & 1 & Placebo\\
16 & 1 & 3 & 2 & 326 & 416 & 0 & Placebo\\
16 & 2 & 1 & 3 & 416 & 514 & 1 & Placebo\\
16 & 2 & 3 & 4 & 416 & 514 & 0 & Placebo\\
16 & 1 & 2 & 1 & 514 & 1200 & 1 & Placebo\\
16 & 1 & 3 & 2 & 514 & 1200 & 0 & Placebo\\
16 & 2 & 1 & 3 & 1200 & 1388 & 0 & Placebo\\
16 & 2 & 3 & 4 & 1200 & 1388 & 1 & Placebo\\
\bottomrule
\end{tabular}
\end{table}

In this example, individuals in either the \textit{Healthy} or \textit{APL} state remain at risk of transitioning to the \textit{Death} state.
Consequently, \textit{APL} is not treated as a terminal event, which is why we must account for left-truncated observed transition times.
The variable \code{Tstart} in the \code{prothr} data already encodes this information.
In addition, \code{prothr} includes not only the observed transitions for each subject but also all possible transitions that were not realized.
This allows us to pass the dataset directly to \code{as\_ped} to build the PED.
Although the data is already in long format, it still lacks the complete set of time intervals covering all distinct event times.
The \code{as\_ped} function in \Rlang-chunk \ref{rchunk:prothr-dataprep} conveniently handles this expansion automatically.
Special care must be taken in case of instantaneous transitions, that is, transitions with zero sojourn time (\code{Tstart = Tstop}).
Instantaneous transitions can originate from data inaccuracies or be an artifact of the data generating process.
In the following illustrative example, for simplicity, we treat instant transitions as data inaccuracies and remove them.

\begin{knitrout}\small
\definecolor{shadecolor}{rgb}{0.961, 0.961, 0.961}\color{fgcolor}\begin{kframe}
\begin{rexample}\label{rchunk:prothr-dataprep}\hfill{}\begin{alltt}
\hlkwd{data}\hldef{(}\hlsng{"prothr"}\hldef{,} \hlkwc{package} \hldef{=} \hlsng{"mstate"}\hldef{)}
\hldef{ped_prothr} \hlkwb{=} \hlkwd{as_ped}\hldef{(}
  \hlkwc{data}       \hldef{= prothr |>}
                \hlkwd{filter}\hldef{(Tstart} \hlopt{!=} \hldef{Tstop) |>}
                \hlkwd{mutate}\hldef{(}\hlkwc{transition} \hldef{=} \hlkwd{factor}\hldef{(}\hlkwd{paste0}\hldef{(from,} \hlsng{"->"}\hldef{, to))),}
  \hlkwc{formula}    \hldef{=} \hlkwd{Surv}\hldef{(Tstart, Tstop, status)}\hlopt{~} \hldef{.} \hlopt{-} \hldef{trans,}
  \hlkwc{transition} \hldef{=} \hlsng{"transition"}\hldef{,}
  \hlkwc{id}         \hldef{=} \hlsng{"id"}\hldef{,}
  \hlkwc{timescale}  \hldef{=} \hlsng{"calendar"}
\hldef{)}
\end{alltt}
\end{rexample}\end{kframe}
\end{knitrout}

The \code{as\_ped} call in \Rlang-chunk \ref{rchunk:prothr-dataprep} transforms the survival data (cf. Table~\ref{tab:prothr-example}) into PED (cf. Table~\ref{tab:prothr-ped-example}) similarly to the single event case.
In order to make it work for multi-state settings, additional specifications must be taken into account:
\begin{compactitem}
  \item the \code{transition} argument contains the name of the column which stores the transition variable in arrow notation, e.g.,~\code{"0->1"}. Alternatively, the transition information can also be passed via the arguments \code{from} and \code{to}.

  \item The choice of time scale governs how the \code{formula} LHS is specified. In the default clock-forward setting (\code{timescale = "calendar"}), time runs continuously from study entry and the LHS must contain the internal left-truncation time in addition to the event time and status indicator, since subjects enter states at different times and are not at risk before doing so. In the clock-reset setting (\code{timescale = "gap"}), time resets to zero at each state transition and no left-truncation time is needed on the LHS.

  \item the \code{cut} argument need not to be specified for the multi-state data transformation, as all observed event times will be used as interval cut points. If \code{cut} is specified, the function will add the specified cut points to the observed event times.

  \item the $\delta_{ijk}$, which will again serve as the outcome of the Poisson regression, are stored in the column \code{ped\_status} and are 1 only in the intervals in which the subject experienced a state transition of type $k$.
  
\end{compactitem}

\begin{table}

\caption{\label{tab:prothr-ped-example}Data excerpt of the \texttt{ped\_prothr} data set, now containing all event times for every id, the time interval, and offset. Event indicator of PED is \texttt{ped\_status}}
\centering
\begin{tabular}[t]{rrrlrrrrll}
\toprule
id & tstart & tend & interval & offset & ped\_status & from & to & treat & transition\\
\midrule
16 & 0 & 4 & (0,4] & 1.39 & 0 & 1 & 2 & Placebo & 1->2\\
16 & 4 & 5 & (4,5] & 0.00 & 0 & 1 & 2 & Placebo & 1->2\\
16 & 5 & 13 & (5,13] & 2.08 & 0 & 1 & 2 & Placebo & 1->2\\
16 & 13 & 16 & (13,16] & 1.10 & 0 & 1 & 2 & Placebo & 1->2\\
16 & 16 & 21 & (16,21] & 1.61 & 0 & 1 & 2 & Placebo & 1->2\\
16 & 21 & 22 & (21,22] & 0.00 & 0 & 1 & 2 & Placebo & 1->2\\
16 & 22 & 23 & (22,23] & 0.00 & 0 & 1 & 2 & Placebo & 1->2\\
16 & 23 & 24 & (23,24] & 0.00 & 0 & 1 & 2 & Placebo & 1->2\\
16 & 24 & 25 & (24,25] & 0.00 & 0 & 1 & 2 & Placebo & 1->2\\
16 & 25 & 27 & (25,27] & 0.69 & 0 & 1 & 2 & Placebo & 1->2\\
\bottomrule
\end{tabular}
\end{table}

For estimation in the competing risk setting, we include a variable \code{cause} which specifies the different event types in order to be able to define cause-specific hazards (cf. Section \eqref{sub:cr-workflow}).
Similarly, in the multi-state example, we include the \code{transition} variable as a time-varying effect (and time-constant baseline effect) to fit different log-hazards for the different transitions, c.f.
\Rlang-chunk \ref{rchunk:prothr-baseline-model}.

\begin{knitrout}\small
\definecolor{shadecolor}{rgb}{0.961, 0.961, 0.961}\color{fgcolor}\begin{kframe}
\begin{rexample}\label{rchunk:prothr-baseline-model}\hfill{}\begin{alltt}
\hldef{pam_msm_base} \hlkwb{=} \hlkwd{pamm}\hldef{(}
  \hldef{ped_status} \hlopt{~} \hlkwd{s}\hldef{(tend,} \hlkwc{by}\hldef{=transition)} \hlopt{+} \hldef{transition,}
  \hlkwc{data} \hldef{= ped_prothr}
  \hldef{)}
\end{alltt}
\end{rexample}\end{kframe}
\end{knitrout}

We use this example to demonstrate that the transition probabilities estimated via a simple PAMM are in close agreement with those obtained from the non-parametric Aalen-Johansen estimator, implemented in the \pkg{etm} package.
In \Rlang-chunk \ref{rchunk:prothr-makenewdata}, similar to the single event case, we employ \pkg{pammtools}'s \code{make\_newdata} function to create a new data set with the desired variable combinations and use \code{add\_trans\_prob} to calculate the transition probabilities according to equation \eqref{eq:msm-transprob}.
It is crucial to group the data by the transition variable, since the transition probabilities depend on the prediction of the transition-specific cumulative hazards.

\begin{knitrout}\small
\definecolor{shadecolor}{rgb}{0.961, 0.961, 0.961}\color{fgcolor}\begin{kframe}
\begin{rexample}\label{rchunk:prothr-makenewdata}\hfill{}\begin{alltt}
\hldef{ndf_base} \hlkwb{=} \hlkwd{make_newdata}\hldef{(}
  \hldef{ped_prothr,}
  \hlkwc{tend}  \hldef{=} \hlkwd{unique}\hldef{(tend),}
  \hlkwc{transition} \hldef{=} \hlkwd{unique}\hldef{(transition)}
  \hldef{) |>}
  \hlkwd{group_by}\hldef{(transition) |>}
  \hlkwd{add_trans_prob}\hldef{(pam_msm_base,} \hlkwc{ci} \hldef{=} \hlnum{TRUE}\hldef{)}
\end{alltt}
\end{rexample}\end{kframe}
\end{knitrout}

\begin{figure}[!ht]
\begin{knitrout}\small
\definecolor{shadecolor}{rgb}{0.961, 0.961, 0.961}\color{fgcolor}

{\centering \includegraphics[width=\maxwidth]{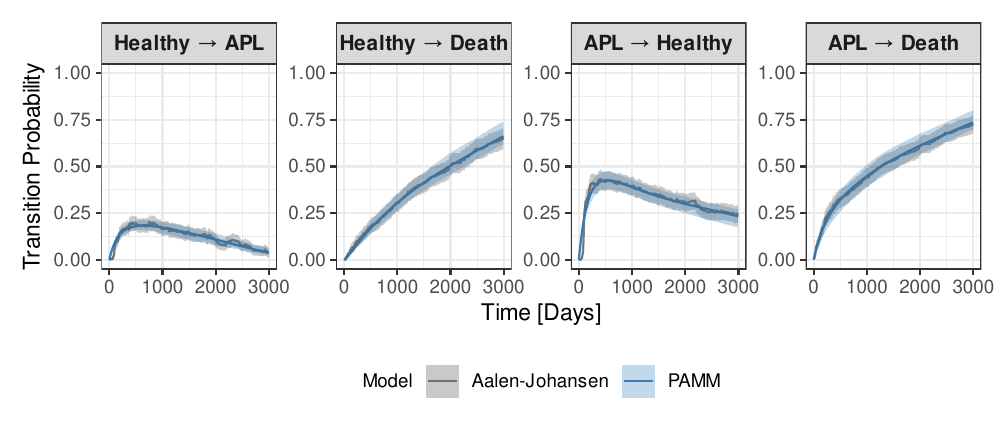} 

}

\end{knitrout}
\caption{Facets contain transition probabilities over time for each state transition. Probabilities are estimated via \pkg{etm}'s non-parametric Aalen-Johansen estimator (AJ, grey) and \pkg{pammtools}'s hazard-based transition probabilities (blue solid). State~2 (low/abnormal prothrombin level) is labelled APL.}
\label{fig:prothr-comparison}
\end{figure}
The results in Figure~\ref{fig:prothr-comparison} show that \pkg{pammtools}'s multi-state extension provides estimates comparable to the established Aalen-Johansen estimator. Quantitatively, over the plotted follow-up (up to 3000 days) the smooth PAMM transition probabilities and the non-parametric AJ estimates differ by at most $0.062$ (maximum absolute deviation) for the forward transitions and death. The largest discrepancy, $0.148$, occurs early (around day $74$) for the recurrent \emph{APL $\rightarrow$ Healthy} back-transition, and is confined to this short initial period: every subject starts in the \emph{Healthy} state, so the transition probability of a (counterfactual) start in \emph{APL} is weakly identified near the time origin, where the non-parametric estimate rises only in discrete steps as the first few back-transitions are observed while the smooth PAMM curve already follows the pooled hazard. Beyond this early transient the two estimators track each other closely, differing by at most $0.031$ over the remainder of the follow-up (after day 300). Note that the AJ reference must be constructed from the subjects' complete transition histories: retaining only each subject's first realised transition would discard every recurrence and bias the non-parametric estimator (see the \code{prothr.etm} construction in the replication code).

Next, we extend our model by including a time-varying treatment effect alongside a time-constant baseline effect, both of which are allowed to vary across transitions (see \Rlang-chunk \ref{rchunk:prothr-treatment-model}).
Note that - in general - not every covariate must be included as a time-varying effect, nor does every covariate need to be transition-specific.
As illustrated in Table~\ref{tab:tvEffectsPAM}, global effects shared across all transitions can be specified by omitting the interaction with the transition indicator.
Likewise, if a covariate is assumed to be time-constant, the interaction with time can be excluded.

\begin{knitrout}\small
\definecolor{shadecolor}{rgb}{0.961, 0.961, 0.961}\color{fgcolor}\begin{kframe}
\begin{rexample}\label{rchunk:prothr-treatment-model}\hfill{}\begin{alltt}
\hldef{pam_msm_treat} \hlkwb{=} \hlkwd{pamm}\hldef{(}
  \hldef{ped_status} \hlopt{~} \hlkwd{s}\hldef{(tend,} \hlkwc{by}\hldef{=}\hlkwd{interaction}\hldef{(treat, transition))} \hlopt{+}
    \hldef{treat}\hlopt{*}\hldef{transition,}
  \hlkwc{data} \hldef{= ped_prothr}
  \hldef{)}
\end{alltt}
\end{rexample}\end{kframe}
\end{knitrout}

In more complex models and especially in multi-state setting with many transition-specific effects, interpreting the summary statistics of a model can become challenging.
One convenient approach is to focus on how these effects affect the outcome distribution.
For this reason, \cite{Amrhein2019StatisticalSignificance} advocate to plot results - showing confidence intervals, cf. Figure~\ref{fig:prothr-comparison} - rather than solely relying on $p$~values and binary classification into statistical significance.

\begin{knitrout}\small
\definecolor{shadecolor}{rgb}{0.961, 0.961, 0.961}\color{fgcolor}\begin{kframe}
\begin{rexample}\label{rchunk:prothr-viz-msm-treat}\hfill{}\begin{alltt}
\hldef{ndf_msm_treat} \hlkwb{=} \hlkwd{make_newdata}\hldef{(}
  \hldef{ped_prothr,}
  \hlkwc{tend}       \hldef{=} \hlkwd{unique}\hldef{(tend),}
  \hlkwc{transition} \hldef{=} \hlkwd{unique}\hldef{(transition),}
  \hlkwc{treat}      \hldef{=} \hlkwd{unique}\hldef{(treat)}
  \hldef{) |>}
  \hlkwd{group_by}\hldef{(treat, transition) |>}
  \hlkwd{add_trans_prob}\hldef{(pam_msm_treat)}

\hldef{p_state_occupation} \hlkwb{=}
  \hlkwd{gg_state_occupation}\hldef{(}
    \hldef{ndf_msm_treat,}
    \hlkwc{group_var} \hldef{=} \hlsng{"treat"}\hldef{,}
    \hlkwc{init_state} \hldef{=} \hlkwd{c}\hldef{(}\hlnum{1}\hldef{,} \hlnum{0}\hldef{,} \hlnum{0}\hldef{)}
    \hldef{)}
\end{alltt}
\end{rexample}\end{kframe}
\end{knitrout}

\begin{figure}[!ht]
\begin{knitrout}\small
\definecolor{shadecolor}{rgb}{0.961, 0.961, 0.961}\color{fgcolor}

{\centering \includegraphics[width=\maxwidth]{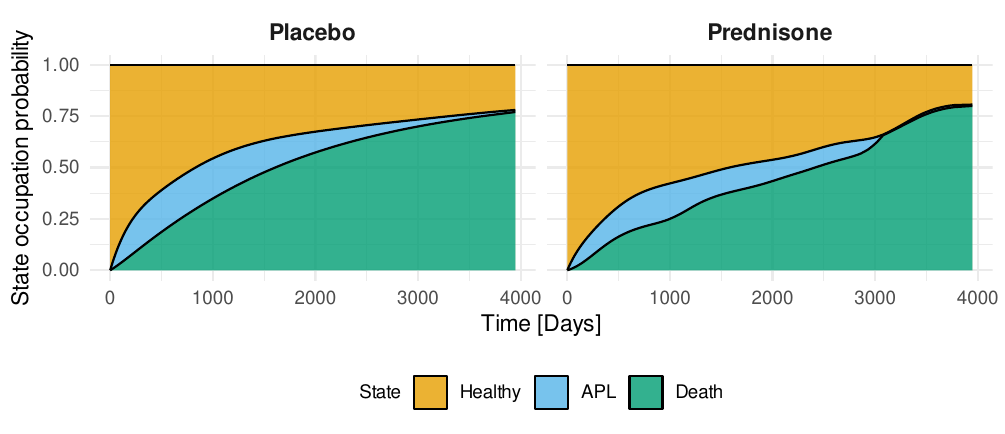} 

}

\end{knitrout}
\caption{Facets contain state occupation probabilities over time for each treatment (Placebo vs. Prednisone). Probabilities are estimated via \pkg{pammtools}'s hazard-based transition probabilities assuming every patient is healthy at the beginning of the study.}
\label{fig:prothr-state-occupation}
\end{figure}

The \pkg{pammtools} package implements many convenience functions for visualizing non-linear covariate effects and specific terms of the fitted model.
For example, in \Rlang-chunk \ref{rchunk:prothr-viz-msm-treat}, the convenience function \code{gg\_state_occupation} plots the state occupation probabilities of the specified \code{group\_var} and the initial state distribution \code{init_state}.
The resulting Figure~\ref{fig:prothr-state-occupation} displays the model-estimated distribution of individuals across all possible states at any given time point, under the assumption that all individuals started in the ``healthy'' state.
The figure shows that the probability of remaining healthy and alive for longer is higher for patients receiving Prednisone than for those receiving Placebo and that especially the probability of death is estimated to be much lower for patients treated with Prednisone than for those receiving a placebo during the first 3,000 days of follow-up.

Additional convenience functions can be found in Tables \ref{tab:ggplot2-extensions} and Table~\ref{tab:ggplot2-convenience} as well as in the \href{https://adibender.github.io/pammtools/reference/index.html#convenience-functions-for-visualization-of-model-outputs-}{vignette on convenience functions for visualization of model outputs}.

\subsection{Restricted mean survival time: A case study of inference for non-linear functionals}
\label{sub:rmst}

The restricted mean survival time (RMST) \citep{zhao2016}, defined as the area under the survival curve, 

\begin{equation}
  RMST(\tau) =  E[\min(T, \tau)] = \int_0^\tau S(s)ds
\end{equation}

offers an intuitive summary of survival over a fixed time horizon (i.e., ``on
average, patients survived $RMST(\tau)$ days of the first $\tau$ days.'').
It has gained attention in recent years as a clinically interpretable alternative for assessing treatment and covariate effects in randomized trials, particularly when the proportional hazards assumption does not hold \citep{Royston2011, Royston2013}.

Classical survival models estimate the restricted mean survival time by first modeling the survival function and then integrating it between $0$ and $\tau$ \citep{Karrison1987, zucker1998}.
With this indirect approach, covariate adjustment is possible when modeling the survival function.
Due to the PAMM assumption of piecewise exponential data, calculating the RMST becomes straightforward using \pkg{pammtools}' post-processing functions: 

\begin{equation}
RMST(\tau) = \sum_{j=1}^{J} S(\kappa_{j-1}) \cdot \frac{1 - e^{-h_j (\kappa_j^* - \kappa_{j-1})}}{h_j},
\end{equation}

with $J$ being the number of intervals such that $\kappa_{J-1} \le \tau < \kappa_{J}$ and $\kappa_j^* = \min(\tau, \kappa_j)$.

\Rlang-Chunk \ref{rchunk:rmst-fct} shows an implementation of the RMST calculation taking prediction data \code{ndf} created with \code{make\_newdata}, estimated log-hazard rates excluding the offset \code{lp} (i.e., the "linear predictor" of a PAMM model), and a time \code{tau} as input.
The function assumes \code{ndf} contains a single covariate profile (one row per interval, ordered by \code{tend}); the leading \code{stopifnot()} enforces this, since \code{cumsum} and \code{lag} would otherwise silently run across distinct profiles.
For multiple profiles, we could wrap the call in \code{dplyr::group\_modify()} or apply it per group.

\begin{knitrout}\small
\definecolor{shadecolor}{rgb}{0.961, 0.961, 0.961}\color{fgcolor}\begin{kframe}
\begin{rexample}\label{rchunk:rmst-fct}\hfill{}\begin{alltt}
\hlcom{# compute RMST for time tau}
\hldef{rmst_lp} \hlkwb{<-} \hlkwa{function}\hldef{(}\hlkwc{ndf}\hldef{,} \hlkwc{lp}\hldef{,} \hlkwc{tau}\hldef{) \{}
  \hlkwd{stopifnot}\hldef{(}
    \hlsng{"rmst_lp expects a single covariate profile (one row per interval)."} \hldef{=}
      \hlkwd{anyDuplicated}\hldef{(ndf}\hlopt{$}\hldef{tend)} \hlopt{==} \hlnum{0L}
  \hldef{)}
  \hldef{ndf |>}
    \hlkwd{mutate}\hldef{(}
      \hlkwc{lp}     \hldef{= lp,}
      \hlkwc{tstart} \hldef{=} \hlkwd{lag}\hldef{(tend,} \hlkwc{default} \hldef{=} \hlnum{0}\hldef{)}
    \hldef{) |>}
    \hlkwd{filter}\hldef{(tstart} \hlopt{<} \hldef{tau) |>}
    \hlkwd{mutate}\hldef{(}
      \hlkwc{intlen}    \hldef{=} \hlkwd{pmin}\hldef{(tend, tau)} \hlopt{-} \hldef{tstart,}
      \hlkwc{hazard}    \hldef{=} \hlkwd{exp}\hldef{(lp),}
      \hlkwc{S_prev}    \hldef{=} \hlkwd{lag}\hldef{(}\hlkwd{exp}\hldef{(}\hlopt{-}\hlkwd{cumsum}\hldef{(hazard} \hlopt{*} \hldef{intlen)),} \hlkwc{default} \hldef{=} \hlnum{1}\hldef{),}
      \hlkwc{rmst_incr} \hldef{= S_prev} \hlopt{*} \hldef{(}\hlnum{1} \hlopt{-} \hlkwd{exp}\hldef{(}\hlopt{-}\hldef{hazard} \hlopt{*} \hldef{intlen))} \hlopt{/} \hldef{hazard}
    \hldef{) |>}
    \hlkwd{summarise}\hldef{(}\hlkwc{rmst} \hldef{=} \hlkwd{sum}\hldef{(rmst_incr)) |>}
    \hlkwd{pull}\hldef{(rmst)}
\hldef{\}}
\end{alltt}
\end{rexample}\end{kframe}
\end{knitrout}

Beyond calculating RMST point estimates (and their differences), we are also interested in inference.
To build confidence intervals for the RMST difference we can simulate \code{nsim} draws from the posterior of our linear predictor and repeat the RMST calculation for all draws (c.f. Chapter \ref{sub:inference}).
We then compute upper and lower $\alpha / 2$ quantiles of these RMST values to obtain confidence bounds.
The function \code{rmst\_diff\_sim} in \Rlang-Chunk \ref{rchunk:rmst-inference} implements this procedure: 
it first obtains the design (or \code{lpmatrix}) for each of the two covariate profiles \code{ndf1} and \code{ndf2}, then draws \code{nsim} coefficient vectors from the (approximate) posterior of the fitted model, and uses these to simulate the corresponding linear predictors (log-hazards, without offsets) for each profile. For each simulated draw, the RMST up to time \code{tau} is calculated for both profiles via \code{rmst\_lp}, and the resulting RMST differences across all draws are summarized by their mean and their 2.5\% and 97.5\% quantiles, yielding a point estimate and confidence interval for the RMST difference between the two profiles.

\begin{knitrout}\small
\definecolor{shadecolor}{rgb}{0.961, 0.961, 0.961}\color{fgcolor}\begin{kframe}
\begin{rexample}\label{rchunk:rmst-inference}\hfill{}\begin{alltt}
\hldef{rmst_diff_sim} \hlkwb{=} \hlkwa{function}\hldef{(}\hlkwc{ndf1}\hldef{,} \hlkwc{ndf2}\hldef{,} \hlkwc{model}\hldef{,} \hlkwc{tau}\hldef{,} \hlkwc{nsim} \hldef{=} \hlnum{1000}\hldef{,} \hlkwc{seed} \hldef{=} \hlnum{1312}\hldef{) \{}
  \hlkwd{set.seed}\hldef{(seed)}
  \hldef{X1} \hlkwb{=} \hlkwd{predict.gam}\hldef{(model,} \hlkwc{newdata} \hldef{= ndf1,} \hlkwc{type} \hldef{=} \hlsng{"lpmatrix"}\hldef{)}
  \hldef{X2} \hlkwb{=} \hlkwd{predict.gam}\hldef{(model,} \hlkwc{newdata} \hldef{= ndf2,} \hlkwc{type} \hldef{=} \hlsng{"lpmatrix"}\hldef{)}
  \hldef{sim_coef_mat} \hlkwb{=} \hldef{mvtnorm}\hlopt{::}\hlkwd{rmvnorm}\hldef{(nsim,} \hlkwc{mean} \hldef{=} \hlkwd{coef}\hldef{(model),} \hlkwc{sigma} \hldef{= model}\hlopt{$}\hldef{Vp)}
  \hldef{sim_fit_mat1} \hlkwb{=} \hlkwd{apply}\hldef{(sim_coef_mat,} \hlnum{1}\hldef{,} \hlkwa{function}\hldef{(}\hlkwc{z}\hldef{) X1} \hlopt{%*%} \hldef{z)}
  \hldef{sim_fit_mat2} \hlkwb{=} \hlkwd{apply}\hldef{(sim_coef_mat,} \hlnum{1}\hldef{,} \hlkwa{function}\hldef{(}\hlkwc{z}\hldef{) X2} \hlopt{%*%} \hldef{z)}
  \hldef{rmst1} \hlkwb{=} \hlkwd{apply}\hldef{(sim_fit_mat1,} \hlnum{2}\hldef{,} \hlkwa{function}\hldef{(}\hlkwc{z}\hldef{)} \hlkwd{rmst_lp}\hldef{(ndf1, z, tau))}
  \hldef{rmst2} \hlkwb{=} \hlkwd{apply}\hldef{(sim_fit_mat2,} \hlnum{2}\hldef{,} \hlkwa{function}\hldef{(}\hlkwc{z}\hldef{)} \hlkwd{rmst_lp}\hldef{(ndf2, z, tau))}
  \hldef{rmst_diff} \hlkwb{=} \hldef{rmst1} \hlopt{-} \hldef{rmst2}

  \hlkwd{data.frame}\hldef{(}
    \hlkwc{rmst_diff_lower} \hldef{=} \hlkwd{quantile}\hldef{(rmst_diff,} \hlnum{0.025}\hldef{) |>} \hlkwd{unname}\hldef{(),}
    \hlkwc{rmst_diff_mean} \hldef{=} \hlkwd{mean}\hldef{(rmst_diff) |>} \hlkwd{unname}\hldef{(),}
    \hlkwc{rmst_diff_upper} \hldef{=} \hlkwd{quantile}\hldef{(rmst_diff,} \hlnum{0.975}\hldef{) |>} \hlkwd{unname}\hldef{()}
  \hldef{)}
\hldef{\}}
\end{alltt}
\end{rexample}\end{kframe}
\end{knitrout}

\Rlang-Chunk \ref{rchunk:rmst-example}, shows the RMST calculation on the previous \code{tumor} example, investigating the effect of complications on death (cf. Section~\ref{sub:singleevent-workflow}), including posterior-simulated confidence intervals based on \Rlang-Chunk \ref{rchunk:rmst-inference}.

\begin{knitrout}\small
\definecolor{shadecolor}{rgb}{0.961, 0.961, 0.961}\color{fgcolor}\begin{kframe}
\begin{rexample}\label{rchunk:rmst-example}\hfill{}\begin{alltt}
\hldef{ndf_yes} \hlkwb{=} \hldef{ped_tumor |>}
  \hlkwd{filter}\hldef{(complications} \hlopt{==} \hlsng{"yes"}\hldef{) |>} \hlkwd{make_newdata}\hldef{(}\hlkwc{tend} \hldef{=} \hlkwd{unique}\hldef{(tend))}
\hldef{ndf_no} \hlkwb{=} \hldef{ped_tumor |>}
  \hlkwd{filter}\hldef{(complications} \hlopt{==} \hlsng{"no"}\hldef{) |>} \hlkwd{make_newdata}\hldef{(}\hlkwc{tend} \hldef{=} \hlkwd{unique}\hldef{(tend))}

\hldef{rmst_diff_tab} \hlkwb{=} \hlkwd{rmst_diff_sim}\hldef{(}\hlkwc{ndf1} \hldef{= ndf_no,} \hlkwc{ndf2} \hldef{= ndf_yes,}
                              \hlkwc{model} \hldef{= pam_tumor_comp,} \hlkwc{tau} \hldef{=} \hlnum{1000}\hldef{)}
\hldef{rmst_diff_tab}
\end{alltt}
\begin{verbatim}
  rmst_diff_lower rmst_diff_mean rmst_diff_upper
1        191.6298       250.9146        314.8796
\end{verbatim}
\end{rexample}\end{kframe}
\end{knitrout}

The results indicate that subjects with complications lose on average 251 days of life between surgery and day 1000 (95\% CI: [191.6, 314.9]).
While estimated hazard rates roughly converge after about 250 days, the difference in RMST at day 1000 is pronounced, driven by the sharp decline in survival among patients with complications shortly after surgery (cf. Figure~\ref{fig:tumor-strata-viz}).


\section{Extended topics}
\label{sec:extended_topics}

Previous sections demonstrated the general and mostly highly similar workflows for PAMMs for single-event, competing risk and multi-state data.
This section showcases more specialized topics that arise in practice and benefit from the additional flexibility provided by \pkg{pammtools}.
PED pre-processing for PAMMs depends not only on the different outcome types, but also on the type of covariates that are present, specifically time-constant (TCC) vs. time-varying (TVC) covariates, as well as the type of effects one wants to estimate (time-constant or time-varying for TCCs, concurrent or cumulative for TVCs).
Specifics depend on the following categorizations:

\begin{compactitem}
  \item TCCs with (potentially time-varying) effects $f(t, x)$ or $f(x)$
  \item TVCs with concurrent effects $f(t, z(t))$
  \item TVCs with cumulative effects $\int_{\tw{}}h(t, \tz, z(\tz))\drm\tz$
\end{compactitem}

The following sections demonstrate these different modelling tasks for single-event time-to-event data.
In addition to these covariate-related extensions, Section~\ref{sub:ic} presents an extension with respect to the \emph{outcome}: the estimation of PAMMs for interval-censored event times via multiple imputation.

Finally, Section~\ref{sub:postprocessing} contains an overview of all \code{add\_*}-functions and convenience functions for visualization, many of which were already used in previous sections.

\subsection{TCCs with potentially time-varying effects}
\label{sub:tcc}

\subsubsection{Pre-processing}
\label{subsub:tcc:preprocessing}
Data pre-processing for TCCs with potentially time-varying effects does not differ from pre-processing single-event time-to-event data and the relevant workflow using \code{as\_ped} and its output is described in \ref{sub:singleevent-workflow}.
Building up on the previous example, the following code snippets give information on already used functionality and introduce additional features of \code{as\_ped}.

\Rlang-chunk \ref{rchunk:tumor-split_ex} illustrates applying \code{as\_ped} for the first two rows for each category of the \code{sex} variable of the \code{tumor} data.
Different to the previous workflow, we define interval cut points by using a very crude 200-day partition of the follow up.

\begin{knitrout}\small
\definecolor{shadecolor}{rgb}{0.961, 0.961, 0.961}\color{fgcolor}\begin{kframe}
\begin{rexample}\label{rchunk:tumor-split_ex}\hfill{}\begin{alltt}
\hldef{tumor_sub} \hlkwb{=} \hldef{tumor |>} \hlkwd{select}\hldef{(}\hlkwd{c}\hldef{(}\hlnum{1}\hldef{,} \hlnum{2}\hldef{,} \hlnum{4}\hldef{,} \hlnum{5}\hldef{)) |>} \hlkwd{group_by}\hldef{(sex) |>} \hlkwd{slice}\hldef{(}\hlnum{1}\hlopt{:}\hlnum{2}\hldef{)}
\hldef{tumor_sub |>} \hlkwd{ungroup}\hldef{() |>} \hlkwd{mutate}\hldef{(}\hlkwc{id} \hldef{=} \hlnum{1}\hlopt{:}\hlnum{4}\hldef{)}
\end{alltt}
\begin{verbatim}
# A tibble: 4 x 5
   days status   age sex       id
  <dbl>  <int> <int> <fct>  <int>
1  1192      0    52 male       1
2    33      1    57 male       2
3   579      0    58 female     3
4   308      1    74 female     4
\end{verbatim}
\begin{alltt}
\hldef{ped_tumor_sub} \hlkwb{=} \hldef{tumor_sub |>}
  \hlkwd{as_ped}\hldef{(}\hlkwd{Surv}\hldef{(days, status)} \hlopt{~}\hldef{.,} \hlkwc{cut} \hldef{=} \hlkwd{seq}\hldef{(}\hlnum{0}\hldef{,} \hlnum{1000}\hldef{,} \hlkwc{by} \hldef{=} \hlnum{200}\hldef{))}
\hldef{ped_tumor_sub |>} \hlkwd{select}\hldef{(}\hlnum{1}\hlopt{:}\hlnum{8}\hldef{)}
\end{alltt}
\begin{verbatim}
# A tibble: 11 x 8
# Groups:   sex [2]
      id tstart  tend interval   offset ped_status   age sex   
 * <int>  <dbl> <dbl> <fct>       <dbl>      <dbl> <int> <fct> 
 1     1      0   200 (0,200]      5.30          0    52 male  
 2     1    200   400 (200,400]    5.30          0    52 male  
 3     1    400   600 (400,600]    5.30          0    52 male  
 4     1    600   800 (600,800]    5.30          0    52 male  
 5     1    800  1000 (800,1000]   5.30          0    52 male  
 6     2      0   200 (0,200]      3.50          1    57 male  
 7     3      0   200 (0,200]      5.30          0    58 female
 8     3    200   400 (200,400]    5.30          0    58 female
 9     3    400   600 (400,600]    5.19          0    58 female
10     4      0   200 (0,200]      5.30          0    74 female
11     4    200   400 (200,400]    4.68          1    74 female
\end{verbatim}
\end{rexample}\end{kframe}
\end{knitrout}
In the \code{as\_ped} call in \Rlang-chunk \ref{rchunk:tumor-split_ex}
\begin{compactitem}
  \item the left hand side (LHS) of the \code{formula} specifies the event time
  and status information.
  \item the right hand side (RHS) of the \code{formula} specifies covariates
  that should be kept after data transformation. This can be useful when the data
  contains many variables but only a few will be used to estimate the hazard.
  As usual, a dot (\code{~ . }) can be used to include all variables.
  \item the follow up is partitioned at the split points $\kappa_j$
  provided through the \code{cut} argument. If the \code{cut} argument is not specified, the function uses all observed event times as interval cut points (cf. \ref{sub:singleevent-workflow}) The start (\code{tstart}) and stop
  (\code{tend}) times are created as well as an \code{interval} column.
  \item the $\delta_{ij}$, which will serve as the outcome of the Poisson
  regression, are stored in the column \code{ped\_status} and are 1 only in the
  interval in which the subject experienced an event (if uncensored), which is
  also the final interval for that subject.
  \item the offset variable is calculated, e.g., subject \code{id = 3}
    was censored at 579 days, therefore
    $o_{i=3,j=3} = \log(\min(579-400, 600-400)) = \log(179) \approx 5.19$.
  \item subjects with event times $t_i > \kappa_J$ will be administratively
  censored at $\kappa_J$ (see \code{id = 1}).
\end{compactitem}

In the \code{as\_ped} call in \Rlang-chunk~\ref{rchunk:tumor-split_ex}, the \code{formula} LHS, RHS, \code{ped\_status}, and offset columns follow the same conventions as described in Section~\ref{sub:singleevent-workflow}. Note that, if  
\code{cut} is specified explicitly, subjects with event times $t_i > \kappa_J$ will be administratively censored at $\kappa_J$ (see \code{id = 1}).
The output data has class \code{ped} (inheriting from \code{data.frame}) and \pkg{pammtools} provides several \code{S3} methods that dispatch on such \code{ped} objects (see Appendix \ref{app:implementation}).\\


\subsubsection{Modelling and interpretation}
\label{subsub:tcc:modelling}

Linearly time-varying effects of time-constant covariates $f(t)\cdot x$ can generally be divided in two groups:

\begin{compactitem}
  \item stratified hazards for a categorical $x$ (cf. Section~\ref{sub:singleevent-workflow}) defining the strata, and
  \item time-varying coefficients for continuous $x$.
\end{compactitem}

We have already seen the first kind in Section~\ref{sub:singleevent-workflow}.
Let's now include all covariates available in the \code{tumor} data, with possibly non-linearly time-varying effects, where the effects of continuous covariates are allowed to vary non-linearly in time, but linearly in the covariate, i.e., $f_p(t)x_p$.
The model specification is given in \Rlang-chunk \ref{rchunk:pam_tumor_tve}.
Note that categorical covariates are included using \code{by = as.ordered(...)}, which (together with \code{ti}) ensures identifiability of the model (cf. \code{?mgcv::factor.smooth}).
No further identifiability constraints are necessary for the linear time-varying effects of numerical covariates \code{age} and \code{charlson\_score}, cf. \code{?mgcv::identifiability}.

\begin{knitrout}\small
\definecolor{shadecolor}{rgb}{0.961, 0.961, 0.961}\color{fgcolor}\begin{kframe}
\begin{rexample}\label{rchunk:pam_tumor_tve}\hfill{}\begin{alltt}
\hldef{pam_tumor_tve} \hlkwb{=} \hlkwd{bam}\hldef{(}
  \hlkwc{formula} \hldef{= ped_status} \hlopt{~} \hlkwd{ti}\hldef{(tend,} \hlkwc{k} \hldef{=} \hlnum{10}\hldef{)} \hlopt{+}
    \hldef{complications} \hlopt{+} \hlkwd{ti}\hldef{(tend,} \hlkwc{by} \hldef{=} \hlkwd{as.ordered}\hldef{(complications),} \hlkwc{k} \hldef{=} \hlnum{10}\hldef{)} \hlopt{+}
    \hldef{metastases}    \hlopt{+} \hlkwd{ti}\hldef{(tend,} \hlkwc{by} \hldef{=} \hlkwd{as.ordered}\hldef{(metastases),} \hlkwc{k} \hldef{=} \hlnum{10}\hldef{)}    \hlopt{+}
    \hldef{sex}           \hlopt{+} \hlkwd{ti}\hldef{(tend,} \hlkwc{by} \hldef{=} \hlkwd{as.ordered}\hldef{(sex),} \hlkwc{k} \hldef{=} \hlnum{10}\hldef{)}           \hlopt{+}
    \hldef{transfusion}   \hlopt{+} \hlkwd{ti}\hldef{(tend,} \hlkwc{by} \hldef{=} \hlkwd{as.ordered}\hldef{(transfusion),} \hlkwc{k} \hldef{=} \hlnum{10}\hldef{)}   \hlopt{+}
    \hldef{resection}     \hlopt{+} \hlkwd{ti}\hldef{(tend,} \hlkwc{by} \hldef{=} \hlkwd{as.ordered}\hldef{(resection),} \hlkwc{k} \hldef{=} \hlnum{10}\hldef{)}     \hlopt{+}
    \hlkwd{s}\hldef{(tend,} \hlkwc{by} \hldef{= charlson_score)} \hlopt{+}
    \hlkwd{s}\hldef{(tend,} \hlkwc{by} \hldef{= age),}
  \hlkwc{data}   \hldef{= ped_tumor,} \hlkwc{family} \hldef{=} \hlkwd{poisson}\hldef{(),} \hlkwc{offset} \hldef{= offset,}
  \hlkwc{method} \hldef{=} \hlsng{"fREML"}\hldef{,} \hlkwc{discrete} \hldef{=} \hlnum{TRUE}\hldef{)}
\end{alltt}
\end{rexample}\end{kframe}
\end{knitrout}

\begin{knitrout}\small
\definecolor{shadecolor}{rgb}{0.961, 0.961, 0.961}\color{fgcolor}\begin{kframe}
\begin{rexample}\label{rchunk:smry_pam_tumor_tve}\hfill{}\begin{alltt}
\hlkwd{summary}\hldef{(pam_tumor_tve)}
\end{alltt}
\begin{verbatim}
...
Parametric coefficients:
                 Estimate Std. Error z value Pr(>|z|)    
(Intercept)       -9.9337     0.3623 -27.422  < 2e-16 ***
complicationsyes   0.3829     0.1231   3.110  0.00187 ** 
metastasesyes      0.2147     0.1183   1.815  0.06957 .  
sexfemale          0.2157     0.1085   1.988  0.04678 *  
transfusionyes     0.2054     0.1156   1.777  0.07561 .  
resectionyes       0.2817     0.1134   2.484  0.01300 *  

Approximate significance of smooth terms:
                                        edf Ref.df Chi.sq  p-value    
ti(tend)                              1.245  1.438  2.076 0.168216    
ti(tend):as.ordered(complications)yes 5.060  5.934 87.429  < 2e-16 ***
ti(tend):as.ordered(metastases)yes    1.001  1.002 11.820 0.000592 ***
ti(tend):as.ordered(sex)female        2.169  2.703  3.764 0.203067    
ti(tend):as.ordered(transfusion)yes   1.000  1.000  2.229 0.135474    
ti(tend):as.ordered(resection)yes     1.000  1.000  2.783 0.095310 .  
s(tend):charlson_score                2.000  2.000 22.386 1.37e-05 ***
...
\end{verbatim}
\end{rexample}\end{kframe}
\end{knitrout}

The model output is presented in \Rlang-chunk \ref{rchunk:smry_pam_tumor_tve}.
The effects of variables \code{metastases}, \code{transfusion} and \code{resection} were estimated as linearly time-varying effects (edf=1), however, they must be interpreted as relative changes (\emph{ceteris paribus} and also taking into account their constant components estimated as parametric coefficients) compared to the baseline log-hazard (\code{(Intercept) + ti(tend)}), which itself is (slightly) non-linear.

The usual visualization of the log-hazard contributions $f_p(t)x_p$ over the follow-up could be used for the interpretation of the estimates.
However, for models with time-varying effects (that are linear in the covariates), an alternative visualization, which is also useful for comparisons to the non-parametric additive Aaalen model \citep{Martinussen2006}, will be used here.
\\

The default visualization of covariate effect estimates for the Aalen model in the \pkg{timereg} package is the so-called cumulative coefficient $\mathcal{B}_p(t) = \int_0^t \beta_p(s)\drm s$.
Since the Aalen model is additive, i.e., $h(t|\bfx) = h_0(t) + \beta_1(t)x_1(t) + \cdots$, this cumulative coefficient can be nicely interpreted as the cumulative hazard difference at time $t$ for a 1 unit increase of the covariate/compared to its reference level (c.p.), i.e., $\mathcal{B}(t) = H(t|x+1)-H(t|x)$.
Thus, to obtain a PAMM analog of the cumulative coefficient, we can calculate the difference between the respective cumulative hazards.
Although $\mathcal{B}(t)$ is not directly estimated for PAMMs as it is for the Aalen model, \pkg{pammtools} provides the function \code{get\_cumu\_coef} that performs these calculations (including simulation based confidence intervals), as illustrated in \Rlang-chunk \ref{rchunk:ex_cumu_coef}. The resulting \code{cumu_hazard} column denotes these cumulative hazard differences, here for a one-unit change in \code{age} and for females versus males (\code{sex}).
\\

\begin{knitrout}\small
\definecolor{shadecolor}{rgb}{0.961, 0.961, 0.961}\color{fgcolor}\begin{kframe}
\begin{rexample}\label{rchunk:ex_cumu_coef}\hfill{}\begin{alltt}
\hldef{pam_tumor_tve |>}
  \hlkwd{get_cumu_coef}\hldef{(ped_tumor,} \hlkwc{terms} \hldef{=} \hlkwd{c}\hldef{(}\hlsng{"age"}\hldef{,} \hlsng{"sex"}\hldef{)) |>}
  \hlkwd{group_by}\hldef{(variable) |>} \hlkwd{slice}\hldef{(}\hlnum{1}\hlopt{:}\hlnum{2}\hldef{)}
\end{alltt}
\begin{verbatim}
# A tibble: 4 x 6
# Groups:   variable [2]
  method variable      time cumu_hazard  cumu_lower cumu_upper
  <chr>  <chr>        <dbl>       <dbl>       <dbl>      <dbl>
1 bam    age              1  0.00000465  0.00000141  0.0000112
2 bam    age              2  0.00000930  0.00000282  0.0000224
3 bam    sex (female)     1 -0.0000220  -0.000102    0.000113 
4 bam    sex (female)     2 -0.0000438  -0.000203    0.000225 
\end{verbatim}
\end{rexample}\end{kframe}
\end{knitrout}

\begin{figure}[!hb]
\begin{knitrout}\small
\definecolor{shadecolor}{rgb}{0.961, 0.961, 0.961}\color{fgcolor}

{\centering \includegraphics[width=\maxwidth]{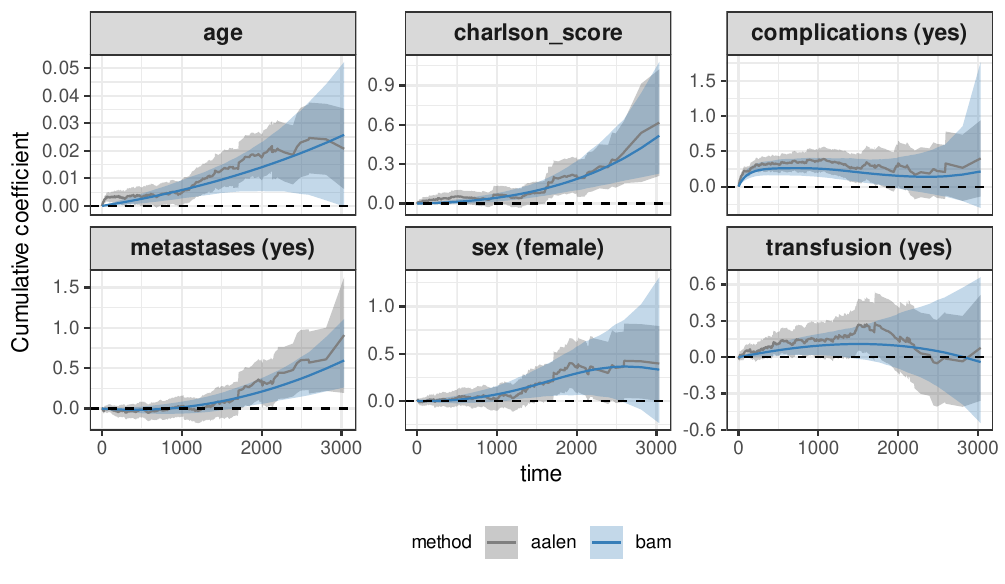} 

}

\end{knitrout}
\caption{Comparison of cumulative coefficients estimated with PAMMs (blue) and
the additive Aalen model (grey) respectively (the effect of resection is not displayed
for conciseness). For PAMMs these are defined as cumulative hazard differences,
e.g., $\mathcal{B}_{\tn{PAMM}}(t):=H(t|\tn{sex = "female"})-H(t|\tn{sex = "male"})$.
Note the different scales on the vertical axes of the panels.}
\label{fig:pam_aalen_comparison}
\end{figure}

The cumulative coefficients of the PAMM and Aalen model are presented in Figure
\ref{fig:pam_aalen_comparison}.
The cumulative hazard difference between a patient with complications (compared to one without, c.p.), increases at the beginning, directly after the operation when complications occurred, while after approximately 500 days, the cumulative hazard difference remains constant (i.e., $\beta_p(t) =f_p(t)\approx 0 \,\forall\, t>500$).
Similarly, the effect of metastases has a plausible interpretation: At $t=0$, as much as possible of the cancerous tissue including metastases is removed, thus the hazard in both groups is almost the same in the beginning, however, the risk of cancer returning after some time due to cancerous tissue that was not removed is higher in patients with metastases, which notably increases their hazard for $t > 1500$ compared to patients without metastases.
For the cumulative coefficients based on PAMMs, confidence intervals were estimated by Monte Carlo estimation based on 1000 draws from the model coefficients' posterior distribution \citep{Argyropoulos2015,Wood2017a}.
Overall, the estimates obtained from this PAMM are fairly close to the estimates obtained from the Aalen model with respect to the cumulative coefficients as well as their confidence intervals, despite the large structural difference between the two models being compared here.

\subsection{TVCs with concurrent (time-varying) effects}
\label{sub:tvc}


\subsubsection{Pre-processing}
\label{subsub:tvc:prep}

Transformation of data containing time-varying covariates involves a little more work, as, usually, the interval split points $\kappa_j$ are now the union of the user-specified or event-time based split points and the time points at which (changes in) the time-varying covariate(s) were recorded.
\\

In this section, we use the \code{pbc} data \citep{Therneau2001}, provided by the \pkg{survival} package (see \code{?pbc} and \Rlang-chunk \ref{rchunk:load_pbc}), ignoring the potentially dependent competing risks, focusing only on the endpoint death (see also \code{vignette("timedep", package="survival")}).
Note that by loading \code{pbc}, two data sets are loaded, the first, \code{pbc}, contains survival information and time-constant covariates (and values of time-varying covariates recorded at beginning of the follow-up) and \code{pbcseq}, which stores information on time-varying covariates.
\\

The variables defining the structure of the data are
\begin{compactitem}
  \item the subject indicator (\code{id}),
  \item the time to event (\code{time}),
  \item the event indicator (\code{status}),
  \item the time of exposure/time at which TVCs were observed (\code{day}).
\end{compactitem}

Note that only the first 312 observations in \code{pbc} also have time-dependent information in \code{pbcseq}, therefore we only use this part of the data.

As shown in \Rlang-chunk \ref{rchunk:load_pbc}, we restrict \code{pbc} to these 312 subjects, recode \code{status} to a binary event indicator for death, and apply a log-transformation to the biomarkers \code{bili} and \code{protime}, retaining only the subject identifier, event time and status, the time-constant covariates \code{trt} through \code{sex}, and the two (log-transformed) biomarkers. The time-varying covariate information in \code{pbcseq} is prepared analogously, applying the same log-transformation to \code{bili} and \code{protime} and retaining only the subject identifier, observation time \code{day}, and the two biomarkers.

\begin{knitrout}\small
\definecolor{shadecolor}{rgb}{0.961, 0.961, 0.961}\color{fgcolor}\begin{kframe}
\begin{rexample}\label{rchunk:load_pbc}\hfill{}\begin{alltt}
\hlcom{# Note that this code loads two data sets, pbc and pbcseq}
\hlkwd{data}\hldef{(}\hlsng{"pbc"}\hldef{,} \hlkwc{package}\hldef{=}\hlsng{"survival"}\hldef{)}
\hldef{pbc} \hlkwb{=} \hldef{pbc |>}
  \hlkwd{filter}\hldef{(id} \hlopt{<=} \hlnum{312}\hldef{) |>}
  \hlkwd{mutate}\hldef{(}\hlkwc{status} \hldef{=} \hlnum{1L}\hlopt{*}\hldef{(status} \hlopt{==} \hlnum{2}\hldef{),}
         \hlkwc{bili} \hldef{=} \hlkwd{log}\hldef{(bili),}
         \hlkwc{protime} \hldef{=} \hlkwd{log}\hldef{(protime)) |>}
  \hlkwd{select}\hldef{(id}\hlopt{:}\hldef{status, trt}\hlopt{:}\hldef{sex, bili, protime)}
\hldef{pbc |>} \hlkwd{slice}\hldef{(}\hlnum{1}\hlopt{:}\hlnum{6}\hldef{)}
\end{alltt}
\begin{verbatim}
  id time status trt      age sex        bili  protime
1  1  400      1   1 58.76523   f  2.67414865 2.501436
2  2 4500      0   1 56.44627   f  0.09531018 2.360854
3  3 1012      1   1 70.07255   m  0.33647224 2.484907
4  4 1925      1   1 54.74059   f  0.58778666 2.332144
5  5 1504      0   2 38.10541   f  1.22377543 2.388763
6  6 2503      1   2 66.25873   f -0.22314355 2.397895
\end{verbatim}
\begin{alltt}
\hldef{pbcseq} \hlkwb{=} \hldef{pbcseq |>}
  \hlkwd{mutate}\hldef{(}\hlkwc{bili} \hldef{=} \hlkwd{log}\hldef{(bili),}
         \hlkwc{protime} \hldef{=} \hlkwd{log}\hldef{(protime)) |>}
  \hlkwd{select}\hldef{(id, day, bili, protime)}
\hldef{pbcseq |>} \hlkwd{slice}\hldef{(}\hlnum{1}\hlopt{:}\hlnum{6}\hldef{)}
\end{alltt}
\begin{verbatim}
  id day        bili  protime
1  1   0  2.67414865 2.501436
2  1 192  3.05870707 2.415914
3  2   0  0.09531018 2.360854
4  2 182 -0.22314355 2.397895
5  2 365  0.00000000 2.451005
6  2 768  0.64185389 2.360854
\end{verbatim}
\end{rexample}\end{kframe}
\end{knitrout}

To combine these data sets and to transform them into the PED format, we again use the \code{as\_ped} function, however, the first argument is a list of data sets and the variables that should be treated as concurrent variables along with the name of the variable recording their measurement times are specified using the \code{concurrent} formula special, as illustrated in \Rlang-chunk \ref{rchunk:ped_concurrent}:\\
\begin{knitrout}\small
\definecolor{shadecolor}{rgb}{0.961, 0.961, 0.961}\color{fgcolor}\begin{kframe}
\begin{rexample}\label{rchunk:ped_concurrent}\hfill{}\begin{alltt}
\hldef{pbc_ped} \hlkwb{=} \hlkwd{as_ped}\hldef{(}
  \hlkwc{data}    \hldef{=} \hlkwd{list}\hldef{(pbc, pbcseq),}
  \hlkwc{formula} \hldef{=} \hlkwd{Surv}\hldef{(time, status)}\hlopt{~}\hldef{sex} \hlopt{+} \hlkwd{concurrent}\hldef{(bili, protime,} \hlkwc{tz_var} \hldef{=} \hlsng{"day"}\hldef{),}
  \hlkwc{id}      \hldef{=} \hlsng{"id"}\hldef{)}

\hldef{pbc_ped |>} \hlkwd{head}\hldef{()}
\end{alltt}
\begin{verbatim}
  id tstart tend  interval    offset ped_status sex     bili  protime
1  1      0   41    (0,41] 3.7135721          0   f 2.674149 2.501436
2  1     41   51   (41,51] 2.3025851          0   f 2.674149 2.501436
3  1     51   71   (51,71] 2.9957323          0   f 2.674149 2.501436
4  1     71   77   (71,77] 1.7917595          0   f 2.674149 2.501436
5  1     77  108  (77,108] 3.4339872          0   f 2.674149 2.501436
6  1    108  110 (108,110] 0.6931472          0   f 2.674149 2.501436
\end{verbatim}
\end{rexample}\end{kframe}
\end{knitrout}

In \Rlang-chunk \ref{rchunk:ped_concurrent} \code{as\_ped}

\begin{itemize}
  \item uses the union of unique event times and all measurement times of the
  TVCs as interval split points,
  \item merges the expanded data set with the data set containing information on
  TVCs by subject ID and time (\code{time} and \code{day}) and
  \item fills in the values of TVCs for any time-points that did not occur in
  \code{tz\_var} by carrying the respective previous value of the TVC forward.
\end{itemize}

The last point of course implies the assumption that the values of the TVCs remain constant between observation points, which can be questionable, especially for longer periods between updates.
If linear or other custom interpolation of time-varying covariates between specific time-points is required instead, the data set containing the TVCs can simply be preprocessed accordingly before it is provided to \code{as\_ped}, but note that PAMMs always assume that hazard rates (and consequently the covariate effects that contribute to them) are constant over each interval $(\kappa_j, \kappa_{j+1}]$.
\\

\subsubsection{Modelling and interpretation}
\label{subsub:tvc:mod}

With the \code{pbc\_ped} data constructed in \Rlang-chunk~\ref{rchunk:ped_concurrent}, fitting a PAMM with concurrent time-varying effects of \code{bili} and \code{protime} requires no further data manipulation.
We first replicate the extended Cox analysis from \code{vignette("timedep", package="survival")} to establish a baseline for comparison.
That analysis demonstrates that using only baseline values of time-varying covariates rather than their full trajectories can lead to substantially different coefficient estimates, underscoring the importance of correctly incorporating TVCs into the model.
\begin{knitrout}\small
\definecolor{shadecolor}{rgb}{0.961, 0.961, 0.961}\color{fgcolor}\begin{kframe}
\begin{rexample}\label{rchunk:pbc-cox}\hfill{}\begin{alltt}
\hldef{pbc2} \hlkwb{<-} \hlkwd{tmerge}\hldef{(pbc, pbc,} \hlkwc{id} \hldef{= id,} \hlkwc{death} \hldef{=} \hlkwd{event}\hldef{(time, status))}
\hldef{pbc2} \hlkwb{<-} \hlkwd{tmerge}\hldef{(pbc2, pbcseq,} \hlkwc{id} \hldef{= id,} \hlkwc{bili} \hldef{=} \hlkwd{tdc}\hldef{(day, bili),}
  \hlkwc{protime} \hldef{=} \hlkwd{tdc}\hldef{(day, protime))}

\hldef{fit_cox_base} \hlkwb{<-} \hlkwd{coxph}\hldef{(}\hlkwd{Surv}\hldef{(time, status} \hlopt{==} \hlnum{1}\hldef{)} \hlopt{~} \hldef{bili} \hlopt{+} \hldef{protime, pbc)}
\hldef{fit_cox_tdc} \hlkwb{<-} \hlkwd{coxph}\hldef{(}\hlkwd{Surv}\hldef{(tstart, tstop, death} \hlopt{==} \hlnum{1}\hldef{)} \hlopt{~} \hldef{bili} \hlopt{+} \hldef{protime, pbc2)}

\hlkwd{rbind}\hldef{(}
  \hlsng{"baseline only"} \hldef{=} \hlkwd{coef}\hldef{(fit_cox_base),}
  \hlsng{"time-dependent"} \hldef{=} \hlkwd{coef}\hldef{(fit_cox_tdc)}
  \hldef{)}
\end{alltt}
\begin{verbatim}
                    bili  protime
baseline only  0.9724914 4.329193
time-dependent 1.2412143 3.983400
\end{verbatim}
\end{rexample}\end{kframe}
\end{knitrout}
The PAMM equivalent is fitted directly on \code{pbc\_ped} using \code{pamm}, with \code{bili} and \code{protime} entering the model as linear concurrent effects and the baseline log-hazard estimated via a smooth term in \code{tend}:
\begin{knitrout}\small
\definecolor{shadecolor}{rgb}{0.961, 0.961, 0.961}\color{fgcolor}\begin{kframe}
\begin{rexample}\label{rchunk:pbc-pam}\hfill{}\begin{alltt}
\hldef{pbc_pam} \hlkwb{<-} \hlkwd{pamm}\hldef{(ped_status} \hlopt{~} \hlkwd{s}\hldef{(tend)} \hlopt{+} \hldef{bili} \hlopt{+} \hldef{protime,} \hlkwc{data} \hldef{= pbc_ped)}

\hlkwd{cbind}\hldef{(}
  \hlkwc{pam} \hldef{=} \hlkwd{coef}\hldef{(pbc_pam)[}\hlnum{2}\hlopt{:}\hlnum{3}\hldef{],}
  \hlkwc{cox} \hldef{=} \hlkwd{coef}\hldef{(fit_cox_tdc)}
  \hldef{)}
\end{alltt}
\begin{verbatim}
             pam      cox
bili    1.266443 1.241214
protime 4.277453 3.983400
\end{verbatim}
\end{rexample}\end{kframe}
\end{knitrout}
The coefficient estimates are very close between the two models, providing reassurance that \pkg{pammtools}'s data transformation correctly captures the concurrent TVC structure.

To visualize the covariate effects, we again use \code{make\_newdata} and \code{add\_term}, which returns the contribution of a specified term to the linear predictor relative to a reference value (here the sample mean, obtained via \code{sample\_info}):
\begin{knitrout}\small
\definecolor{shadecolor}{rgb}{0.961, 0.961, 0.961}\color{fgcolor}\begin{kframe}
\begin{rexample}\label{rchunk:pbc-pam-viz}\hfill{}\begin{alltt}
\hldef{reference} \hlkwb{<-} \hlkwd{sample_info}\hldef{(pbc_ped)}

\hldef{bili_df} \hlkwb{<-} \hldef{pbc_ped} \hlopt{%>%}
  \hlkwd{ungroup}\hldef{()} \hlopt{%>%}
  \hlkwd{make_newdata}\hldef{(}\hlkwc{bili} \hldef{=} \hlkwd{seq_range}\hldef{(bili,} \hlkwc{n} \hldef{=} \hlnum{100}\hldef{))} \hlopt{%>%}
  \hlkwd{add_term}\hldef{(pbc_pam,} \hlkwc{term} \hldef{=} \hlsng{"bili"}\hldef{,} \hlkwc{reference} \hldef{= reference)} \hlopt{%>%}
  \hlkwd{mutate}\hldef{(}\hlkwc{cox} \hldef{=} \hlkwd{predict}\hldef{(fit_cox_tdc, .,} \hlkwc{type} \hldef{=} \hlsng{"term"}\hldef{)[,} \hlsng{"bili"}\hldef{])}

\hldef{protime_df} \hlkwb{<-} \hldef{pbc_ped |>} \hlkwd{ungroup}\hldef{()} \hlopt{%>%}
  \hlkwd{make_newdata}\hldef{(}\hlkwc{protime} \hldef{=} \hlkwd{seq_range}\hldef{(protime,} \hlkwc{n} \hldef{=} \hlnum{100}\hldef{))} \hlopt{%>%}
  \hlkwd{add_term}\hldef{(pbc_pam,} \hlkwc{term} \hldef{=} \hlsng{"protime"}\hldef{,} \hlkwc{reference} \hldef{= reference)} \hlopt{%>%}
  \hlkwd{mutate}\hldef{(}\hlkwc{cox} \hldef{=} \hlkwd{predict}\hldef{(fit_cox_tdc, .,} \hlkwc{type} \hldef{=} \hlsng{"term"}\hldef{)[,} \hlsng{"protime"}\hldef{])}
\end{alltt}
\end{rexample}\end{kframe}
\end{knitrout}
Figure~\ref{fig:pbc-tvc-viz} compares the estimated partial effects from the PAMM and the extended Cox model.
Both effects are on the log scale since \code{bili} and \code{protime} were log-transformed prior to analysis.
The two approaches yield near-identical effect estimates across the full range of both covariates, with the PAMM providing slightly smoother curves and explicit pointwise confidence bands.
Elevated bilirubin and prolonged prothrombin time are both associated with a substantially increased log-hazard, in line with their known roles as markers of liver disease severity \citep{Therneau2001}.
We stress, however, that \code{bili} and \code{protime} are \emph{internal} (endogenous) time-varying covariates -- biomarkers generated by the same disease process that drives the event -- rather than external exposures. The concurrent hazard model above is a valid descriptive association model and serves here to verify that \pkg{pammtools} reproduces the extended-Cox fit, but the resulting coefficients should not be read as causal disease-severity effects: for internal covariates the hazard and the covariate path are jointly determined, so the usual caveats on modeling and predicting survival with internal time-varying covariates apply \citep[Ch.~6.3]{kalbfleisch2002}. The extended Cox comparison shares this limitation.

\begin{figure}[!h]
\begin{knitrout}\small
\definecolor{shadecolor}{rgb}{0.961, 0.961, 0.961}\color{fgcolor}

{\centering \includegraphics[width=\maxwidth]{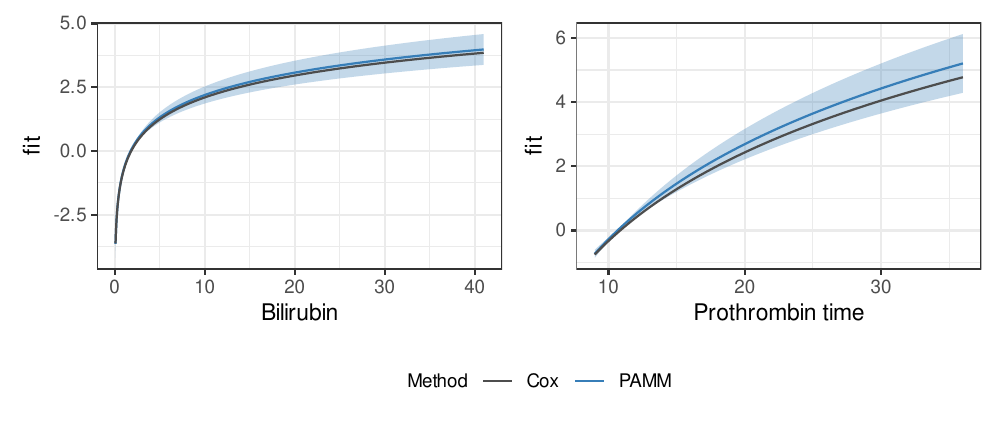} 

}

\end{knitrout}
\caption{\label{fig:pbc-tvc-viz} Estimated partial effects of log-bilirubin (left) and log-prothrombin time (right) on the log-hazard, comparing the PAMM (blue, with pointwise 95\% CI) and the extended Cox model (grey). Both covariates are displayed on their original scale. Effects are shown relative to the sample mean of each covariate.}
\end{figure}


\subsection{TVCs with cumulative effects}
\label{sub:elra}

\subsubsection{Pre-processing}
\label{subsub:elra:prep}

Some additional effort is required to create PEDs with TVCs that will be modeled as cumulative effects.
If \code{mgcv::gam} is used for estimation, we need to construct \emph{covariate matrices} for each TVC with a cumulative effect, as well as additional covariate matrices representing either time and/or time of exposure and/or the latency of exposure and the lag-lead matrix defining the time window $\tw{}$, see also \code{?mgcv::linear.functional.terms} for technical details.
\\

Let's consider a model with one cumulative effect $g(\bfz, t)$ of TVC $\bfz$, such that a general representation of the cumulative effect is given by
\begin{equation}\label{eq:elra}
g(\bfz, t) = \int_{\tw{}}h(t, \tz, z(\tz))\drm\tz
\end{equation}

In \eqref{eq:elra}
\begin{compactitem}
  \item the tri-variate function $h(t,\tz,z(\tz))$ defines the so-called
    \emph{partial effects} of the TVC $z(\tz)$ observed at exposure time \tz\ on the hazard at time $t$ \citep{Bender2018a}. Other specifications commonly used in the literature are special cases of the general partial effect definition given above, e.g.,
    \begin{compactitem}
      \item $h(t-\tz)z(\tz)$ is the WCE model of \cite{Sylvestre2009} and
      \item $h(t-\tz, z(\tz))$ corresponds to the distributed lag non-linear model (DLNM) of \cite{Gasparrini2014}
    \end{compactitem}
  \item the cumulative effect $g(\bfz,t)$ at follow-up time $t$ is the integral
  of the partial effects over exposure times \tz\ contained within \tw{}
  \item \tw{} denotes the \emph{lag-lead window} (or window of effectiveness).
  The most common definition is $\tw{}=\{t_{z}:t\geq t_{z}\}$,
  which means that all exposures that were observed prior to $t$ or at $t$ can affect the hazard at time $t$.
\end{compactitem}

Thus, when transforming the data to a format suitable to fit such effects using \code{mgcv::gam}, the required covariate matrices will be created depending on
\begin{compactitem}
  \item the specific definition of the partial effect $h()$,
  \item the grid of exposure times \tz\ and
  \item the lag-lead window \tw{}
\end{compactitem}
\vspace{1em} As before, the \code{as\_ped} function can be used to transform the data into the right format by extending the RHS of the \code{formula} using the formula special \code{cumulative}.
The most important arguments to \code{cumulative} are:\\
\begin{compactitem}
  \item[\code{...}:] a place holder where the individual components (variables)
  of the partial effects can be specified. See Table~\ref{tab:partial} for a
  selection of possible partial effect specifications and how to represent them
  in \code{cumulative}. Section~\ref{subsub:elra:mod} describes the respective formula syntax in \code{mgcv::gam}.
  \item[\code{tz\_var}:] the name of the variable that contains exposure times \tz\ of TVC $\bfz$
  \item[\code{ll\_fun}:] a boolean function of follow-up time $t$ and exposure time \tz,
  which defines \tw{} in Equation~\eqref{eq:elra} (see also Figure~\ref{fig:LL_demo})
\end{compactitem}

For illustration of the data transformation using \code{as\_ped} and \code{cumulative}, consider the simulated data \code{simdf\_elra} contained in \pkg{pammtools} (see example in \code{?sim\_pexp} for data generation):
\begin{knitrout}\small
\definecolor{shadecolor}{rgb}{0.961, 0.961, 0.961}\color{fgcolor}\begin{kframe}
\begin{alltt}
\hlkwd{data}\hldef{(}\hlsng{"simdf_elra"}\hldef{,} \hlkwc{package} \hldef{=} \hlsng{"pammtools"}\hldef{)}
\hldef{simdf_elra |>} \hlkwd{slice}\hldef{(}\hlnum{1}\hlopt{:}\hlnum{3}\hldef{)}
\end{alltt}
\begin{verbatim}
# A tibble: 3 x 9
     id   time status     x1    x2 tz1        z.tz1      tz2        z.tz2     
  <int>  <dbl>  <int>  <dbl> <dbl> <list>     <list>     <list>     <list>    
1     1  3.22       1  1.59  4.61  <int [10]> <dbl [10]> <int [11]> <dbl [11]>
2     2 10          0 -0.530 0.178 <int [10]> <dbl [10]> <int [11]> <dbl [11]>
3     3  0.808      1 -2.43  3.25  <int [10]> <dbl [10]> <int [11]> <dbl [11]>
\end{verbatim}
\end{kframe}
\end{knitrout}

It contains
\begin{compactitem}
\item the follow-up time $t$ (\code{time}),
\item the event indicator (\code{status}, censoring only occurs at the end of the follow up at $t=10$),
\item two time constant covariates $x_1$ (\code{x1}) and $x_2$ (\code{x2}) and
\item two TVCs $\bfz_1$ (\code{z1.tz1}), $\bfz_2$ (\code{z2.tz2}) observed at two different
exposure time grids $t_{z_1}$ (\code{tz1}) and $t_{z_2}$ (\code{tz2}).
\end{compactitem}

Let's further assume that two different lag-lead windows $\tw{1}=\{t_{z_1}:t\geq t_{z_1}\}$ and $\tw{2}=\{t_{z_2}:t\geq t_{z_2}+2\}$ (the latter defined by \code{ll\_2 = function(t, tz) { t >= tz + 2}}) are associated with the cumulative effects of the respective TVCs.
The latter corresponds to a lag time (i.e., a delayed effect of exposure) of 2 days, so, for example, the value of $z_2(3)$ only affects the hazard for follow-up times $t \geq 5$.
\\

Table~\ref{tab:partial} shows a selection of partial effect specifications for this setting and the respective specification using the formula special \code{cumulative}.
Note that
\begin{compactitem}
 \item the variable representing follow-up time $t$ in \code{cumulative}
 (here \code{time}) must match the time variable specified on the LHS
 of the formula  (\code{Surv(time, status)}) provided to \code{as\_ped}

\item if the latency $t-\tz$ should be used instead of $\tz$,
the variables representing exposure time \tz\ (here \code{tz1} and \code{tz2})
must be wrapped within \code{latency()}

\item by default, $\tw{}$ is defined as \code{function(t, tz) {t >= tz}}, thus
for \tw{1} there is no need to specify the lag-lead window explicitly.
To define a custom lag-lead window, provide the respective function to the \code{ll\_fun}
argument in \code{cumulative} (see \code{ll\_2} in Table~\ref{tab:partial})

\item \code{cumulative} does not distinguish between partial effects
$h(t-\tz,z(\tz))$ and $h(t-\tz)z(\tz)$ as the required data transformations are
identical

\item more than one $\bfz$ variable can be provided to \code{cumulative}, which
can be convenient if multiple covariates share time components and
will be integrated over the same lag-lead windows

\item multiple \code{cumulative} terms with different exposure times $t_{z_1}$,
$t_{z_2}$ and/or different lag-lead windows for different covariates $z_1$, $z_2$
can be specified, as illustrated in Table~\ref{tab:partial}

\item to tell \code{cumulative} which of the variables provided is the exposure
time \tz, the \code{tz\_var} argument must be specified within each
\code{cumulative} term. The follow-up time component $t$ (\code{time}) will be
recognized from the LHS of the formula
\end{compactitem}

\begin{table}[!hb]
\caption{A selection of possible partial effect specifications and the usage of
\code{cumulative} to create matrices needed to estimate different types of
cumulative effects of $\bfz_1$ and $\bfz_2$.}
\label{tab:partial}
\resizebox{\textwidth}{!}{%
\begin{tabular}{L{4cm}L{12cm}}
\toprule
cumulative effect(s) & data transformation (\code{pammtools})\\
\midrule
$\int_{\mathcal{T}_1}h(t-t_{z_1}, z_1(t_{z_1}))\drm \tz$ & \code{cumulative(latency(tz1), z1.tz1, tz\_var="tz1")}\\[1em]
\midrule
$\int_{\mathcal{T}_1}h(t, t - t_{z_1}, z_1(t_{z_1}))\drm \tz$ & \code{cumulative(time, latency(tz1), z1.tz1, tz\_var="tz1")}\\[1em]
\midrule
$\int_{\mathcal{T}_1}h(t, t_{z_1}, z_1(t_{z_1}))\drm \tz$   & \code{cumulative(time, tz1, z1.tz1, tz\_var="tz1")}\\[1em]
\midrule
$\int_{\mathcal{T}_1}h(t, t_{z_1}, z_1(t_{z_1}))\drm t_{z_1} + \int_{\mathcal{T}_2}h(t-t_{z_2}, z_2(t_{z_2}))\drm t_{z_2}$ & \code{cumulative(time, tz1, z1.tz1, tz\_var="tz1") + cumulative(latency(tz2), z2.tz2, tz\_var="tz2", ll\_fun=ll\_2)}\\
\bottomrule
\end{tabular}%
}
\end{table}

One possible data transformation call for the \code{simdf\_elra} data is given in \Rlang-chunk~\ref{rchunk:ped_simdf}.

\begin{knitrout}\small
\definecolor{shadecolor}{rgb}{0.961, 0.961, 0.961}\color{fgcolor}\begin{kframe}
\begin{rexample}\label{rchunk:ped_simdf}\hfill{}\begin{alltt}
\hldef{ped_simdf} \hlkwb{=} \hldef{simdf_elra |>} \hlkwd{as_ped}\hldef{(}\hlkwd{Surv}\hldef{(time, status)}\hlopt{~} \hldef{x1} \hlopt{+} \hldef{x2} \hlopt{+}
    \hlkwd{cumulative}\hldef{(time,} \hlkwd{latency}\hldef{(tz1), z.tz1,} \hlkwc{tz_var}\hldef{=}\hlsng{"tz1"}\hldef{)} \hlopt{+}
    \hlkwd{cumulative}\hldef{(}\hlkwd{latency}\hldef{(tz2), z.tz2,} \hlkwc{tz_var}\hldef{=}\hlsng{"tz2"}\hldef{),} \hlkwc{cut} \hldef{=} \hlnum{0}\hlopt{:}\hlnum{10}\hldef{)}
\hldef{p_laglead} \hlkwb{=} \hlkwd{gg_laglead}\hldef{(ped_simdf)}

\hlkwd{str}\hldef{(ped_simdf)}
\end{alltt}
\begin{verbatim}
...
 $ interval    : Factor w/ 10 levels "(0,1]","(1,2]",..: 1 2 3 4 1 2 3 4 5 6 ...
 $ offset      : num  0 0 0 -1.53 0 ...
 $ ped_status  : num  0 0 0 1 0 0 0 0 0 0 ...
 $ x1          : num  1.59 1.59 1.59 1.59 -0.53 ...
 $ x2          : num  4.612 4.612 4.612 4.612 0.178 ...
 $ time_tz1_mat: int [1:1004, 1:10] 0 1 2 3 0 1 2 3 4 5 ...
 $ tz1_latency : num [1:1004, 1:10] 0 0 1 2 0 0 1 2 3 4 ...
 $ z.tz1_tz1   : num [1:1004, 1:10] -2.014 -2.014 -2.014 -2.014 -0.978 ...
  ..- attr(*, "dimnames")=List of 2
  .. ..$ : NULL
  .. ..$ : chr [1:10] "z.tz11" "z.tz12" "z.tz13" "z.tz14" ...
 $ LL_tz1      : num [1:1004, 1:10] 0 1 1 1 0 1 1 1 1 1 ...
 $ tz2_latency : num [1:1004, 1:11] 5 6 7 8 5 6 7 8 9 10 ...
 $ z.tz2_tz2   : num [1:1004, 1:11] -0.689 -0.689 -0.689 -0.689 0.693 ...
  ..- attr(*, "dimnames")=List of 2
  .. ..$ : NULL
  .. ..$ : chr [1:11] "z.tz21" "z.tz22" "z.tz23" "z.tz24" ...
 $ LL_tz2      : num [1:1004, 1:11] 1 1 1 1 1 1 1 1 1 1 ...
 - attr(*, "breaks")= int [1:11] 0 1 2 3 4 5 6 7 8 9 ...
 - attr(*, "id_var")= chr "id"
 - attr(*, "intvars")= chr [1:6] "id" "tstart" "tend" "interval" ...
...
\end{verbatim}
\end{rexample}\end{kframe}
\end{knitrout}
The newly created matrix valued variables have
\begin{itemize}
\item different number of columns (10 vs. 11), reflecting the different exposure time grids
($t_{z_1,1}=1,\ldots, t_{z_1,10}=10$ and $t_{z_2, 1}=-5,\ldots, t_{z_2,11}=5$).

\item different components, depending on the partial effect and
\code{cumulative} specification, respectively. Thus, for \code{z.tz1} a time
matrix \code{time\_tz1\_mat} was created as well as a latency matrix \code{tz1\_latency},
whereas only the latency matrix \code{tz2\_latency} was created for the partial
effects associated with \code{z.tz2}.
\item different lag-lead specifications, which can be extracted and visualized
using convenience functions \code{get\_laglead} and \code{gg\_laglead}.
Applied to a \code{ped} object, they retrieve the lag-lead definition  used
during data transformation (cf. Figure~\ref{fig:gg_laglead}). More complex
specifications of $\tw{}$ can be generated easily (cf. Figure~\ref{fig:LL_demo}),
where a lead time of $\tlead = 5$ is included in addition to a lag time of $\tlag=2$.
\end{itemize}

\begin{figure}[!h]
\begin{knitrout}\small
\definecolor{shadecolor}{rgb}{0.961, 0.961, 0.961}\color{fgcolor}

{\centering \includegraphics[width=\maxwidth]{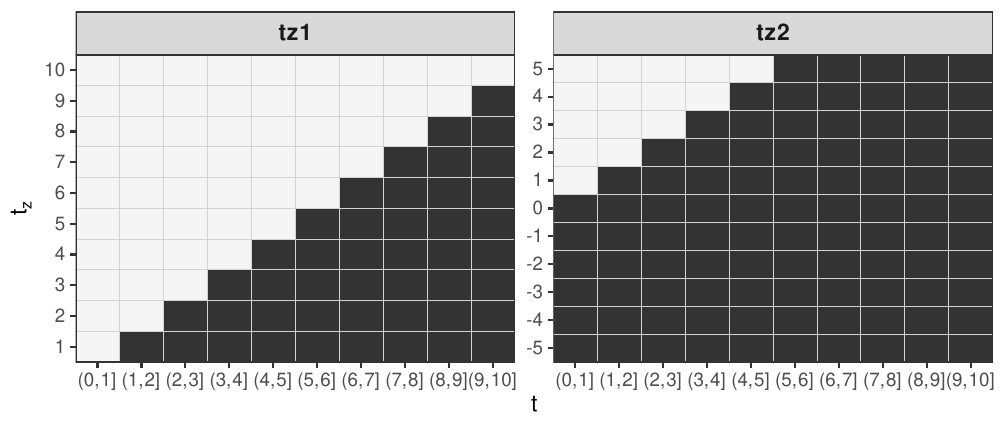} 

}

\end{knitrout}
\caption{Lag-lead windows created by \code{as\_ped} in
\Rlang-chunk \ref{rchunk:ped_simdf}. When viewed row-wise, the black squares
indicate the intervals at which the respective exposure times $\tz$ can affect
the hazard.
For example, in the left panel, exposure at time $\tz = 5$ can affect
the hazard in intervals $(5,6]$ through $(9,10]$ (\code{as\_ped} is conservative
and $t\geq \tz$ is only true if the relationship is true for the interval
start time). When viewed column-wise,
one can obtain the exposure times contained within $\tw{}$. For example,
$\mathcal{T}_1(t=5)=\mathcal{T}_1((\kappa_{j-1}=4, \kappa_j=5]) = \{t_{z_1}=1,\ldots,t_{z_1}=4\}$.}
\label{fig:gg_laglead}
\end{figure}

\begin{figure}[!htbp]
\begin{knitrout}\small
\definecolor{shadecolor}{rgb}{0.961, 0.961, 0.961}\color{fgcolor}\begin{kframe}
\begin{alltt}
\hldef{my_ll_fun} \hlkwb{=} \hlkwa{function}\hldef{(}\hlkwc{t}\hldef{,} \hlkwc{tz}\hldef{,} \hlkwc{tlag} \hldef{=} \hlnum{2}\hldef{,} \hlkwc{tlead} \hldef{=} \hlnum{5}\hldef{) \{}
  \hldef{t} \hlopt{>=} \hldef{tz} \hlopt{+} \hldef{tlag} \hlopt{&} \hldef{t} \hlopt{<} \hldef{tz} \hlopt{+} \hldef{tlag} \hlopt{+} \hldef{tlead}
\hldef{\}}
\hlkwd{gg_laglead}\hldef{(}\hlnum{0}\hlopt{:}\hlnum{10}\hldef{,} \hlkwc{tz}\hldef{=}\hlopt{-}\hlnum{5}\hlopt{:}\hlnum{5}\hldef{,} \hlkwc{ll_fun} \hldef{= my_ll_fun)} \hlopt{+}
  \hldef{theme_jss} \hlopt{+} \hlkwd{theme}\hldef{(}\hlkwc{legend.position} \hldef{=} \hlsng{"None"}\hldef{)}
\end{alltt}
\end{kframe}

{\centering \includegraphics[width=\maxwidth]{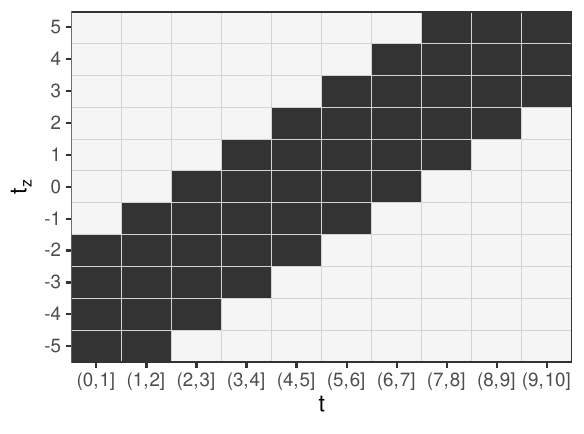} 

}

\end{knitrout}
\caption{Illustration of a more complex definition of the lag-lead window
$\tw{}$ with $\tlag = 2$ and $\tlead=5$. For example, exposure at time
$\tz=-1$, starts to affect the hazard at time $t = \tz + \tlag = -1 + 2 = 1$,
i.e., interval $(1, 2]$, as $t$ in the specification of the lag-lead function
refers to the start time of the interval. Similarly, exposure at time $\tz$
lasts until $t = \tz + \tlag + \tlead = -1 + 2 + 5 = 6$, i.e., interval $(5,6]$.
Note that we used the condition $t < \tz + \tlag + \tlead$ to ensure that
the condition is true for the end time of the interval.}
\label{fig:LL_demo}
\end{figure}


\subsubsection{Modelling and interpretation}
\label{subsub:elra:mod}

As a worked example, we estimate a weighted cumulative exposure (WCE) effect, i.e.\ the special case $h(t-\tz)z(\tz)$ of \eqref{eq:elra} in which the partial effect depends on the latency $t-\tz$ only. We simulate $n = 2000$ event times (about $13\%$ administrative censoring at $t=10$) from

\begin{equation}\label{eq:sim-elra}
h_i(t\mid\bfz_i) = \exp\left(
  \beta_0 + f_0(t) - 0.5\,x_{1,i} + \sqrt{x_{2,i}} +
  \int_{\tw{}} h(t-\tz)\,z_i(\tz)\drm\tz \right),
\end{equation}
with weight function $h(t-\tz) = 1.5\,\phi(t-\tz;\,6,\,2.5)$ (a Gaussian peaking at latency $6$), exposure $z(\tz)$ following an auto-regressive process and lag-lead window $\tw{} = \{\tz: 0 \leq t-\tz \leq 12\}$; data generation is detailed in the \href{https://adibender.github.io/pammtools/articles/simulations.html#sim:tdc}{simulation vignette}. The latency specification is obtained by wrapping the exposure time in \code{latency()} (cf.\ Table~\ref{tab:partial}); data transformation and estimation are shown in \Rlang-chunk~\ref{rchunk:wce-fit}.

\begin{knitrout}\small
\definecolor{shadecolor}{rgb}{0.961, 0.961, 0.961}\color{fgcolor}\begin{kframe}
\begin{rexample}\label{rchunk:wce-fit}\hfill{}\begin{alltt}
\hldef{ped_wce} \hlkwb{=} \hldef{sim_wce |>} \hlkwd{as_ped}\hldef{(}\hlkwd{Surv}\hldef{(time, status)} \hlopt{~} \hldef{x1} \hlopt{+} \hldef{x2} \hlopt{+}
    \hlkwd{cumulative}\hldef{(}\hlkwd{latency}\hldef{(tz), z.tz,} \hlkwc{tz_var} \hldef{=} \hlsng{"tz"}\hldef{,} \hlkwc{ll_fun} \hldef{= ll_fun),}
  \hlkwc{cut} \hldef{=} \hlkwd{seq}\hldef{(}\hlnum{0}\hldef{,} \hlnum{10}\hldef{,} \hlkwc{by} \hldef{=} \hlnum{0.25}\hldef{))}
\hldef{mod_wce} \hlkwb{=} \hlkwd{gam}\hldef{(ped_status} \hlopt{~} \hlkwd{s}\hldef{(tend)} \hlopt{+} \hlkwd{s}\hldef{(x1)} \hlopt{+} \hlkwd{s}\hldef{(x2)} \hlopt{+}
    \hlkwd{s}\hldef{(tz_latency,} \hlkwc{by} \hldef{= z.tz} \hlopt{*} \hldef{LL),}
  \hlkwc{data} \hldef{= ped_wce,} \hlkwc{family} \hldef{=} \hlkwd{poisson}\hldef{(),} \hlkwc{offset} \hldef{= offset,} \hlkwc{method} \hldef{=} \hlsng{"REML"}\hldef{)}
\hlkwd{summary}\hldef{(mod_wce)}
\end{alltt}
\begin{verbatim}
...
Approximate significance of smooth terms:
                          edf Ref.df Chi.sq p-value    
s(tend)                 7.224  8.120  667.9  <2e-16 ***
s(x1)                   1.004  1.009  835.0  <2e-16 ***
s(x2)                   4.281  5.273  393.6  <2e-16 ***
s(tz_latency):z.tz * LL 5.802  6.504  906.9  <2e-16 ***
...
\end{verbatim}
\end{rexample}\end{kframe}
\end{knitrout}

\begin{figure}[!htbp]
\begin{knitrout}\small
\definecolor{shadecolor}{rgb}{0.961, 0.961, 0.961}\color{fgcolor}\begin{kframe}
\begin{alltt}
\hldef{p_partial} \hlkwb{=} \hlkwd{gg_slice}\hldef{(ped_wce, mod_wce,} \hlkwc{term} \hldef{=} \hlsng{"z.tz"}\hldef{,}
    \hlkwc{tz_latency} \hldef{= lat_grid,} \hlkwc{z.tz} \hldef{=} \hlkwd{c}\hldef{(}\hlnum{1}\hldef{),} \hlkwc{LL} \hldef{=} \hlkwd{c}\hldef{(}\hlnum{1}\hldef{))} \hlopt{+}
  \hlkwd{geom_line}\hldef{(}\hlkwc{data} \hldef{= part_truth,} \hlkwd{aes}\hldef{(tz_latency, fit),} \hlkwc{linetype} \hldef{=} \hlnum{2}\hldef{)} \hlopt{+}
  \hlkwd{labs}\hldef{(}\hlkwc{x} \hldef{=} \hlkwd{expression}\hldef{(t} \hlopt{-} \hldef{t[z]),} \hlkwc{y} \hldef{=} \hlkwd{expression}\hldef{(}\hlkwd{h}\hldef{(t} \hlopt{-} \hldef{t[z])))}
\hldef{p_cumu} \hlkwb{=} \hlkwd{gg_cumu_eff}\hldef{(ped_wce, mod_wce,} \hlkwc{term} \hldef{=} \hlsng{"z.tz"}\hldef{,} \hlkwc{z1} \hldef{=} \hlnum{1}\hldef{)} \hlopt{+}
  \hlkwd{geom_line}\hldef{(}\hlkwc{data} \hldef{= cumu_truth,} \hlkwd{aes}\hldef{(tend, cumu_eff),} \hlkwc{linetype} \hldef{=} \hlnum{2}\hldef{)} \hlopt{+}
  \hlkwd{labs}\hldef{(}\hlkwc{x} \hldef{=} \hlsng{"t"}\hldef{,} \hlkwc{y} \hldef{=} \hlkwd{expression}\hldef{(}\hlkwd{g}\hldef{(}\hlkwd{bold}\hldef{(z), t)))}
\hldef{p_partial} \hlopt{+} \hldef{p_cumu}
\end{alltt}
\end{kframe}

{\centering \includegraphics[width=\maxwidth]{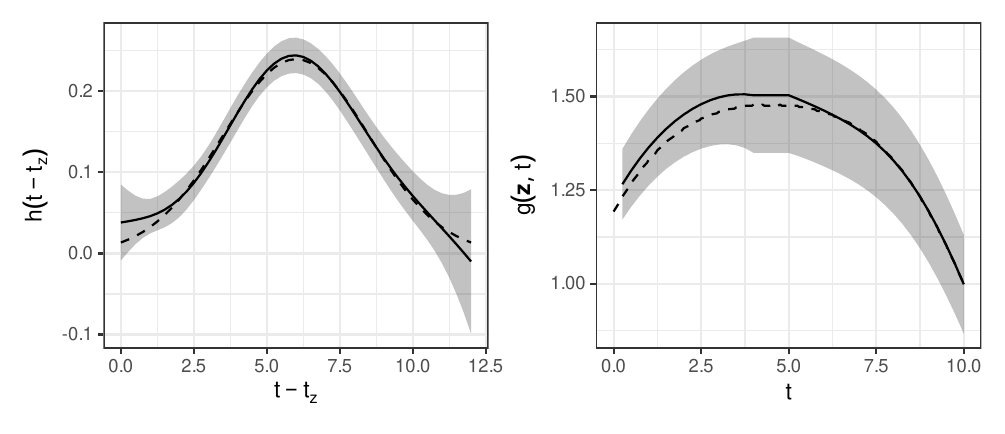} 

}

\end{knitrout}
\caption{Estimated weighted cumulative exposure effect (solid) with pointwise
$95\%$ confidence intervals and true effect (dashed). Left: partial effect
$\hat h(t-\tz)$ as a function of the latency $t-\tz$. Right: resulting cumulative
effect $\hat g(\bfz, t)$ for $z(\tz) \equiv 1$.}
\label{fig:sim-elra-est}
\end{figure}

In this single simulated illustrative data set, the estimated partial effect and the resulting cumulative effect closely follow the true functions, with the pointwise confidence intervals covering the truth across the lag-lead window (Figure~\ref{fig:sim-elra-est}). A systematic evaluation of the bias and coverage of cumulative-effect estimators under varying designs is beyond the scope of this software article and is studied elsewhere \citep{Bender2018a}.


\subsection{Interval-censored data}
\label{sub:ic}

In many applications, units are only examined intermittently, so the event time is not observed exactly but only known to lie in an interval $(L_i, R_i]$ between the last event-free and the first positive inspection.
Such interval-censored event times cannot be accommodated by simply adjusting the PED transformation, since the equivalence of the PEM likelihood to a Poisson likelihood (Section~\ref{sub:singleevent}) requires exactly observed or right-censored event times.
A common workaround -- treating, e.g., each censoring interval's midpoint as an exact event time -- retains the standard workflow, but the resulting model treats these imputed event times as known: confidence intervals become too narrow and inference invalid \citep{wiegrebe2025msm}, and if the censoring intervals are wide relative to the dynamics of the hazard, point estimates are biased as well.

\pkg{pammtools} instead implements a full multiple imputation (MI) approach \citep[cf.][for MI in Cox models under interval censoring]{pan2000}.
Since the piecewise constant hazard implies a closed-form conditional distribution of the true event time within the censoring interval $(L_i, R_i]$ given covariates $\bfx_i$, exact event times can be drawn from this distribution under a fitted PAMM, with coefficients drawn from their posterior (Section~\ref{sub:inference}) to propagate estimation uncertainty into the imputations \citep[``proper'' MI;][]{rubin1987}.
Each of the $m$ completed data sets is then transformed and re-fit with the standard right-censored pipeline, using one fixed set of cut points $\kappa_j$ across all imputations.
Parametric coefficients are pooled by Rubin's rules, with the median-$p$ rule for tests of smooth terms \citep{bolt2022}.
For derived quantities, the \code{add\_*} functions combine the simulation-based posterior draws (Section~\ref{sub:inference}) across the $m$ fits, so that all resulting intervals reflect both within- and between-imputation uncertainty.
Because the first batch of imputations is drawn from a preliminary fit based on the censoring-interval midpoints, imputation and re-fitting can also be iterated (argument \code{iter}), in the spirit of ``poor man's data augmentation'' \citep{weitanner1990} and iterative imputation algorithms such as \pkg{mice} \citep{vanbuuren2011}; doing so quickly attenuates the dependence on the initial fit and is strongly recommended for sparser inspection schedules that result in wide censoring intervals.

\Rlang-chunk \ref{rchunk:ic-sim} illustrates this workflow with simulated data: the reference group (\code{trt = 0}) has constant log-hazard $-2.2$, while the log-hazard of the treatment group additionally increases linearly over time ($0.15\cdot t$), a strong violation of the proportional hazards assumption.
The convenience function \code{add\_inspections} turns the exact simulated event times into interval-censored observations by drawing per-subject inspection times with mean gap \mbox{\code{1/rate}~$= 2.5$}, a rather sparse inspection schedule.
The resulting columns \code{L} and \code{R} follow the encoding of \code{survival::Surv(L, R, type = "interval2")}: $L = R$ denotes an exact event time, $R = \infty$ a right-censored, $L = 0$ (with finite $R$) a left-censored, and $0 < L < R < \infty$ an interval-censored observation.

\begin{knitrout}\small
\definecolor{shadecolor}{rgb}{0.961, 0.961, 0.961}\color{fgcolor}\begin{kframe}
\begin{rexample}\label{rchunk:ic-sim}\hfill{}\begin{alltt}
\hlkwd{set.seed}\hldef{(}\hlnum{3}\hldef{)}
\hldef{df_ic}  \hlkwb{=} \hlkwd{data.frame}\hldef{(}\hlkwc{trt} \hldef{=} \hlkwd{factor}\hldef{(}\hlkwd{rbinom}\hldef{(}\hlnum{1000}\hldef{,} \hlnum{1}\hldef{,} \hlnum{0.5}\hldef{)))}
\hldef{sim_ic} \hlkwb{=} \hlkwd{sim_pexp}\hldef{(}\hlopt{~ -}\hlnum{2.2} \hlopt{+} \hlnum{0.15} \hlopt{*} \hldef{t} \hlopt{*} \hldef{(trt} \hlopt{==} \hlnum{1}\hldef{), df_ic,}
  \hlkwc{cut} \hldef{=} \hlkwd{seq}\hldef{(}\hlnum{0}\hldef{,} \hlnum{10}\hldef{,} \hlkwc{by} \hldef{=} \hlnum{0.1}\hldef{))}
\hlkwd{set.seed}\hldef{(}\hlnum{7}\hldef{)}
\hldef{icd} \hlkwb{=} \hlkwd{add_inspections}\hldef{(sim_ic,} \hlkwc{rate} \hldef{=} \hlnum{0.4}\hldef{,} \hlkwc{max_time} \hldef{=} \hlnum{10}\hldef{)}
\hldef{icd |>} \hlkwd{select}\hldef{(id, trt, true_time, L, R) |>} \hlkwd{head}\hldef{(}\hlkwc{n} \hldef{=} \hlnum{3}\hldef{)}
\end{alltt}
\begin{verbatim}
# A tibble: 3 x 5
     id trt   true_time     L      R
  <int> <fct>     <dbl> <dbl>  <dbl>
1     1 0         10    10    Inf   
2     2 1          6.89  4.18   8.46
3     3 0         10    10    Inf   
\end{verbatim}
\end{rexample}\end{kframe}
\end{knitrout}

In this example, $809$ of the $1000$ subjects are interval- or left-censored, with a median censoring-interval width of $3.2$ time units.
\code{pamm\_ic} runs the entire MI workflow described above (\Rlang-chunk \ref{rchunk:ic-fit}): models are specified as usual, except that the data transformation arguments (here: the \code{interval2} response specification and the \code{cut} points) are passed along with the model formula.
The \code{summary} reports the pooled coefficient table and smooth term tests as well as the estimated \emph{fraction of missing information} (FMI), i.e., the proportion of each estimate's total uncertainty that is attributable to the unobserved exact event times.
The FMI diagnoses how strongly the interval censoring affects different parts of the model -- here, the smooth baseline hazards are affected far more than the parametric treatment effect.
Large values mean that the censoring genuinely dominates the respective estimate's uncertainty; they also call for a larger number of imputations $m$ to stabilize the pooled results, although moderate $m$ often retains high relative efficiency even then.

\begin{knitrout}\small
\definecolor{shadecolor}{rgb}{0.961, 0.961, 0.961}\color{fgcolor}\begin{kframe}
\begin{rexample}\label{rchunk:ic-fit}\hfill{}\begin{alltt}
\hlkwd{set.seed}\hldef{(}\hlnum{11}\hldef{)}
\hldef{fit_ic} \hlkwb{=} \hlkwd{pamm_ic}\hldef{(}
  \hlkwd{Surv}\hldef{(L, R,} \hlkwc{type} \hldef{=} \hlsng{"interval2"}\hldef{)} \hlopt{~} \hldef{trt,}
  \hlkwc{data}          \hldef{= icd,}
  \hlkwc{model_formula} \hldef{= ped_status} \hlopt{~} \hldef{trt} \hlopt{+} \hlkwd{s}\hldef{(tend,} \hlkwc{by} \hldef{= trt),}
  \hlkwc{cut}           \hldef{=} \hlkwd{seq}\hldef{(}\hlnum{0}\hldef{,} \hlnum{10}\hldef{,} \hlkwc{by} \hldef{=} \hlnum{0.5}\hldef{),}
  \hlkwc{m}             \hldef{=} \hlnum{20}\hldef{,}
  \hlkwc{iter}          \hldef{=} \hlnum{3}\hldef{)}
\hlkwd{summary}\hldef{(fit_ic)}
\end{alltt}
\begin{verbatim}
Pooled PAMM summary (multiple imputation for interval-censored data)

Task           : single event 
Family         : poisson 
Model          : ped_status ~ trt + s(tend, by = trt) 
Imputations    : 20 (proper) 
Subjects       : 1000 ( 809 interval/left-censored ); PED rows per fit: 11131 

Parametric coefficients (Rubin-pooled estimates & SEs, median-p):
            Estimate Std. Error z value Pr(>|z|)    
(Intercept)  -2.1960     0.0577  -38.05  < 2e-16 ***
trt1          0.5508     0.0777    7.09  8.1e-14 ***
---
Signif. codes:  0 '***' 0.001 '**' 0.01 '*' 0.05 '.' 0.1 ' ' 1

Approximate significance of smooth terms (mean edf, median-p over imputations):
              edf Ref.df statistic p-value    
s(tend):trt0 2.10   2.42      3.57    0.28    
s(tend):trt1 2.18   2.54     90.30  <2e-16 ***
---
Signif. codes:  0 '***' 0.001 '**' 0.01 '*' 0.05 '.' 0.1 ' ' 1

Fraction of missing information (FMI):
Parametric coefficients:
              FMI
(Intercept) 0.049
trt1        0.109

Smooth terms (five-number summaries over training PED rows):
               Min    Q1 Median    Q3   Max
s(tend):trt0 0.129 0.129  0.246 0.473 0.809
s(tend):trt1 0.237 0.240  0.240 0.603 0.810

Standard errors include within- + between-imputation variance (Rubin's rules);
p-values are medians over imputations.
\end{verbatim}
\end{rexample}\end{kframe}
\end{knitrout}

Post-processing works as for all other PAMMs: the \code{add\_*} functions accept the pooled fit directly and combine the posterior draws of all $m$ imputation fits (\Rlang-chunk \ref{rchunk:ic-pred}).

\begin{knitrout}\small
\definecolor{shadecolor}{rgb}{0.961, 0.961, 0.961}\color{fgcolor}\begin{kframe}
\begin{rexample}\label{rchunk:ic-pred}\hfill{}\begin{alltt}
\hldef{ped_ic} \hlkwb{=} \hlkwd{as_ped}\hldef{(icd,} \hlkwd{Surv}\hldef{(L, R,} \hlkwc{type} \hldef{=} \hlsng{"interval2"}\hldef{)} \hlopt{~} \hldef{trt,}
  \hlkwc{cut} \hldef{=} \hlkwd{seq}\hldef{(}\hlnum{0}\hldef{,} \hlnum{10}\hldef{,} \hlkwc{by} \hldef{=} \hlnum{0.5}\hldef{))}
\hldef{ndf_ic} \hlkwb{=} \hldef{ped_ic |>} \hlkwd{make_newdata}\hldef{(}\hlkwc{tend} \hldef{=} \hlkwd{unique}\hldef{(tend),} \hlkwc{trt} \hldef{=} \hlkwd{unique}\hldef{(trt)) |>}
  \hlkwd{group_by}\hldef{(trt)}
\hlkwd{set.seed}\hldef{(}\hlnum{1}\hldef{)}
\hldef{surv_ic_mi}  \hlkwb{=} \hldef{ndf_ic |>} \hlkwd{add_surv_prob}\hldef{(fit_ic,} \hlkwc{nsim} \hldef{=} \hlnum{500}\hldef{)}
\hldef{surv_ic_mid} \hlkwb{=} \hldef{ndf_ic |>}
  \hlkwd{add_surv_prob}\hldef{(fit_ic}\hlopt{$}\hldef{init_fit,} \hlkwc{ci_type} \hldef{=} \hlsng{"sim"}\hldef{,} \hlkwc{nsim} \hldef{=} \hlnum{500}\hldef{)}
\end{alltt}
\end{rexample}\end{kframe}
\end{knitrout}

\begin{figure}[!htbp]
\begin{knitrout}\small
\definecolor{shadecolor}{rgb}{0.961, 0.961, 0.961}\color{fgcolor}

{\centering \includegraphics[width=\maxwidth]{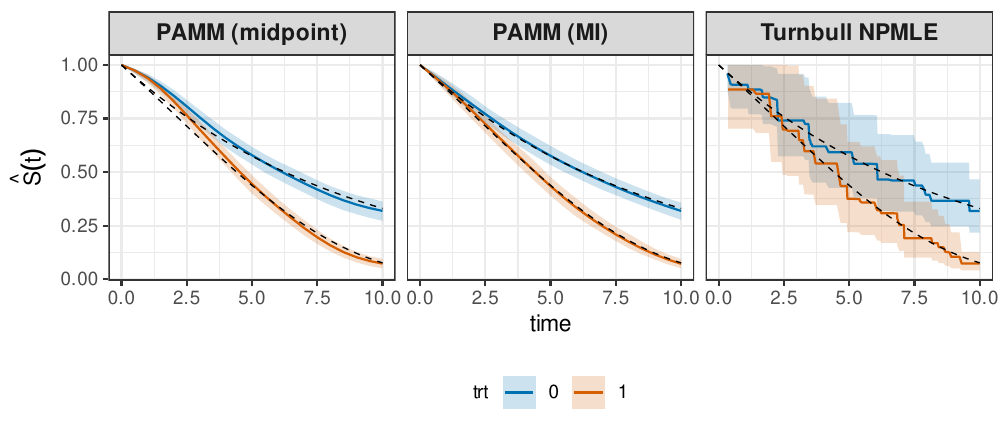} 

}

\end{knitrout}
\caption{Estimated survival probabilities with pointwise 95\% confidence intervals for the interval-censored simulated data, true survival functions added as dashed lines. Left: PAMM treating censoring-interval midpoints as exact event times (the initial fit, \code{fit\_ic\$init\_fit}). Center: PAMM with multiple imputation (\code{fit\_ic}). Right: non-parametric Turnbull estimator.}
\label{fig:ic-surv}
\end{figure}

Figure~\ref{fig:ic-surv} compares the resulting survival probability estimates to the non-parametric maximum likelihood estimator for interval-censored data \citep[Turnbull NPMLE;][]{turnbull1976}, as computed by \code{survival::survfit}, and to the true survival functions of the data generating process.
For this single simulated data set under a sparse inspection schedule, the midpoint-based fit lies visibly away from the true survival curves and its narrow confidence intervals exclude the truth over much of the follow-up, whereas the MI estimates stay close to the truth with wider intervals that reflect the imputation uncertainty, and the (consistent) Turnbull estimator has visibly wider steps while -- as a non-parametric estimator of the marginal survival function per stratum -- providing neither hazard rates nor covariate effects.

For interval-censored competing risks data, \code{pamm\_ic\_cr} provides the analogous workflow based on cause-specific hazards: event times are imputed from the conditional cause-specific sub-density implied by the fitted model while retaining the observed event causes, and missing causes can be sampled from the fitted cause-specific hazards \citep[cf.][for a related MI approach based on subdistribution hazards]{delord2016}.
Interval censoring in recurrent-event or general multi-state settings is not yet supported, as imputed transition times alter the at-risk windows of subsequent transitions, requiring joint imputation along each subject's path.
Further details, additional diagnostics, and guidance on choosing $m$ and \code{iter} are provided in the package's vignette on \href{https://adibender.github.io/pammtools/articles/interval-censored.html}{interval-censored data}.


\subsection{Post-processing}
\label{sub:postprocessing}
For communicating and checking the results of complex time-to-event models, post-processing is necessary and works in the same way for all previously mentioned use cases.
The generally suggested workflow is to create a data set with the required covariate specifications, typically using \code{make\_newdata}, then use one of the \code{add\_*} functions to calculate quantities of interest, and, finally, use either \code{ggplot} or one of \pkg{pammtools}' convenience functions to visualize the quantities of interest.

The following tables contain an overview of required functions provided by \pkg{pammtools}.

\subsubsection{Creating prediction data sets}
\pkg{pammtools} provides several functions that facilitate the creation of data sets with customized covariate specifications:

\begin{longtable}{
|>{\raggedright\arraybackslash}p{3cm}
|>{\raggedright\arraybackslash}p{10cm}|}

\caption{Overview of convenience functions to create data.} \\

\hline
\textbf{Fct.-Name} & \textbf{Description} \\
\hline
\endfirsthead

\hline
\textbf{Fct.-Name} & \textbf{Description} \\
\hline
\endhead

\hline
\multicolumn{2}{r}{\textit{Continued on next page}} \\
\hline
\endfoot

\hline
\endlastfoot

\code{int\_info} & 
Given interval breaks points, returns data frame with information on interval start time, interval end time, interval length and a factor variable indicating the interval (left open intervals). If an object of class \code{ped} is provided, extracts unique interval information from object.\\
\hline
\code{ped\_info} & 
Given an object of class \code{ped}, returns data frame with one row for each interval containing interval information, mean values for numerical variables and modal values for non-numeric variables in the data set. \\
\hline
\code{combine\_df} & 
Create a data frame of all row-wise combinations of multiple data frames (Cartesian join), similar to \code{base::expand.grid()} \\
\hline
\code{make\_newdata} & 
This functions provides a flexible interface to create a data set that can be plugged in as \code{newdata} argument to a suitable  \code{predict} function (or similar). The function is particularly useful in combination with one of the \code{add\_*} functions.  \\
\hline

\end{longtable}


\subsubsection{Adding hazards, cumulative incidence functions, and transition probabilities}
\label{subsub:post:add}

For single time-to-first-event data, using these flexibly created new data sets, we employ \pkg{mgcv}'s \code{predict} function to calculate estimated log-hazards as well as secondary quantities like survival probabilities. Throughout, these survival probabilities are \emph{covariate-conditional} and computed from the start of follow-up: for a given covariate profile in \code{newdata}, \code{add\_surv\_prob} returns $\hat S(t\mid\bfx) = \exp(-\hat\Lambda(t\mid\bfx))$ with $\hat\Lambda$ accumulated from $t=0$, i.e.\ the probability of surviving to $t$ for a subject with that profile who is event-free at time $0$. They are conditional on the covariate profile but not marginalized over any covariate distribution; a marginal (population-averaged) survival curve would require averaging $\hat S(t\mid\bfx)$ over a sample of profiles. Conditioning on being event-free at a later start time $t_0>0$ (i.e.\ $\hat S(t\mid\bfx, T>t_0) = \hat S(t\mid\bfx)/\hat S(t_0\mid\bfx)$) is obtained by restricting the accumulation to intervals after $t_0$.

\begin{longtable}{
|>{\raggedright\arraybackslash}p{3cm}
|>{\raggedright\arraybackslash}p{10cm}|}

\caption{Overview of functions to add quantities of interest.} \\

\hline
\textbf{Fct.-Name} & \textbf{Description} \\
\hline
\endfirsthead

\hline
\textbf{Fct.-Name} & \textbf{Description} \\
\hline
\endhead

\hline
\multicolumn{2}{r}{\textit{Continued on next page}} \\
\hline
\endfoot

\hline
\endlastfoot

\code{add\_hazard} & 
Add predicted (cumulative) hazard to data set (and approximate pointwise CIs). Their width can be controlled via the \code{se\_mult} argument. The method by which the CI are calculated can be specified by \code{ci\_type}. This is a wrapper around \code{predict.gam}. If \code{reference} is specified, (log-)hazard ratios relative to the  \code{reference} are calculated.\\
\hline
\code{add\_cumu\_hazard} & 
Add predicted cumulative hazard to data set (and approximate pointwise CIs), obtained by numerically integrating hazards obtained by \code{add\_hazard}.\\
\hline
\code{add\_surv\_prob} & 
Add survival probability estimates (and CIs \code{surv\_lower}, \code{surv\_upper}).\\
\hline
\code{add\_term} & 
Adds the contribution of a specific term to the linear predictor to the data specified by \code{newdata}. \\
\hline
\code{add\_cif} & 
Add cumulative incidence function to data (and CIs \code{cif\_lower}, \code{cif\_upper}).\\
\hline
\begin{tabular}[t]{@{}l@{}}\code{add\_}\\\code{counterfactual\_}\\\code{transitions}\end{tabular} & 
If data only contains only transition that actually took place, this function adds additional rows for every transition that would have been possible at every interval for each subject in the data. \\
\hline
\code{add\_trans\_prob} & 
Add transition probabilities to data (and CIs \code{trans\_lower}, \code{trans\_upper}). \\
\hline

\end{longtable}


\subsubsection{Visualization, convenience functions, and ggplot-extensions}
\label{subsub:post:vis}

After the \code{add\_*} functions add the calculated quantities of interested directly to the data, it can then be used for visualizations.
To facilitate creating these plots, \pkg{pammtools} contains specialized \pkg{ggplot} extensions:

\begin{longtable}{
|>{\raggedright\arraybackslash}p{3cm}
|>{\raggedright\arraybackslash}p{10cm}|}

\caption{Overview of \pkg{ggplot2} extensions.}
\label{tab:ggplot2-extensions} \\

\hline
\textbf{Fct.-Name} & \textbf{Description}\\
\hline
\endfirsthead

\hline
\textbf{Fct.-Name} & \textbf{Description} \\
\hline
\endhead

\hline
\multicolumn{2}{r}{\textit{Continued on next page}} \\
\hline
\endfoot

\hline
\endlastfoot

\code{geom\_hazard} & 
(Cumulative) (Step-) Hazard Plots. \code{geom\_hazard} is an extension of the \code{geom\_line}, and is optimized for (cumulative) hazard plots.\\
\hline
\code{geom\_stephazard} & 
Piecewise constant hazards plots.\\
\hline
\code{geom\_surv} & 
Survival probability plots. \\
\hline
\code{geom\_stepribbon} & 
Step ribbon plots. \code{geom\_stepribbon} is an extension of the \code{geom\_ribbon}, and is optimized for Kaplan-Meier plots with pointwise confidence intervals or a confidence band. The default \code{direction}-argument \code{"hv"} is appropriate for right-continuous step functions like the hazard rates etc returned by \code{pammtools}.\\
\hline

\end{longtable}

Additionally, \pkg{pammtools} also provides convenience functions to plot without data post-processing:

\begin{longtable}{
|>{\raggedright\arraybackslash}p{4cm}
|>{\raggedright\arraybackslash}p{9cm}|}

\caption{Overview of convenience functions for plotting.}
\label{tab:ggplot2-convenience} \\

\hline
\textbf{Fct.-Name} & \textbf{Description} \\
\hline
\endfirsthead

\hline
\textbf{Fct.-Name} & \textbf{Description} \\
\hline
\endhead

\hline
\multicolumn{2}{r}{\textit{Continued on next page}} \\
\hline
\endfoot

\hline
\endlastfoot

\code{gg\_smooth} & 
Given a gam model this convenience function returns a plot of all smooth terms contained in the model. If more than one smooth is present, the different smooth are faceted.\\
\hline
\code{gg\_tensor} & 
Given a gam model this convenience function returns a \code{ggplot2} object depicting 2d smooth terms specified in the model as heat/contour plots. If more than one 2d smooth term is present individual terms are faceted.\\
\hline
\code{gg\_re} & 
Plot Normal QQ plots for random effects.\\
\hline
\code{gg\_fixed} & 
Forest plot of fixed coefficients.\\
\hline
\code{gg\_partial} & 
Visualize effect estimates for specific covariate combinations.\\
\hline
\code{gg\_slice} & 
Flexible, high-level plotting function for (non-linear) effects conditional on further covariate specifications and potentially relative to a comparison specification.\\
\hline
\code{gg\_state\_occupation} & 
Given transition probabilities resulting from the \code{add\_trans\_prob} call this convenience function returns a \code{ggplot2} object depicting state occupation probabilities depending on the specified initial distribution.\\
\hline

\end{longtable}

\section{Comparison with existing software}
\label{sec:comparison}

Table~\ref{tab:pkg-comparison} compares \pkg{pammtools} against five packages that partially overlap in scope; \pkg{dlnm} \citep{Gasparrini2011}, \pkg{flexsurv} \citep{Jackson2016flexsurv}, \pkg{mstate} \citep{dewreede2011mstate}, \pkg{rstpm2} \citep{Liu2018rstpm2}, and \pkg{casebase} \citep{Bhatnagar2022casebase}.

\begin{table}[!h]
\centering
\caption{Feature comparison of \proglang{R} packages for survival analysis.
\yes{} = built-in support; \lmark{} = limited/indirect support (e.g.,\ only via an external back-end or requiring manual data restructuring); \no{} = no support. Ratings are assessed per feature in isolation among the packages compared here; not all combinations of features marked \yes{} are necessarily supported jointly (see footnotes and the Outlook in Section~\ref{sec:pammtools:discussion}).}
\label{tab:pkg-comparison}
\small
\resizebox{\textwidth}{!}{%
\begin{tabular}{l*{6}{>{\centering\arraybackslash}p{1.6cm}}}
\toprule
Feature
  & \pkg{pammtools}
  & \pkg{flexsurv}
  & \pkg{mstate}
  & \pkg{rstpm2}
  & \pkg{casebase}
  & \pkg{dlnm} \\
\midrule
Time-varying effects
  & \yes & \yes & \lmark & \yes & \yes & \no \\
Time-varying covariates
  & \yes & \lmark & \yes & \lmark & \yes & \no \\
Cumulative effects
  & \yes$^{b}$ & \no & \no & \no & \no & \yes \\
Competing risks
  & \yes & \yes & \yes & \yes & \yes & \no \\
Multi-state models
  & \yes & \yes & \yes & \yes & \no & \no \\
Recurrent events
  & \yes & \no & \lmark & \lmark & \lmark & \no \\
Left truncation
  & \yes & \yes & \yes & \yes & \no & \no \\
Interval censoring
  & \yes$^{a}$ & \yes & \no  & \yes & \no & \no \\
Penalized/smooth estimation
  & \yes & \lmark & \no & \yes & \lmark & \yes \\
Deterministic estimates
  & \yes & \yes & \yes & \yes & \no & \yes \\
Fully Bayesian inference
  & \lmark$^{c}$ & \no & \no & \no & \no & \no \\
Simulation-based CIs
  & \yes & \yes & \no & \no & \yes & \no \\
Tidyverse-compatible API
  & \yes & \no & \no & \no & \lmark & \no \\
Package version$^{d}$
  & 0.8.0 & 2.3.2 & 0.3.3 & 1.7.1 & 0.10.6 & 2.4.10 \\
\bottomrule
\multicolumn{7}{l}{\footnotesize $^{a}$~via multiple imputation, for single-event and competing risks data only (Section~\ref{sub:ic})}\\
\multicolumn{7}{l}{\footnotesize $^{b}$~cumulative (weighted-exposure) effects are developed and validated primarily for single-event outcomes;}\\
\multicolumn{7}{l}{\footnotesize \phantom{$^{b}$~}their combination with multi-state/recurrent-event models is not yet validated (cf.\ Section~\ref{sec:pammtools:discussion})}\\
\multicolumn{7}{l}{\footnotesize $^{c}$~indirect: fully Bayesian (MCMC/HMC) inference is available by passing the PED to external back-ends}\\
\multicolumn{7}{l}{\footnotesize \phantom{$^{c}$~}(\code{mgcv::jagam} with \pkg{rjags}, or \pkg{brms}), not through a native \pkg{pammtools} interface}\\
\multicolumn{7}{l}{\footnotesize $^{d}$~the \pkg{pammtools} version is printed dynamically and is the one used to produce this article; the}\\
\multicolumn{7}{l}{\footnotesize \phantom{$^{d}$~}remaining entries are the package versions current at the time of writing (cf.\ the \code{sessionInfo()} in Appendix~\ref{app:sessioninfo})}\\
\end{tabular}%
}
\end{table}

All packages except \pkg{dlnm} support non-proportional hazards in some form. In \pkg{flexsurv}, time-varying effects are achieved by including covariates on the ancillary shape or scale parameters of the spline model rather than through a direct interaction with time. \pkg{mstate} accommodates non-proportionality via stratified or transition-specific Cox models but does not provide smooth time-varying coefficients. Time-varying covariates are handled natively in \pkg{pammtools} and \pkg{casebase} via counting-process data formats; \pkg{flexsurv} and \pkg{rstpm2} require prior data restructuring into this format without dedicated helper functions.

Among the packages compared here, \pkg{pammtools} is the only one to implement weighted cumulative exposure effects (ELRA) natively. \pkg{dlnm} provides cross-basis functions for distributed lag non-linear models, but operates as a model-building toolkit rather than a self-contained survival analysis package, relying on external back-ends such as \code{survival::coxph} or \code{mgcv::gam} \citep{Wood2017a} for estimation.

\pkg{pammtools}, \pkg{flexsurv}, \pkg{mstate}, and \pkg{rstpm2} all support competing risks and multi-state models; \pkg{casebase} covers competing risks but not general multi-state models. Recurrent events are not a primary focus of any of the comparison packages: \pkg{mstate} and \pkg{rstpm2} can accommodate them by casting the problem as a multi-state model, while \pkg{casebase} provides experimental support for recurrent event intensities.

Left truncation is supported by \pkg{pammtools}, \pkg{flexsurv}, \pkg{mstate}, and \pkg{rstpm2}. Interval censoring can be handled by \pkg{flexsurv} and \pkg{rstpm2}, which maximize the corresponding interval-censored likelihoods directly, and in \pkg{pammtools}, which uses multiple imputation (cf.\ Section~\ref{sub:ic}). Neither feature is directly provided by \pkg{casebase} or \pkg{dlnm}.

\pkg{pammtools} and \pkg{rstpm2} offer full penalized spline or GAM-based smooth estimation. \pkg{flexsurv} provides spline-based baseline hazards via the Royston--Parmar model \citep{royston2002} but without penalization. \pkg{mstate} is built on semi-parametric Cox models and has no built-in penalized smoothing. \pkg{casebase} supports GAM-based smoothing through the \code{family = "gam"} option in \code{casebase::fitSmoothHazard}. \pkg{dlnm} supports spline bases (\code{ns}, \code{bs}) and penalized smoothing via an \code{mgcv::gam} back-end \citep{Wood2017a}.

\pkg{casebase} estimates are stochastic unless the case-base sample is fixed; increasing the base-to-case ratio reduces Monte Carlo variation. All other packages yield deterministic estimates -- with the exception of \pkg{pammtools}' interval-censoring workflow, whose estimates necessarily depend on the imputation draws (cf.\ Section~\ref{sub:ic}). Note also that \pkg{pammtools} can easily be used with Bayesian MCMC- or HMC-based inference back-ends like \code{mgcv::jagam} using \pkg{rjags} \citep{rjags} or \pkg{brms} \citep{brms} using \pkg{RStan} \citep{rstan}, see also the \href{https://adibender.github.io/pammtools/articles/bayesian.html}{vignette on Bayesian PAMM inference.} 

\pkg{pammtools}, \pkg{flexsurv}, and \pkg{casebase} support simulation-based confidence intervals: \pkg{pammtools} and \pkg{flexsurv} via posterior/parametric-bootstrap simulation, \pkg{casebase} via Monte Carlo integration for absolute risk estimates. \pkg{mstate} provides Greenwood and Aalen-type standard errors alongside a dedicated bootstrap function (\pkg{msboot}), but not posterior simulation. \pkg{rstpm2} relies on the delta method.

Among the packages compared here, \pkg{pammtools} is the only one with a fully tidy, \pkg{tidyverse}-compatible API \citep{Wickham2019}, including \pkg{broom}-style output and \pkg{ggplot2}-based \citep{Wickham2016a} visualization helpers. \pkg{casebase} produces data-frame output and \pkg{ggplot2}-compatible plots but does not integrate with \pkg{broom} \citep{robinson2014broom}.

\section{Discussion}
\label{sec:pammtools:discussion}
\subsection{Summary}
The \Rlang\ package \pkg{pammtools} provides a comprehensive framework for the estimation, interpretation, and visualization of flexible time-to-event regression models based on GAMMs. Compared to existing \Rlang{} packages, \pkg{pammtools} enables extremely flexible modeling of both time-varying and cumulative effects within a unified framework. While other packages focus on specific aspects such as parametric survival models, multi-state processes, or distributed lag structures, \pkg{pammtools} integrates data transformation, model estimation, and prediction across a very broad range of settings in one consistent workflow.

In particular, Section~\ref{sec:practical_workflow} demonstrates how time-to-event data in most conceivable settings, including complex multi-state scenarios, can be transformed into a format suitable for such analyses.
Special attention was given to the modeling and interpretation of time-varying effects and cumulative effects (cf. Section~\ref{sec:extended_topics}).
In addition, \pkg{pammtools} supports simulating time-to-event data with complex time-varying and cumulative effects, thereby facilitating research and development of novel approaches on complex time-to-event models.

\subsection{Limitations}
Incorporating interval-censoring directly into the PEM likelihood while preserving its equivalence to a Poisson likelihood remains an open methodological problem -- ignoring it invalidates confidence intervals and inference even where point estimates remain relatively robust \citep{wiegrebe2025msm}, and biases point estimates as well once censoring intervals are wide relative to the dynamics of the hazard (cf.\ Section~\ref{sub:ic}).
\pkg{pammtools} therefore handles interval-censored data via the multiple imputation approach described in Section~\ref{sub:ic}, which aims to restore approximately calibrated inference -- relative to treating imputed event times as exact -- at the cost of $m$-fold re-estimation, but is currently limited to single-event and competing risks settings; interval-censored recurrent-event and multi-state data are not yet supported. A systematic assessment of the resulting bias and coverage across designs is left to future methodological work.

Even though the PED format created by the \code{as\_ped} function could be provided to any method able to optimize a covariate-conditional Poisson likelihood with offsets, most post-processing functions and convenience plot functions are customized to work with the \Rlang\ package \pkg{mgcv}.
Although much effort went into making the respective functions robust and generally applicable, these efforts are limited by the fact that the estimation process is currently performed outside of \pkg{pammtools}. That said, \cite{benderGeneralMachineLearning2021a} demonstrate benchmark evaluations and provide code for using e.g., gradient-boosted trees \citep{xgboost} to fit PAMMs (see also the package vignette on \href{https://adibender.github.io/pammtools/articles/xgboost-backend.html}{PAMM inference using gradient-boosted trees.}). 

\subsection{Outlook}
Future development of \pkg{pammtools} will focus on three main areas.

First, the multiple imputation approach for interval-censored data (Section~\ref{sub:ic}) will be extended beyond the single-event and competing risks settings. Recurrent-event and general multi-state processes require imputations to be drawn jointly along each subject's path, since each imputed transition time changes the at-risk windows of subsequent transitions; developing and validating such joint imputation schemes is ongoing work.

Second, several modeling features currently mature for single-event data will be extended to the full event-history setting. In particular, the specification and estimation of complex cumulative and exposure-lag-response effects has so far been developed and validated primarily for single-event outcomes; future releases will improve this functionality for multi-state and recurrent-event models.

Third, while \pkg{pammtools} currently targets \pkg{mgcv} as its estimation back-end, the PED format is fully compatible with any method capable of optimizing Poisson likelihoods with offsets. The \pkg{mlr3proba} package \citep{sonabend2021mlr3proba} already provides a rich ecosystem of survival learners. Further development will provide tighter integration between \pkg{pammtools} and \pkg{mlr3proba}, enabling users to leverage the flexible data transformation and post-processing infrastructure of \pkg{pammtools} together with the broad range of learners available in \pkg{mlr3proba} for systematic model comparison and evaluation across the entire range of time-to-event and event history modeling tasks.

Feedback, bug reports and feature requests are welcome at \newline
\href{https://github.com/adibender/pammtools/issues}{https://github.com/adibender/pammtools/issues} 
or by contacting the authors directly.


\clearpage \bibliography{bibliography}


\appendix

\section{Implementation Details}
\label{app:implementation}

This appendix summarizes the internal architecture of \pkg{pammtools}, the
\code{ped} class hierarchy, the dispatch logic of \code{as_ped} across survival
tasks, and practical guidance on performance and scalability.

\subsection{Package Architecture}
\label{app:architecture}

The package is organized along the three stages of the modeling pipeline shown
in Figure~\ref{fig:pammtools}: data transformation, model fitting, and
post-processing.

\begin{compactitem}
\begin{sloppypar}
\item \emph{Data transformation} (\code{R/as-ped.R}, \code{R/split-data.R}, \code{R/tdc-utils.R}, \code{R/formula-specials.R}). 
  Converts input data frames or nested lists into the PED format. Handles right-censored,
  left-truncated, competing risks and multi-state data, and parses the
  \code{concurrent()} and \code{cumulative()} formula specials that encode
  time-varying covariates and exposure-lag-response associations.
\end{sloppypar}
\item \emph{Estimation} (\code{R/pammfit.R}). The \code{pamm} function is a
  thin wrapper around \code{mgcv::gam} that sets \code{family = poisson()},
  uses \code{offset = log(intlen)} and forwards the call to \code{mgcv}. PAMM
  models returned by \code{pamm} are therefore specialized \code{gam}/\code{bam} 
  objects with class vector \code{c("pamm", "gam", "glm", "lm")} and inherit all
  of \pkg{mgcv}'s smoothing, selection, and inference machinery.
\begin{sloppypar}
\item \emph{Post-processing} (\code{R/add-functions.R},
  \code{R/make-newdata.R}, \code{R/predict.R}, \code{R/cumulative-effect.R},
  \code{R/multi-state-helpers.R}). Provides the \code{add\_hazard},
  \code{add\_cumu\_hazard}, \code{add\_surv\_prob}, \code{add\_cif},
  \code{add\_trans\_prob}, and \code{add\_term} families. These take a fitted
  model and a \code{newdata} grid constructed via \code{make\_newdata} and
  return the original grid augmented with the requested quantities and (where
  appropriate) confidence intervals computed using one of the three approaches
  of Section~\ref{sub:inference}.
\item \emph{Visualization} (\code{R/ggplot-extensions.R},
  \code{R/gg\_state\_occupation.R}, \code{R/geom-hazard.R},
  \code{R/convenience-plots.R}). \code{ggplot2}-based helpers for hazard,
  survival, CIF and state-occupation displays, plus \code{tidy}/\code{glance}
  methods for \pkg{broom} integration.
\end{sloppypar}
\item \emph{Simulation} (\code{R/sim-pexp.R}, \code{R/rpexp.R}). Building
  blocks for simulating piecewise-exponential survival times with arbitrary
  time-varying and cumulative effects, used both in the package's own tests
  and for the ELRA examples in Section~\ref{sub:elra}.
\end{compactitem}

\subsection{The ped Class Hierarchy}
\label{app:ped-class}

The PED format is represented as a \code{data.frame} (technically a
\code{tibble}) carrying additional S3 classes and attributes that record the
information needed for post-processing:

\begin{compactitem}
\item \code{ped}: the base class for single-event PED data. Mandatory columns
  are a subject identifier, \code{tstart}, \code{tend}, \code{interval},
  \code{intlen}, and a binary status indicator $\delta_{ij}$. Attributes
  include \code{breaks} (the interval cut-points $\kappa_j$),
  \code{intvars}, \code{time_var}, \code{status_var}, \code{id_var}, and
  \code{trafo_args} (the call used to create the object, which makes the
  transformation reproducible if data changes).
\item \code{ped_cr} and \code{ped_cr_list}: competing-risk extensions. The
  former is a stacked long-format frame with a \code{cause} factor;
  the latter is a list of cause-specific \code{ped} objects, used when
  cause-specific hazards are fit separately. A \code{ped_cr_union} variant
  marks the union representation used internally for joint CIF predictions.
\item \code{fped} / \code{nested_fdf}: ``functional'' PED objects for
  cumulative-effect models. Time-varying covariates that enter through
  \code{cumulative()} are stored as list-columns of exposure histories and
  lag-lead matrices alongside the standard interval columns.
\item Multi-state PED data uses the base \code{ped} class with an additional
  \code{transition} column; transitions are dispatched downstream by the
  \code{add\_trans\_prob} family.
\end{compactitem}

Post-processing functions dispatch on these classes so that calls like
\code{add\_hazard(newdata, model)} produce the appropriate quantity for the
underlying data type without requiring the user to specify the task
explicitly.

\subsection{Dispatch of as-ped Across Survival Tasks}
\label{app:as-ped-dispatch}

\code{as\_ped} is an S3 generic with methods for \code{data.frame},
\code{nested_fdf}, \code{list}, \code{ped}, and \code{pamm}. The
\code{data.frame} method routes the call to one of three internal pipelines
based on inspection of its arguments:

\begin{compactenum}
\item If \code{transition} is supplied, the call is routed to
  \code{as_ped_multistate}, which expands the data by transition and
  interval and returns a \code{ped} with an additional \code{transition}
  column.
\item Otherwise, the number of distinct event types in the response is
  determined via \code{get\_event\_types}. If more than one non-censoring
  event type is found, \code{as_ped_cr} is called and returns a
  \code{ped_cr} or \code{ped_cr_list}, depending on the \code{combine}
  argument.
\item In all other cases (standard single-event or left-truncated data), the
  formula is rewritten via \code{get_ped_form} and the split is performed by
  \code{split_data}, which delegates to \code{survival::survSplit} for the
  actual interval expansion.
\end{compactenum}

Left truncation (delayed entry) is handled through the counting-process
formula syntax: a three-argument response \code{Surv(entry, exit, status)}
signals that subjects enter the risk set only at \code{entry} rather than at
time zero. \code{survSplit} understands this start--stop form, and
\code{split_data} drops all intervals lying entirely before a subject's entry
time, so that the subject contributes no rows---and hence no risk
time---to those intervals. For the interval in which a subject first becomes
at risk, the exposure offset is computed from the actual entry time,
$\log(t_{\text{exit}} - t_{\text{entry}})$, rather than from the interval
boundary, giving the correct at-risk duration in the piecewise-exponential
likelihood. As a result, a PAMM fit to data prepared this way reproduces the
partial-likelihood estimates of a Cox model for left-truncated data (see the
\code{left-truncation} package vignette for a numerical comparison with
\code{eha::coxreg}). Note that the counting-process syntax is intended for
creating left-truncated data and is not combined with the cumulative-effect
specials (\code{cumulative()}/\code{concurrent()}) of
Section~\ref{sub:elra}.

Cumulative-effect specifications are detected during this pre-processing
step: presence of \code{cumulative()} or \code{concurrent()} terms in the
formula triggers nesting of the relevant exposure history columns via the
\code{nested_fdf} method. The resulting object carries the lag-lead window
function (\code{ll_fun}) and integration weights as attributes that
\code{pamm} reads when constructing the model frame.

The dispatch is intentionally driven by the data rather than by a task
argument: the user calls \code{as_ped(data, formula)} regardless of whether
the underlying problem is single-event, competing risks, multi-state, or
involves cumulative effects.

\subsection{Performance and Scalability}
\label{app:performance}

The PED transformation expands a data set of $n$ subjects into roughly
$n \cdot \bar{J}$ rows, where $\bar{J}$ is the average number of intervals
each subject contributes. Three practical implications follow:

\begin{compactitem}
\item \emph{Choice of cut-points.} By default, \code{as\_ped} uses all unique
  observed event times as interval boundaries, which gives the finest grid
  but the largest data set. For large $n$, supplying a coarser grid via the
  \code{cut} argument (e.g.\ deciles of the event-time distribution, or a
  fixed-width grid) reduces memory and run time substantially with only a
  small loss in flexibility, since smooth baseline and time-varying effects
  are estimated by penalized splines instead of interval-wise, and usually do not
  change that much across short periods of time. Note that coarsening applies only to standard single-event data (\code{Surv} with two arguments); for counting-process data (\code{Surv} with three arguments, covering multi-state and left-truncated data), \code{as\_ped} always re-injects observed event times into the grid, so a user-supplied \code{cut} cannot reduce the grid below the set of observed event times.
\item \emph{Estimation back-end.} For PED data sets exceeding a few hundred
  thousand rows we recommend switching from \code{mgcv::gam} to
  \code{mgcv::bam}, which uses a discretized covariate representation and
  block-wise updates and scales much better in both $n$ and the number of
  smooth terms. \code{pamm} accepts an \code{engine} argument that toggles
  between the two.
\item \emph{Cumulative effects.} Models with \code{cumulative()} terms
  further expand the data by the number of exposure measurements per
  subject; an ELRA model on $n$ subjects with $Q$ exposure times produces on
  the order of $n \cdot \bar{J} \cdot Q$ design-matrix entries per
  cumulative term. Restricting the lag-lead window through \code{ll_fun}
  (Section~\ref{sub:elra}) is the most effective way to control data set sizes 
  for these models.
\end{compactitem}

For inference, the simulation-based approach (Section~\ref{sub:inference})
scales linearly in the number of posterior draws and the size of the
prediction grid. In our experience, $N = 200$ to $500$ draws keep the \emph{Monte-Carlo}
error of the interval bounds on CIFs, transition probabilities, and RMST
differences small relative to their width (for approximately Gaussian
functionals, the simulation error of the empirical $2.5\%/97.5\%$ quantiles
is then still roughly $6$--$10\%$ of the interval half-width), with larger
$N$ needed for extreme tail quantiles or when reported bounds must be
stable to more digits. This concerns only the
stability of the quantile estimates for a given posterior and is distinct
from the frequentist coverage of the intervals, which is a property of the
posterior itself (Section~\ref{sub:inference}). The simulation
loop is embarrassingly parallel and can be wrapped with \pkg{future} or
\pkg{furrr} if needed.

The package is covered by unit tests in \code{tests/testthat/}, which
exercise the data transformations, the \code{add\_*} family, and the
multi-state and cumulative-effect machinery. Numerical accuracy of derived
quantities (CIFs, transition probabilities) is verified against the
Aalen-Johansen estimator from \pkg{etm} and the analytical 
piecewise-exponential integrals used in the recursions of
Sections~\ref{sub:cr} and~\ref{sub:eha}.

\subsection{Reproducibility and Session Information}
\label{app:sessioninfo}

This article is a literate (\pkg{knitr}) document: every number, table, and
figure is produced by the embedded \Rlang{} code when the source
\code{pammtools-jss.Rnw} is compiled, so the manuscript and its results are
generated from a single source. The document is built with
\begin{verbatim}
Rscript -e "knitr::knit('pammtools-jss.Rnw', output='pammtools-jss.tex')"
latexmk -pdf -interaction=nonstopmode pammtools-jss.tex
\end{verbatim}
as documented in the repository \code{README}. A full compile from a cold
\pkg{knitr} cache takes on the order of a few minutes on a current laptop; the
most expensive steps are the interval-censoring multiple-imputation fit
(\code{pamm\_ic} with \code{m = 20}, \code{iter = 3}) and the large competing-risks
\code{bam} fit. Intermediate results are cached under \code{preprint-v4-cache/};
this cache is a build artifact and is not required for verification -- deleting
it and recompiling reproduces all results from scratch. The exact software
environment used to produce this version of the article is recorded below.

\begin{knitrout}\small
\definecolor{shadecolor}{rgb}{0.961, 0.961, 0.961}\color{fgcolor}\begin{kframe}
\begin{verbatim}
R version 4.6.1 (2026-06-24)
Platform: x86_64-pc-linux-gnu
Running under: Ubuntu 24.04.4 LTS

Matrix products: default
BLAS:   /usr/lib/x86_64-linux-gnu/blas/libblas.so.3.12.0 
LAPACK: /usr/lib/x86_64-linux-gnu/lapack/liblapack.so.3.12.0  LAPACK version 3.12.0

locale:
 [1] LC_CTYPE=en_US.UTF-8       LC_NUMERIC=C              
 [3] LC_TIME=de_DE.UTF-8        LC_COLLATE=en_US.UTF-8    
 [5] LC_MONETARY=de_DE.UTF-8    LC_MESSAGES=en_US.UTF-8   
 [7] LC_PAPER=de_DE.UTF-8       LC_NAME=C                 
 [9] LC_ADDRESS=C               LC_TELEPHONE=C            
[11] LC_MEASUREMENT=de_DE.UTF-8 LC_IDENTIFICATION=C       

time zone: Europe/Berlin
tzcode source: system (glibc)

attached base packages:
[1] stats     graphics  grDevices utils     datasets  methods   base     

other attached packages:
 [1] timereg_2.0.7    mstate_0.3.3     ggforce_0.5.0    kableExtra_1.4.0
 [5] etm_1.1.2        purrr_1.2.2      tidyr_1.3.2      dplyr_1.2.1     
 [9] pammtools_0.8.0  mgcv_1.9-4       nlme_3.1-170     survival_3.8-9  
[13] patchwork_1.3.2  ggplot2_4.0.3    knitr_1.51      

loaded via a namespace (and not attached):
 [1] gtable_0.3.6        xfun_0.57           lattice_0.22-9     
 [4] numDeriv_2016.8-1.1 vctrs_0.7.3         tools_4.6.1        
 [7] generics_0.1.4      parallel_4.6.1      tibble_3.3.1       
[10] highr_0.12          pkgconfig_2.0.3     Matrix_1.7-5       
[13] data.table_1.18.4   checkmate_2.3.4     RColorBrewer_1.1-3 
[16] S7_0.2.2            lifecycle_1.0.5     scam_1.2-22        
[19] compiler_4.6.1      farver_2.1.2        stringr_1.6.0      
[22] textshaping_1.0.5   codetools_0.2-20    htmltools_0.5.9    
[25] lazyeval_0.2.3      prodlim_2026.03.11  Formula_1.2-5      
[28] pillar_1.11.1       MASS_7.3-66         iterators_1.0.14   
[31] foreach_1.5.2       parallelly_1.47.0   lava_1.9.1         
[34] tidyselect_1.2.1    digest_0.6.39       mvtnorm_1.3-7      
[37] stringi_1.8.7       future_1.70.0       listenv_0.10.1     
[40] labeling_0.4.3      splines_4.6.1       polyclip_1.10-7    
[43] fastmap_1.2.0       grid_4.6.1          cli_3.6.6          
[46] magrittr_2.0.5      utf8_1.2.6          future.apply_1.20.2
[49] withr_3.0.2         scales_1.4.0        backports_1.5.1    
[52] rmarkdown_2.31      globals_0.19.1      otel_0.2.0         
[55] evaluate_1.0.5      pec_2025.06.24      viridisLite_0.4.3  
[58] rlang_1.2.0         Rcpp_1.1.1-1.1      glue_1.8.1         
[61] tweenr_2.0.3        xml2_1.5.2          svglite_2.2.2      
[64] rstudioapi_0.18.0   R6_2.6.1            systemfonts_1.3.2  
\end{verbatim}
\end{kframe}
\end{knitrout}


\end{document}